\documentstyle[epsf]{mn}

\def\DIRFIGS{FIGS}

\newcommand{\cidfig}[6]{
    \protect\centerline{
    \epsfxsize=#1\epsffile[#2 #3 #4 #5]{#6}
    }}

\def\ni{\noindent}                                       
\def\et{et\thinspace al.\ }                                 
 
\def\gapprox{_>\atop{^\sim}}       
 
\def\kms{km\thinspace s$^{-1}$}                        
 
\newcommand{\ET}[1]{\times 10^{#1}}             

\def\sec{$^{\prime\prime}$}
\def\arcmin{$^{\prime}$}

\def\flunit{erg$\,$cm$^{-2}\,$s$^{-1}\,$\AA$^{-1}$}    

\title[The Stellar Content of Active Galaxies]
      {The Stellar Content of Active Galaxies}

\author[R. Cid Fernandes, T. Storchi-Bergmann \& H. R. Schmitt]
       {Roberto Cid Fernandes Jr.$^{1\mbox{$\ddagger$}}$, 
        Thaisa Storchi Bergmann$^2$\thanks{Visiting Astronomer,
       Cerro Tololo Interamerican Observatory, operated by the
       Association of Universities for Research in Astronomy, Inc.,
       under contract with the National Science 
       Foundation.}$^{\mbox{$\ddagger$}}$\
        and 
        Henrique R. Schmitt$^2$\thanks{CNPq
        fellow.}\thanks{E-mail: 
          cid@fsc.ufsc.br (RCF); thaisa@if.ufrgs.br (TSB);
          schmitt@if.ufrgs.br (HRS).} \\
        $^1$ Departamento de F\'{\i}sica, CFM - UFSC, Campus
  Universit\'ario - Trindade, Caixa Postal: 476. CEP: 88040-900
  Florian\'opolis, SC, Brazil. \\ 
        $^2$ Instituto de F\'{\i}sica - UFRGS, Caixa Postal: 15051.
  CEP: 91501-970 Porto Alegre, RS, Brazil.}

\begin{document}
\maketitle

    \begin{abstract}

    We present the results of a long-slit spectroscopic study of 39
active and 3 normal galaxies. Stellar absorption features, continuum
colors and their radial variations are analyzed in an effort to
characterize the stellar population in these galaxies and detect the
presence of a featureless continuum underlying the starlight
spectral component. Spatial variations of the equivalent widths of
conspicuous absorption lines and continuum colors are detected in
most galaxies. Star-forming rings, in particular, leave clear
fingerprints in the equivalent widths and color profiles. We find
that the stellar populations in the inner regions of active galaxies
present a variety of characteristics, and cannot be represented by a
single starlight template. Dilution of the stellar lines by an
underlying featureless continuum is detected in most broad-lined
objects, but little or no dilution is found for the most of the 20
type 2 Seyferts in the sample. Color gradients are also ubiquitous.
In particular, all but one of the observed Seyfert 2s are redder at
the nucleus than in its immediate vicinity. Possible consequences of
these findings are briefly outlined.

    \end{abstract}


\section{Introduction}

\label{sec:Introduction}

Starlight constitutes a substantial fraction of the light gathered
in optical spectra of active galactic nuclei (AGN), particularly in
low luminosity objects such as LINERs and Seyfert galaxies, where
the stellar component is often dominant. Quantifying and removing
the contribution of the stellar population is one of the first and
most critical steps in the analysis of AGN spectra, as the shape of
the resulting continuum is strongly dependent on the adopted stellar
population spectrum.

The most widely used technique for starlight subtraction consists of
using an appropriately chosen spectrum of a normal galaxy as a
template for the stellar component. The template spectrum is scaled
to match as closely as possible the stellar absorption features in
the observed spectrum, and the residual after its subtraction is
taken as the stellar-free, pure AGN spectrum. This technique has
been extensively applied since it was first introduced by Koski
(1978), in his early study of Seyfert 2s (e.g., Stauffer 1982,
Phillips, Charles \& Baldwin 1983, Malkan \& Filippenko 1983,
Filippenko \& Sargent 1988, 1983, Miller \& Goodrich 1990, Ho,
Filippenko \& Sargent 1993, Kay 1994, Tran 1995a). Alternatively,
the shape and strength of the starlight component can be estimated
by means of stellar population-synthesis techniques. This approach
was followed by Keel (1983), who built synthetic templates using a
library of stellar spectra plus assumptions about the mass function
and star-formation history. Synthetic templates can also be obtained
combining a spectral library of star-clusters, a method first used
by Bica (1988) in his analysis of the stellar populations of normal
galaxies. Bonatto, Bica \& Alloin (1989), Storchi-Bergmann, Bica \&
Pastoriza (1990) used this technique to determine the integrated
stellar content of the bulge of active galaxies.

The starlight evaluation procedure yields the stellar and AGN fluxes
as a function of wavelength. The AGN spectrum, of course, is not
known {\it a priori}, but the starlight subtraction is usually done
in a way to produce a smooth residual continuum. This assumption is
based on the fact that the continuum is essentially featureless in
objects like bright Seyfert 1s and QSOs, where the AGN component
dominates. It is customary to characterize the relative intensity of
the starlight and featureless continuum (hereafter FC) components by
their fractions ($f_\star$\ and $f_{{\rm FC}}$) of the total flux at
a given wavelength. These fractions depend on the chosen template,
the aperture and the actual contrast between the AGN and stellar
spectral components. The effect of the FC is to {\it dilute} the
stellar absorption lines in the nuclear spectrum, and it is
precisely the degree of dilution with respect to the starlight
template which determines $f_\star$\ and $f_{{\rm FC}}$.

In this paper we explore a different approach to estimate the
dilution of absorption lines caused by the presence of a FC. Our
method makes use of the radial information available through high
signal to noise long-slit spectroscopy of a large sample of objects.
By studying the variation of the equivalent width (W) of conspicuous
absorption lines as a function of distance to the nucleus, we are
able to determine how much dilution (if any) occurs in the nucleus
with respect to extra-nuclear regions, presumably not affected by
the FC. In a way, this method amounts to using the off-nucleus
spectrum of a galaxy as its own starlight template, a starlight
evaluation technique which has been a sucessifully applied in some
previous works (Fosbury \et 1978, Schmitt, Storchi-Bergmann \&
Baldwin 1994, Winge \et 1995, Storchi-Bergmann \et 1997). Continuum
color gradients are also examined, complementing the study of
absorption line gradients and bringing in information about the
reddening of the nuclear regions. 

A second goal of this paper is to provide an homogeneous database
for the study of the stellar populations in active galaxies. As
discussed above, in previous works the interest in the stellar
population has focused in removing its contribution from the nuclear
spectrum. The Ws and continuum colors measured in this work allow a
characterization of the global properties of the stellar populations
as close as a few hundred of parsecs from the nucleus, offering the
prospect of a comparative study between the stellar content of
normal and active galaxies and between different types of AGN. 

This investigation touches upon a number of key issues in current
AGN research, such as: (1) the question of starlight evaluation and
subtraction techniques; (2) the characterization of the stellar
population in active galaxies; and (3) the nature of the FC in
Seyfert 2 galaxies and its implications for unified models. In the
current paper we focus on the presentation of the data, the analysis
of the stellar population features and their radial variations.
Future communications will explore the consequences and
interpretation of our results in full detail.

This paper is organized as follows. In \S\ref{sec:Observations} we
present the sample galaxies and describe the observations. The
method of analysis (continuum determination and W measurement)  is
discussed in \S\ref{sec:Method}. The results are presented in
\S\ref{sec:Results} and discussed in \S\ref{sec:Discussion}.
Finally, \S\ref{sec:Conclusions} summarizes our main results.

\section{Observations}

\label{sec:Observations}

The sample comprises 20 Seyfert 2s, 6 Seyfert 1s, 7 LINERs and 5
radio galaxies, as well as 4 `normal' nuclei. With the exception of
the radio galaxies, all of them are bright nearby objects.
Table~\ref{tab:sample} describes the sample properties.  Long-slit
spectra of these galaxies have been collected in three epochs using
the same technique and instrumentation such that we have formed an
homogeneous database, which can be used for a comparative study of
the different kinds of galaxies.

The observations were carried out using the Cassegrain Spectrograph
with the 4m telescope of the Cerro Tololo Inter-American
Observatory. The spectral range covered is $\sim 3500$--7000 \AA, at
a resolution of 5--8 \AA. A slit width corresponding to 2\sec\ in
the sky was oriented along the position angle (hereafter PA) which
allowed the best coverage of the extended emission (although in this
work we concentrate on the properties of the continua and stellar
absorption features). A log of the observations is presented in
Table 2, where we also list the parallactic angle, air mass during
the observations, pixel size and spatial scale (evaluated with H$_0
= 75$~\kms$\,$Mpc$^{-1}$). It can be noticed that, only for a few
galaxies, namely 3C33, CGCG420-015 and NGC~4303, the effect of
differential refraction is important, displacing the light at 3700
\AA\ by $\sim 1$\sec\ relative to the light at 5000 \AA, so that
some distortion of the continuum slope can be expected in the
individual extractions for these objects, if the continuum presents
significant variations along the slit. The reduction of the spectra
was performed using standard techniques in IRAF. After flux and
wavelength calibration, the redshifts were determined from the
narrow emission lines present and the spectra were converted to
rest-frame units. 

The long-slit used corresponded to 5\arcmin\ in the sky, and sky
subtraction was performed by the fit of a polynomium to the outer
regions of the frame and then subtracting its contribution to the
hole frame using the task background in IRAF.  The most extented
galaxies, presenting enough S/N ratio to allow measure the spectral
features up to approximately 100\sec, were NGC1097, the most
extended, followed by NGC1672. For these galaxies the sky level
adopted may have been overestimated. In order to evaluate an upper
limit to the effect, we have compared the flux level of the adopted
sky to the flux of the inner regions. We have concluded that its
value amounts to less than one percent when compared with the flux
of the nucleus, but rising with distance from the nucleus. At
approximately 25\sec, its contribution is about 10\%. Assuming that
the `sky' flux is actually galaxy flux, the flux level of the
sky-subtracted spectrum will be underestimated by at most 10 percent
at 25\sec\ and more outwards. Thus the measurements of these two
galaxies are subjected to this uncertainty farther than 25\sec\ from
the nucleus. The effect should be negligible for the other less
extended galaxies.

\begin{table*}
\begin{centering}
\begin{tabular}{llrrrlrrr}
\multicolumn{8}{c}{ SAMPLE PROPERTIES } \\ \hline 
%
NAME & ALT. NAME & $\alpha$\ (1950) & $\delta$\ (1950) & TYPE & MORPH. & $v$ & B$_{To}$&E(B-V)$_G$ \\ \hline
Mrk~348  &NGC~262     &00 46 05&31 41 04&2&SA(s)0$/$a&4669&13.94&0.060 \\
3~C~33	&Pks~0106+13 &01 06 12&13 02 33&NLRG&E&17230&19.50&0.025 \\
NGC~526a&ESO~352-IG66&01 21 37&-35 19 32&1.9&S0 pec&5762&14.50&0.000 \\
Mrk~573	&UGC~1214    &01 41 23&02 05 56&2&SAB(rs)0$^+$&5161&13.57&0.008 \\
IC~1816	&ESO~355-G25 &02 29 48&-36 53 29&2&SA:(r:)a&5086&13.66&0.000 \\
NGC~1097&ESO~416-G20 &02 44 11&-30 29 01&Lin-1&SB(r'l)b&1193&9.92&0.020 \\
ESO~417-G6&MCG-05-08-006&02 54 18&-32 23 00&2&(R)SA0$/$a&4792&14.15&0.000 \\
NGC~1326&ESO~357-G026&03 22 01&-36 38 24&Lin&SB(rl)0$/$a&1244&11.25&0.000 \\
Mrk~607	&NGC~1320    &03 22 18&-03 13 03&2&Sa:sp&2716&13.31&0.018 \\
NGC~1358&MCG-1-10-003&03 31 11&-05 15 24&2&SAB(r)0$/$a&3980&12.70&0.025 \\
NGC~1386&ESO~358-G35 &03 34 51&-36 09 47&2&SB(s)0$^+$&741&12.12&0.000 \\
NGC~1433&ESO~249-G014&03 40 27&-47 22 48&Lin&SB(rs)ab&920&11.37&0.000 \\
Pks~0349-27  &GSP~022 &03 49 32&-27 53 29&NLRG&E&19190&16.80&0.000 \\
NGC~1598&ESO~202-G26 &04 27 08&-47 53 29&Lin&SAB(s)c&4939&13.44&0.000 \\
NGC~1672&ESO~118-G043&04 44 55&-59 20 18&Lin&SB(r)bc&1155&10.25&0.000 \\
CGCG~420-015&        &04 50 47&03 58 47&2&E$/$S0&8811&15.00&0.070 \\
ESO~362-G8 &MCG-06-12-009&05 09 20&-34 27 12&2&S0?&4616&13.51&0.005 \\
ESO~362-G18 &MCG-05-13-017&05 17 44&-32 42 30&1&S0$/$a&3603&13.58&0.000 \\
PICTOR A &ESO~252-GA018&05 18 24&-45 49 43&BLRG&SA0$^0$:pec&10308&15.94&0.225 \\
Pks~0634-20 &IRAS~06343-2032&06 34 24&-20 32 19&NLRG&E&16320&17.58&0.405 \\
Pks~0745-19 &                    &07 45 17&-19 10 15&NLRG&E&35940&18.00&--- \\
Mrk~1210&UGC~4203&08 01 27&05 15 22&2&S?&3910&14.21&0.018 \\
NGC~2935&ESO~565-G23&09 34 26&-20 54 12&Normal&SAB(s)b&2072&11.81&0.028 \\
NGC~3081&ESO~499-IG31&09 57 10&-22 35 06&2&SAB(r)0$/$a&2164&12.59&0.033 \\
Mrk~732	&IC~2637&11 11 14& 09 51 33&1.5&E pec&8670&13.89&0.005 \\
IRAS~11215-2806&        &11 21 35&-28 96 46&2&&4047&13.00&0.088 \\
MCG-05-27-013&         &11 24 55&-28 59 00&2&SB(r)a?&7263&13.71&0.063 \\
NGC~4303&UGC~7420&12 19 22&04 45 03&Lin&SAB(rs)bc&1486&13.68&0.000 \\
MCG-02-33-034&         &12 49 35&-13 08 36&1&&4386&15.00&0.015 \\
FAIRALL~316&ESO~269-G12&12 53 50&-46 39 18&2&S0?&4772&13.81&0.185 \\
NGC~5135&ESO~444-G32&13 22 57&-29 34 26&2-SB&SB(l)ab&3959&12.37&0.058 \\
NGC~5248	&UGC~8616&13 35 02&09 08 23&Normal&SAB(rs)bc&1128&10.63&0.000 \\
NGC~5643	&ESO~272-G16&14 29 28& -43 57 12&2&SAB(rs)c&1066&10.23&0.125 \\
NGC~6221	&ESO~138-G03&16 48 26&-59 08 00&SB&SB(s)bc pec&1368&9.77&0.220\\
NGC~6300	&ESO~101-G25&17 12 18&-62 45 54&2&SB(rs)b&997&10.20&0.120 \\
NGC~6814&MCG-02-50-001&19 39 56&-10 26 33&1&SAB(rs)bc&1676&10.96&0.150 \\
IC~4889&ESO~185-G14&19 41 18&-54 27 54&Normal&E&2473&12.06&0.045 \\
NGC~6860&ESO~143-G09&20 04 29&-61 14 42&1&SB(r)ab&4385&13.68&0.020 \\
NGC~6890	&MCG-07-41-023&20 14 49&-44 57 48&2&SB(r)ab&2459&12.82&0.008 \\
NGC~7130	&IC~5135&21 45 20&-35 11 07&2-SB&Sa pec&4850&12.88&0.000 \\
NGC~7213&ESO~288-G43&22 06 08&-47 24 45&Lin-1&SA(s)0$^0$&1767&11.13&0.000 \\
NGC~7582&ESO~291-G16&23 15 38&-42 38 36&2-SB&SB(s)ab&1551&10.83&0.000 \\ \hline
\end{tabular}
    \caption{Column (5) gives the activity class of the galaxy, where SB
means Starburst. Column (7) gives the radial velocity relative to the local 
group (in \kms), while column (8) lists the total blue magnitude of the
galaxies and column (9) contains the foreground galactic value of E(B-V) 
(extracted from the Nasa Extragalactic Database).
    \label{tab:sample}
    }
\end{centering}
\end{table*}

\begin{table*}
\begin{centering}
\begin{tabular}{llrrrrrrr}
\multicolumn{9}{c}{ LOG OF OBSERVATIONS } \\ \hline 
NAME & DATE & EXP. TIME & PA [$^\circ$] & AIR  & $\phi$\ [$^\circ$] & 
  PIXEL & SCALE & COMMENTS \\ 
     &      & (sec)     &               & MASS & & & pc/arcsec & \\ \hline
Mrk~348	&6/7 Dec. 94 &1800 &170 &2.20 &170 &0.9    &302&  \\
3~C~33	&6/7 Dec. 94 &1800 &65  &1.43 &167 &0.9    &1114& \\
NGC~526a &5/6 Dec. 94 &3600 &124 &1.02 &69  &0.9   &372&  \\
Mrk~573	&6/7 Dec. 94 &1800 &125 &1.64 &131 &0.9    &334&  \\
IC~1816	&6/7 Jan. 94 &1800 &90  &1.07 &82  &1      &328& \\
NGC~1097 &6/7 Jan. 94 &1800 &139 &1.01 &88 &1      &77& \\
ESO~417-G6&6/7 Dec. 94 &3600 &155 &1.01 &69 &0.9   &310& \\
NGC~1326 &5/6 Jan. 94 &1800 &77 &1.15 &88 &1       &80& \\
Mrk~607	&6/7 Jan. 94 &900 &135 &1.22 &140 &1       &176& \\
NGC~1358 &6/7 Jan. 94 &900 &145 &1.23 &136 &1      &257& \\
NGC~1386 &6/7 Jan. 94 &1800 &169 &1.12 &89 &1      &48& \\
NGC~1433 &6/7 Dec. 94 &1200 &19  &1.05 &19 &0.9    &59& \\
Pks~0349-27 &5/6 Dec. 94 &3600 &82 &1.04 &103 &0.9 &1240& \\
NGC~1598 &6/7 Jan 94 &1800 &123 &1.13 &65 &1       &319& \\
NGC~1672 &5/6 Jan 94 &1800 &94  &1.32 &63 &1       &75& \\
CGCG~420-015 &6/7 Jan 94 &900 &40 &1.46 &140 &1    &570& \\
ESO~362-G8 &6/7 Jan. 94 &1800 &165 &1.17 &88 &1    &298& \\
ESO~362-G18  &6/7 Jan. 94 &1800 &55 &1.19 &82 &1   &233& \\
PICTOR A&6/7 Dec. 94 &1800 &71 &1.15 &71 &0.9      &666& \\
Pks~0634-20  &5/6 Dec. 94 &3600 &139 &1.07 &129 &0.9 &1055& \\
Pks~0745-19  &6/7 Dec. 94 &1800 &150 &1.03 &149 &0.9 &2323& \\
Mrk~1210 &5/6 Jan. 94 &1800 &163 &1.25 &163 &1     &253& \\
NGC~2935 &5/6 Jan. 94 &1800 &0   &1.02 &182 &1     &134& \\
NGC~3081 &28/29 May 92 &1200 &73 &1.25 &70 &0.9    &140& seeing 3$^{\prime\prime}$ \\
Mrk~732	 &6/7 Jan. 94 &1800 &55 &1.50 &31 &1       &560& \\
IRAS~11215-2806 &5/6 Jan. 94 &900 &145 &1.02 &76 &1 &262& \\
MCG-05-27-013 &6/7 Jan. 94 &1800 &0 &1.20 &78 &1   &470& \\
NGC~4303 &28/29 May 92 &1800 &112 &1.69 &50 &0.9   &96& \\
MCG-02-33-034 &6/7 Jan. 94 &1800 &25 &1.24 &55 &1  &284& \\
FAIRALL~316 &6/7 Jan. 94 &1800  &100 &1.29 &102 &1 &308& \\
NGC~5135 &29/30 May 92 &1800 &30 &1.25 &100 &0.9   &256& \\
NGC~5248 &28/29 May 92 &1800 &146 &1.58 &144 &0.9  &73& \\
NGC~5643 &28/29 May 92 &1800 &90 &1.19 &75 &0.9    &69& \\
NGC~6221 &28/29 May 92 &1800 &5  &1.24 &56 &0.9    &88& clouds \\
NGC~6300 &28/29 May 92 &1800 &124 &1.19 &18 &0.9   &63& clouds \\
NGC~6814 &29/30 May 92 &1800 &161 &1.06 &170 &0.9  &108& clouds \\
IC~4889  &28/29 May 92 &1800 &0 &1.12 &29 &0.9     &160& clouds \\
NGC~6860 &28/29 May 92 &1800 &70 &1.21 &36 &0.9    &283& clouds \\
NGC~6890 &29/30 May 92 &1800 &153 &1.03 &12 &0.9   &159& clouds \\
NGC~7130 &29/30 May 92 &900 &143 &1.01 &50 &0.9    &314& \\
NGC~7213 &29/30 May 92 &1200 &50 &1.04 &0 &0.9     &114& \\
NGC~7582 &6/7  Jan. 94 &300 &67 &1.53 &83 &1       &100& \\ \hline
\end{tabular}
    \caption{
    Details of the observations. Column (4) lists the slit position 
angle (PA). The parallactic angle ($\phi$) is listed in column (6).
Column (7) lists the pixel size in arcsec, while column (8) lists
the spatial scale in pc arcsec$^{-1}$.
    \label{tab:log_obs}
    }
\end{centering}
\end{table*}

One-dimensional spectra were extracted from the two-dimensional frames.
The number of extractions varied from 4 for PKS0745-19 to as many as 26
for NGC 1097, totaling 491 spectra for the whole sample. The spatial
coverage ranged between $\pm$3 and $\pm$80\sec\ from the nucleus, while
the width of the extractions was typically 2\sec\ in the brighter
nuclear regions, gradually increasing towards the fainter outer regions
to guarantee enough signal. Given that the seeing during the
observations was typically 1.5\sec, some of the narrow extractions are
slightly oversampled, resulting in a smoothing of the sharpest spectral
variations.

The S/N ratio at 5650~\AA\ ranges from $\sim$\ 5 to 100, with an
average of 40 for all extractions. Even in the blue region, where the
noise is usually higher, the spectra are of excellent quality, with a
mean S/N ratio of 25 at 4200~\AA.  The high quality of our data has
allowed the measurement of stellar population features in spectra
extracted up to several kpc's from the nuclei. There is no similar
database with this quality for so many galaxies in the literature.

\begin{figure*}
    \cidfig{17.8cm}{55}{175}{575}{710}{\DIRFIGS/FIG1.ps}
    \caption{
    Selected examples of the spatially resolved spectra of four of
the sample galaxies. The nuclear spectra are indicated by thick
lines. The angular distances from the nuclei are indicated, positive
values corresponding to West (W, NW or SW) or South direction (in
the case when the observed PA is 0$^{\circ}$). The spectra are
normalized at 5870~\AA\ and shifted vertically by one unit to
facilitate comparison. Residual sky emission lines have been ticked
out.
    }
    \label{fig:example_spectra}
\end{figure*}

Fig.~\ref{fig:example_spectra} shows representative spectra of four
of the sample galaxies for several positions along the slit.  Notice
that absorption features are well detected in all spectra, as well
as variations of spectral characteristics with distance from the
nuclei. Signatures of young stellar populations, in particular, are
clearly present in several of the spectra shown:  over the whole
inner 8\sec\ of Mrk~732, more conspicuously at 4\sec\ NE ($r =
-4$\sec\ in Fig. 1); at $\sim 8$\sec\ from the nucleus in IC~1816,
and at $\sim 8$--12\sec\ in NGC~1097 (corresponding to the locus of
the nuclear ring; Storchi-Bergmann, Wilson \& Baldwin 1996). 
Further examples of the spatially resolved spectra are shown in
Fig.~\ref{fig:ZOOMED_spectra}.

\section{Method of analysis}

\label{sec:Method}

Our main goal is to study the run of the stellar absorption features
with distance from the nucleus. To this end, the analysis of each of
the sample spectra consisted of (1) determining the
`pseudo'-continuum in selected pivot-points (3780, 4020, 4510, 4630,
5313, 5870 and 6080 \AA) and (2) measuring the equivalent widths of
the CaII K and H, CN, G-band, MgI$+$MgH and NaI absorption lines
integrating the flux with respect to the pseudo continuum.

The continuum and W measurements followed the method outlined by
Bica \& Alloin (1986a,b), Bica (1988) and Bica, Alloin \& Schmitt
(1994), and subsequently used in several studies of both normal and
emission line galaxies (e.g., Bonatto, Bica \& Alloin 1989,
Storchi-Bergmann, Bica \& Pastoriza 1990, Jablonka, Alloin \& Bica
1990, Storchi-Bergmann, Kinney \& Challis 1995b, McQuade, Calzetti
\& Kinney 1995). This method consists of determining a
pseudo-continuum in a few pivot-wavelengths and integrating the flux
difference with respect to this continuum in pre-defined wavelength
windows (Table~\ref{tab:lambda_windows}) to determine the Ws. The
pivot wavelengths used in this work are based on those used by the
above authors and were chosen to avoid, inasmuch as possible,
regions of strong emission or absorption features. The use of a
compatible set of pivot points and wavelength windows is important
because it will allow a detailed quantitative analysis of the
stellar population---via synthesis techniques using Bica's (1988)
spectral library of star clusters---to be presented in a forthcoming
paper. Here we will use our measurements only to look for broad
trends in the properties of the stellar populations, and to identify
signs of a nuclear FC (see below).

\begin{table}
\begin{centering}
\begin{tabular}[h]{lr}
\multicolumn{2}{c}{ LINE WINDOWS } \\ \hline 
CaII K                & 3908--3952 \\
CaII H$+$H$\epsilon$  & 3952--3988 \\
CN                    & 4150--4214 \\
G band                & 4284--4318 \\
MgI$+$MgH             & 5156--5196 \\
NaI                   & 5880--5914 \\ \hline
\end{tabular}
    \caption{
    Wavelength windows used to measure the equivalent widths of the
absorption lines.
    \label{tab:lambda_windows}
    }
\end{centering}
\end{table}

The determination of the continuum has to be done interactively,
taking into account the flux level, the noise and minor wavelength
calibration uncertainties as well as anomalies due to the presence
of emission lines. The 5870 and 6080 \AA\ points, in particular, are
sometimes buried underneath HeI~5876 and [Fe~VII]6087~\AA\ emission
lines (for instance, in the nuclear regions of IC 1816, Mrk 348 and
NGC 3081). In such cases the placement of the continuum was guided
by adjacent wavelength regions. In a few cases, it was also
necessary to make `cosmetic' corrections when a noise spike was
present in the wavelength window used to compute the W. We
emphasize, however, that such anomalies were very rare. In the
majority ($> 90$\%) of cases the excellent quality of the spectra
allowed a precise determination of the continua and Ws.

\begin{figure}
    \cidfig{8.5cm}{60}{175}{395}{705}{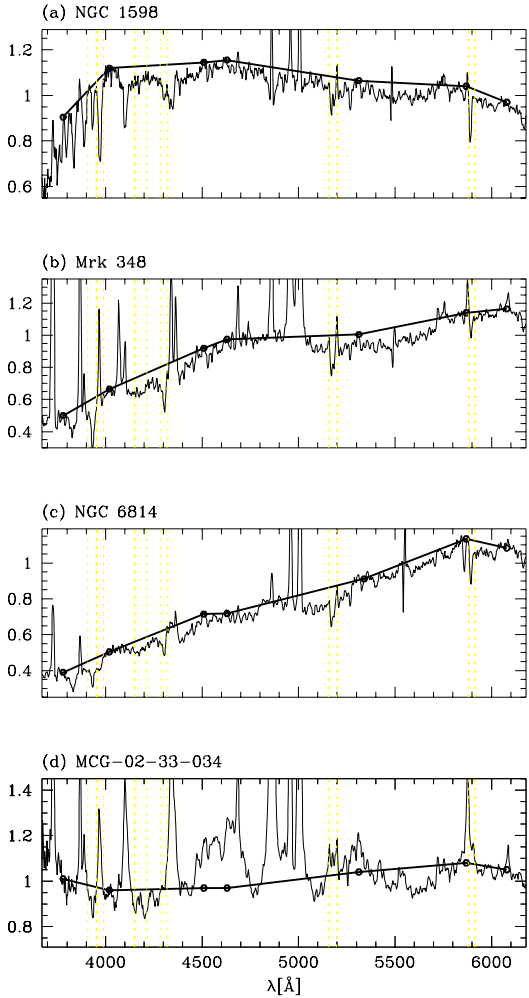}
    \caption{
    Illustration of the continuum determination procedure for the
nuclear spectra of the LINER NGC~1598 (a), the Seyfert 2 Mrk~348 (b),
and the Seyfert 1 galaxies NGC 6814 (c) and MCG~-02-33-034 (d).
The continuum pivot points are
marked by the filled circles. Dashed vertical lines indicate the
wavelength windows used to measure the equivalent widths of Ca~K and
H, CN, G band, Mg and Na lines.
    }
    \label{fig:example_method}
\end{figure}

Fig.~\ref{fig:example_method} illustrates the application of the
method to four of the sample spectra. The spectrum in
Fig.~\ref{fig:example_method}a corresponds to the nucleus of the
LINER NGC 1598. The spectrum is almost purely stellar, posing no
difficulties to the determination of the pseudo continuum. Note that
the continuum runs well above the actual spectrum except in the
pivot points. Fig.~\ref{fig:example_method}b shows the nuclear
spectrum of Mrk~348, a type 2 Seyfert. Here, as for all Seyfert 2s
in the sample, the continuum can still be placed with little
ambiguity, despite the presence of many emission lines. As in many
other cases, the Ca~H line is filled with [NeIII] and H$\epsilon$\
emission in the central regions, but not in outer positions, an
effect which must be taken into account when interpreting W(Ca~H)
spatial gradients. Note also that, as in
Fig.~\ref{fig:example_method}d, the 5870 and 6080~\AA\ points are
underneath HeI and [Fe~VII] emission, respectively. This is not a
problem, as the continuum can still be determined from the adjacent
regions.

Whilst for most LINERs, Seyfert 2s and normal galaxies the placement
of the continuum is straightforward, this is not the case in the
nuclear regions of strong Seyfert 1s, where the numerous broad lines
and intense non-stellar continuum complicate the analysis. This is
illustrated in Fig.~\ref{fig:example_method}d, where we plot the
nuclear spectrum of MCG~-02-33-034. The 4510, 4630 and 5313~\AA\
points, in particular, are all underneath Fe~II blends, making it
impossible to determine them accurately. None of the absorption
lines in Fig.~\ref{fig:example_method}d, with the possible exception
of Ca~K is free of contamination. The Mg line, for instance, is
filled up by FeII and [NI]5200~\AA\ emission, while the Na window is
contamined by broad HeI 5876~\AA. It is clear that in extreme cases
like this no reliable value for the Ws of the absorption lines can
be derived with our or any other method. The presence of such a
strong nuclear FC should appear as a dip at $r = 0$\ in the radial
profiles of the Ws, and this is indeed seen in Fig.~\ref{fig:i309}.
However, given the difficulties in defining a continuum, the depth
of this dip is unreliable. At any rate, the only other galaxy in the
present sample with a spectrum as complex as MCG~-02-33-034 is the
Broad Line Radio Galaxy Pictor A (Fig.~\ref{fig:pic}). For less
extreme Seyfert 1s like NGC 6814 (Fig.~\ref{fig:example_method}c),
the continuum can be defined with little ambiguity.

The four examples in Fig.~\ref{fig:example_method} span essentially
all types of spectra found in the sample. Overall, we found that the
method works well for most spectra, with the exception of the
nuclear regions of strong Seyfert 1s. The comparison of the Ws and
continuum colors in the nuclear and outer regions is a useful tool
to detect the presence of a FC, particularly in Seyfert 2s and
LINERs, where the FC, if present, is not as conspicuous as in type 1
objects. Indeed, one of our major goals is to verify whether Seyfert
2s and LINERs exhibit signs of such a nuclear FC, which should cause
a dilution of the absorption lines with respect to those measured
outside the nucleus.

\subsection{Consistency and Error Analysis}

\label{sec:Error}

The positioning of the continuum can be somewhat subjective,
particularly in the noisier spectra, given that it has to be carried
out interactively. In order to verify whether our measurements are
consistent with those of Bica (1988) we have measured the continuum
fluxes and Ws of the 15 templates corresponding to his spectral
groups E1--E8 and S1--S7. The difference between his and our
continuum measurements is of the order of 1\%. For the Ws the
difference is typically $\approx 0.5$~\AA\ and always smaller than
1~\AA, corresponding to an agreement at a 5--10\% level. This
compatibility is important, since it insures that we can apply
stellar population characterization techniques similar to those
employed by Bica.

In the discussion of our results (Section \ref{sec:Results}) we will
compare the Ws and continua of the galaxies in the present sample
with templates S1 to S7 and E1 to E8 of Bica (1988) as a guide to 
obtain a broad characterization of the stellar populations. (As
pointed out in the previous section, a full quantitative analysis of
the stellar populations will be presented in a forthcoming paper.)
These templates correspond to increasing contributions of young
components, from S1 and E1, to S7 and E8. Templates S1 to S3 and E1
to E6 are composed of old ($> 5\ET{9}$\ yr) and intermediate age
stars ($5\ET{8}$--$10^9$\ yr), differing mostly on the contribution
from stars of different metallicities. From S4 to S7 and E7 to E8 we
have an increasing contribution from young stars ($< 10^8$\ yr),
which can reach as much as 65\% for S7 and 20\% for E8 of the flux
at 5870~\AA.

The Ws measured in this work are subjected to two different sources
of error: (1) the noise present in the spectra, and (2) the
uncertainty in the placement of the continuum. 

The error in the Ws due to noise was evaluated using standard
propagation of errors. The noise level was computed from the rms
dispersion in a 60--100~\AA\ region free of strong emission or
absorption lines. Two windows, one in the 4200~\AA\ region and the
other around 5650~\AA, were used for this purpose. The noise in the
first window was used to compute the errors in the Ws of Ca~K and H,
CN and G-band, whereas the latter window was used for the Mg and Na
lines. The resulting errors are typically 0.5~\AA\ for the `blue'
lines (Ca~K to G-band), and 0.3~\AA\ for Mg and Na. The errors in
the central regions are usually smaller than these average values,
since the noise is lower there. The Ca~K line in NGC~7213, for
instance, has an error of $\sim 0.3$~\AA\ in the central 5\sec,
gradually increasing to $\sim 0.7$~\AA\ for $r > 22$\sec.

The uncertainties in the placement of the continuum are less
straightforward to evaluate, since, as explained above, the
continuum is not automatically defined, but determined
interactively. To compute the error in the Ws associated with this
source of uncertainty we have re-analyzed the spectra of 4 galaxies
(42 extractions in total), defining upper and lower values of the
continuum at the pivot-points. The mean difference between these
values and the adopted one (all evaluated in the middle of the
line-window), was taken as a measure of the error in the continuum
flux, which was then propagated to the Ws. We found that the
resulting errors are typically 30--50\% larger than the errors due
to noise alone. As expected, the uncertainty in the positioning of
the continuum increases with decreasing S/N. Since it is not
practical to measure lower and upper continua for all 491 spectra,
we have used this scaling between the S/N ratio and the continuum
errors measured above to estimate the errors due to the positioning
of the continuum for the remaining objects.

The combined error in the Ws due to noise and continuum positioning,
added in quadrature, are typically 0.5~\AA\ for Ca K, Ca H and the
G-band, 0.4~\AA\ in Mg and Na, and 1.0~\AA\ for the CN line. The
larger error in the CN reflects the fact that this feature is very
shallow, being essentially a measure of the local height of the
pseudo continuum above the actual spectrum. In fact, it is often
difficult to identify the CN feature in the spectra (e.g.,
Fig.~\ref{fig:example_method}). Indeed, the CN band is only
pronounced in very metal rich populations, such as templates S1 and
E1 of Bica (1988). We nevertheless keep this line in our analysis,
since our measured Ws are consistent with those found by Bica and in
most cases its radial variations follow those of the other
absorption lines. 

In the discussion of our results, we shall put more emphasis on the
Ca~K, G-band and Mg features, since the Ca~H line is often
contaminated by emission and the CN line is subjected to larger
errors. The Na line is not a good diagnostic of stellar population
since it is partially produced in the interstellar medium of the
host galaxy. This line can, however, be used as an indication of the
presence of dust (Bica \& Alloin 1986, Bica \et 1991). In the
atribution of template types to the different spectra a larger
weight is given to Ca~K, since this is often the strongest line,
less affected by errors. As discussed in section~\ref{sec:Indiv},
sometimes different lines indicate different template types, which
simply indicates a different mixture of stellar populations than
found in Bica's (1988) grid of spectral templates.

The uncertainty in the determination of the Ws limits the degree of
dilution measurable from our data. The effect of a FC contributing a
fraction $f_{{\rm FC}}$\ of the total flux in a given wavelength is
to decrease the W of an absorption line in this wavelength by this
same factor: $\Delta {\rm W} = f_{{\rm FC}} {\rm W}$. Therefore, the
minimum amount of dilution that can be safely measured from the
radial variations of a line with an W of, say, 10 \AA\ and an error
of 0.5 \AA\ is $\sim 5$\%. From the results presented in the next
section we estimate that this method of evaluating the contribution
of a FC works well for $f_{{\rm FC}}$ $\gapprox$ 10\% for the
present data set as a whole, though smaller lower limits are reached
in the best spectra. 

\section{Results}

\label{sec:Results}

The main results of the data analysis are presented in Figs.~3--44,
where we plot the radial variations of the Ws, colors and surface
brightness for all the galaxies in the sample. The upper panel shows
the Ws of Ca~K (solid line) and Ca~H (dashed) against the distance 
from the nucleus ($r$). Similarly, the second panel shows the Ws of
the G-band (solid) and CN (dashed), while the third panel shows Mg
(solid) and Na (dashed). Negative values of the W indicate emission,
as often found in the Ca~H plots due to contamination by [NeIII] and
H$\epsilon$. The Mg and Na lines are also contaminated by emission
in some cases. Apart from such cases (discussed in detail in
Section~\ref{sec:Indiv}), a dip in the W plots indicates dilution
of the absorption lines by a FC or the presence of a young stellar
population. The fourth panel shows the color profiles, mapped by the
ratios of the 5870 to 4020~\AA\ (solid line) and 4630 to 3780~\AA\
(dashed line) pseudo-continuum fluxes. (Note that the left hand
scale in this plot corresponds to $F_{5870} / F_{4020}$, whereas the
values in right correspond to $F_{4630} / F_{3780}$.) Dips in the
color plots indicate a blue continuum, whereas peaks indicate a red
continuum, due to an old stellar population and/or reddening. The
colors have been corrected by foreground galactic reddening using
the values of E(B-V) in Table~\ref{tab:sample}. This correction,
besides being small, does not affect the shape of the curves. 
Finally, to give a measure of the flux level in each position the
bottom panels show the run of the surface brightness in the
5870~\AA\  pseudo continuum with distance from the nucleus.

\begin{figure}
    \cidfig{8cm}{30}{145}{515}{700}{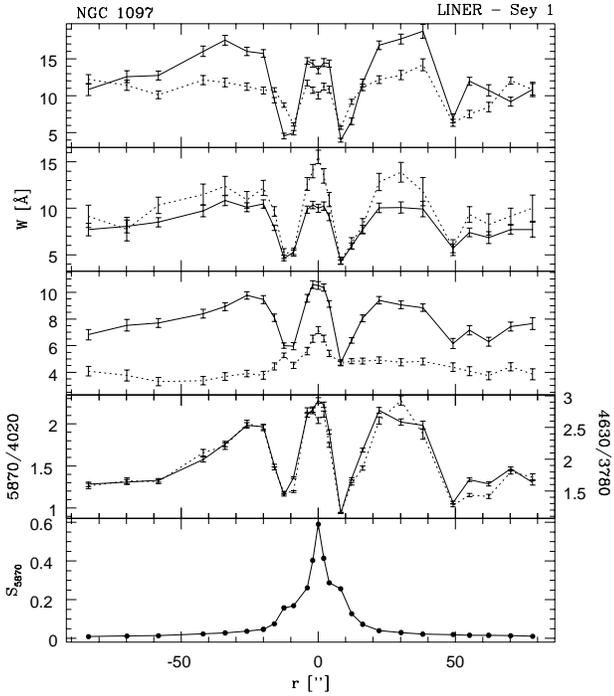}
    \caption{
    Radial variations of the equivalent widths (W), colors and
`surface brightness' for NGC 1097. The upper panel plots the Ws of
CaII~K (solid line) and CaII~H (dashed), while the second panel
shows G-band (solid), CN (dashed), and the third shows MgI$+$MgH
(solid) and NaI (dashed). The fourth panel shows the radial
variations of the continuum colors, mapped with the $F_{5870} /
F_{4020}$\ (solid line, left hand scale) and $F_{4630} / F_{3780}$\
(dashed line, right hand scale) flux ratios. Red is up and blue is
down in this plot. The bottom panel shows the run of the surface
brightness at 5870~\AA\ (in units of \flunit per arcsec$^2$) along
the slit.
    }
    \label{fig:n1097}
\end{figure}

In order to better illustrate the results presented in Figs.~3--44,
Fig.~\ref{fig:ZOOMED_spectra} shows the spatially resolved spectra
of five galaxies, representing the five classes of objects studied,
zoomed into the wavelength regions of the absorption lines analysed
in this work. This figure also serves as a further illustration of
the determination of the pseudo-continuum and of the quality of the
current database.

We start our discussion of the results with the cases of NGC~1097,
NGC~1433 and NGC~1672. The presence of star-forming rings in these
galaxies serves as a good illustration of the consistency of our
measurements and as a guide to the interpretation of the W and
color gradients in other objects. NGC 1326, 2935, 4303 and 5248 also
present star-forming rings, and much of the the discussion below
also applies to these objects. 

NGC~1097, a nearby spiral galaxy, has a low luminosity nucleus,
characterized by a LINER like emission line spectrum, surrounded by
a ring of HII regions (Sersic \& Pastoriza 1965, 1967, Phillips \et
1984). In addition to its LINER spectrum, the broad, double peaked,
variable  Balmer lines in the nucleus of NGC~1097 are a clear
signature of the presence of an active nucleus in this galaxy
(Storchi-Bergmann, Baldwin \& Wilson 1993;
Fig.~\ref{fig:example_spectra}). The low Ws and blue colors typical
of the young stellar populations of the ring are remarkably well
depicted in Fig.~\ref{fig:n1097} by the dips at $|r| \sim 10$\sec\
in the radial profiles of the Ws and colors. Despite its active
nucleus, the optical continuum shows little evidence of a blue
featureless continuum of the kind found in type 1 Seyferts and QSOs.
This is seen in Fig.~\ref{fig:n1097}, which shows that the Ws in the
nucleus of NGC 1097 are not significantly diluted with respect to
their values inside or outside the ring. On the contrary, the CN and
Mg lines seem to be stronger in the nucleus than outside it, while
the Na line clearly peaks at $r = 0$. The Ca~K and H lines, however,
do present a central dip, albeit of a low amplitude. Comparing the
nuclear spectrum with the $r = -4$\ and $+4$\sec\ extractions we
find variations of 7 and 11\% in the Ws of Ca K and H respectively
(\S\ref{sec:Dilution}), which indicates that an AGN continuum, if
present, contributes less than $\sim 10$\% of the light in this
wavelength region. Considering that the error in these Ws is $\sim
5$\%, the evidence for a diluting FC in NGC 1097 is marginal.

The Ws in NGC 1097 can be represented by a S2--S3 template at the
nucleus, S6--S7 at the ring and S3 outwards. The continuum ratios
are consistent with the spectral templates inferred from the Ws,
with the exception of the nuclear region, where the continuum is
redder, more typical of a S1 template. This could be interpreted as
evidence of reddening of the nuclear regions, as is also indicated
by the increase of W(Na) towards the nucleus (Storchi-Bergmann \et
1995a). This difference in template types can be accounted for with
$E(B-V) \approx 0.15$. Storchi-Bergmann, Kinney \& Challis (1995b)
obtained a large aperture spectrum of this object
(10\sec$\times$20\sec), including the nucleus and part of the
starforming ring. Their measured Ws can be represented by a S5
template, which is comparable to our results in the inner 10\sec.
The IUE spectrum of NGC1097 is dominated by the ring, showing
several absorption lines typical of young stars (Kinney \et 1993).

NGC~1433 has been classified as a Seyfert 2 by V\'eron-Cetty \&
V\'eron (1986), though our spectrum presents emission line ratios
more typical of LINERs. It also contains a star-forming ring $\sim
8$\sec\ from its nucleus (Buta 1986). As in the case of NGC~1097,
the ring leaves a clear finger print in the W and color profiles
(Fig.~\ref{fig:n1433}). No dilution of the absorption lines or
blueing of the continuum is seen in the nuclear spectrum. Both the
Ws and continuum ratios indicate similar spectral templates: S3 at
the nucleus, S5 at the ring and S3 outwards. As in NGC~1097, the
general decrease of the Ws and bluer colors towards large distances
seen in Fig.~\ref{fig:n1433} correspond to the stellar population of
the disk of the galaxy. 

NGC~1672 is a Sersic-Pastoriza galaxy, whose nucleus has  already
been classified as Starburst (Garcia-Vargas \et 1990, Kinney \et
1993), Seyfert 2 (Mouri \et 1989) and LINER (D\'{\i}az 1985,
Storchi-Bergmann \et 1996b). Our nuclear spectrum is consistent with
a LINER classification. Its W and color profiles
(Fig.~\ref{fig:n1672}) are very similar to those of the previous two
galaxies. Both Ws and continuum ratios indicate a S4 template at the
nucleus, S7 at the $r = 5$\sec\ ring, S4 outside the ring,
decreasing to values typical of S7 template in the outer regions.

The spatio-spectral variations in NGC~1672 are also seen in
Fig.~\ref{fig:ZOOMED_spectra}a. The correspondence between
Figs.~\ref{fig:n1672} and \ref{fig:ZOOMED_spectra}a is clear. The
star forming ring, with its shallower absorption lines and bluer
continuum, is visible in the $r = -8$, $+ 4$\ and $+8$\sec\
extractions plotted in Fig.~\ref{fig:ZOOMED_spectra}a. The blueing
of the spectra towards the outer regions of the galaxy is also
illustrated. The $+67$\sec\ extraction, in particular, crosses
an HII region.

These three examples demonstrate that the method employed in this
study is able to detect variations of the stellar populations as a
function of distance to the nucleus. Star-forming rings, in
particular, leave unambiguous signatures in the W and color
profiles. 

Since ``dilution'' of the metal lines by a young stellar population
is detectable, we also expect to be able to measure dilution by an
AGN continuum. This is best illustrated in the case of Seyfert 1
objects, such as MCG-02-33-034 (Fig.~\ref{fig:i309}), ESO 362-G18
(Fig.~\ref{fig:e18}), NGC 6860 (Fig.~\ref{fig:n6860}), NGC 6814
(Fig.~\ref{fig:n6814}), and the Broad Line Radio Galaxy Pictor A
(Fig.~\ref{fig:pic}).



\clearpage

\begin{figure}
    \cidfig{7.5cm}{30}{150}{430}{705}{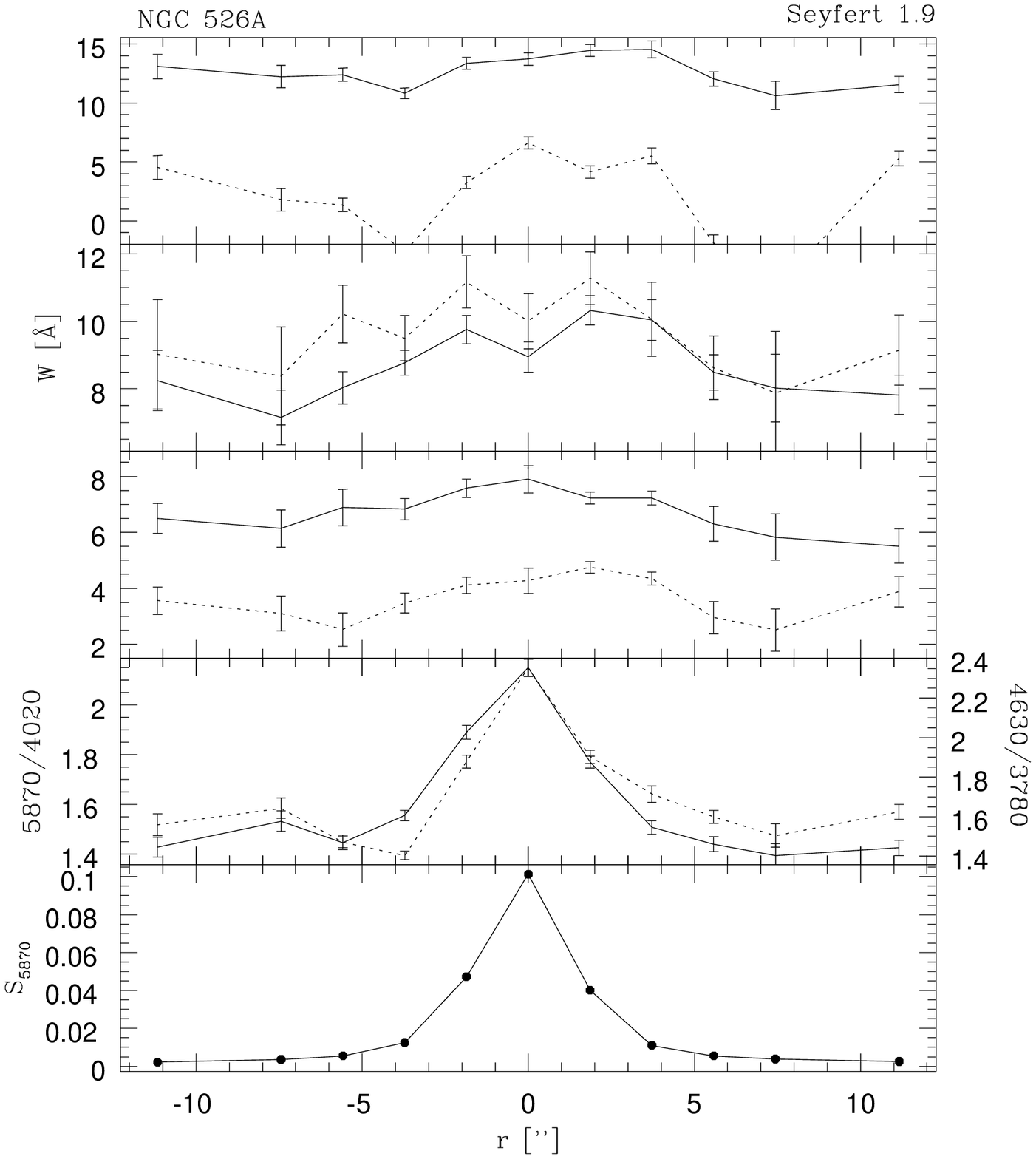}
    \caption{Same as Fig.\ 3.   \label{fig:n526a} }
\end{figure}

\begin{figure}
    \cidfig{7.5cm}{30}{150}{430}{705}{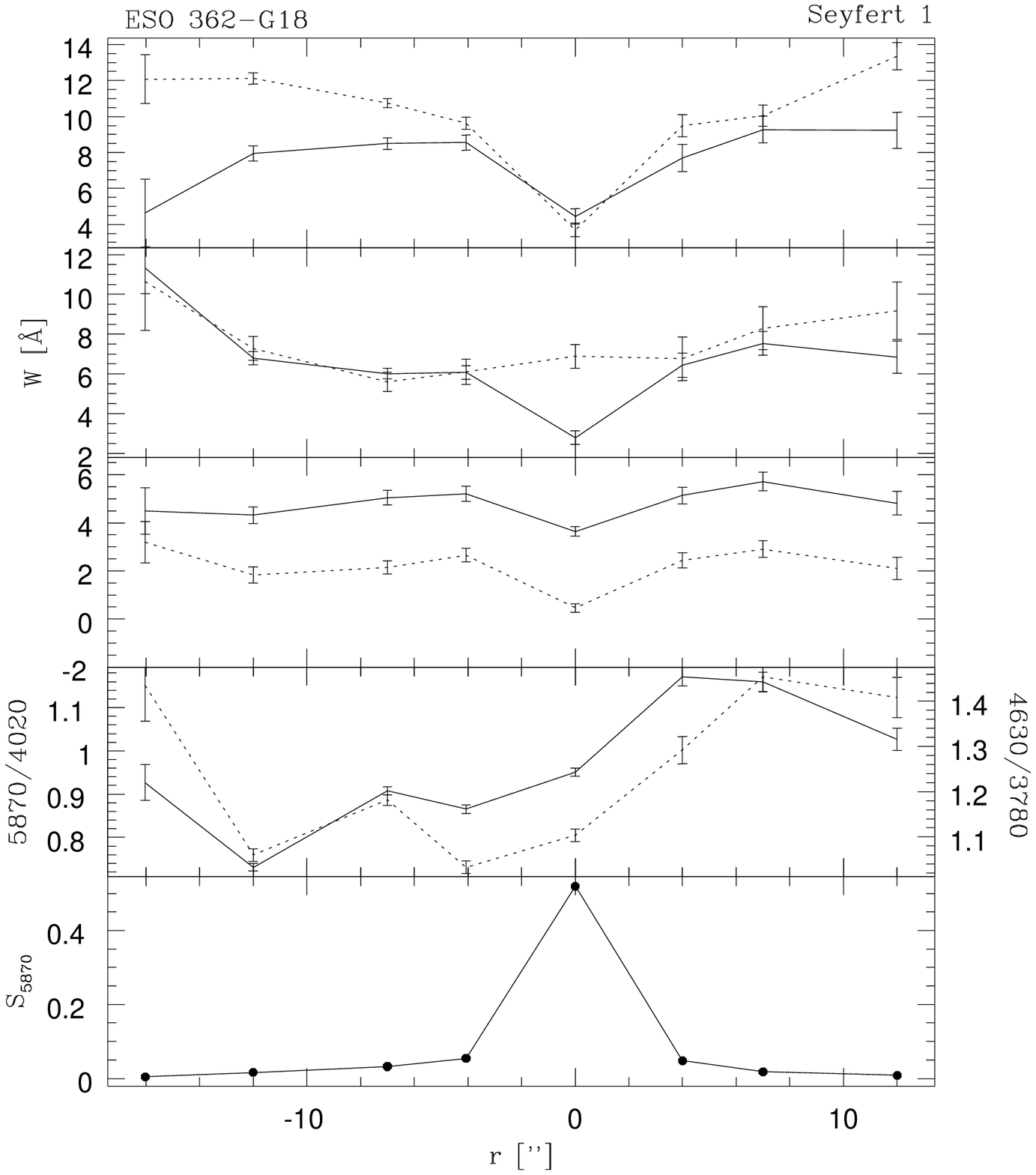}
    \caption{Same as Fig.\ 3.   \label{fig:e18} }
\end{figure}

\begin{figure}
    \cidfig{7.5cm}{30}{150}{430}{705}{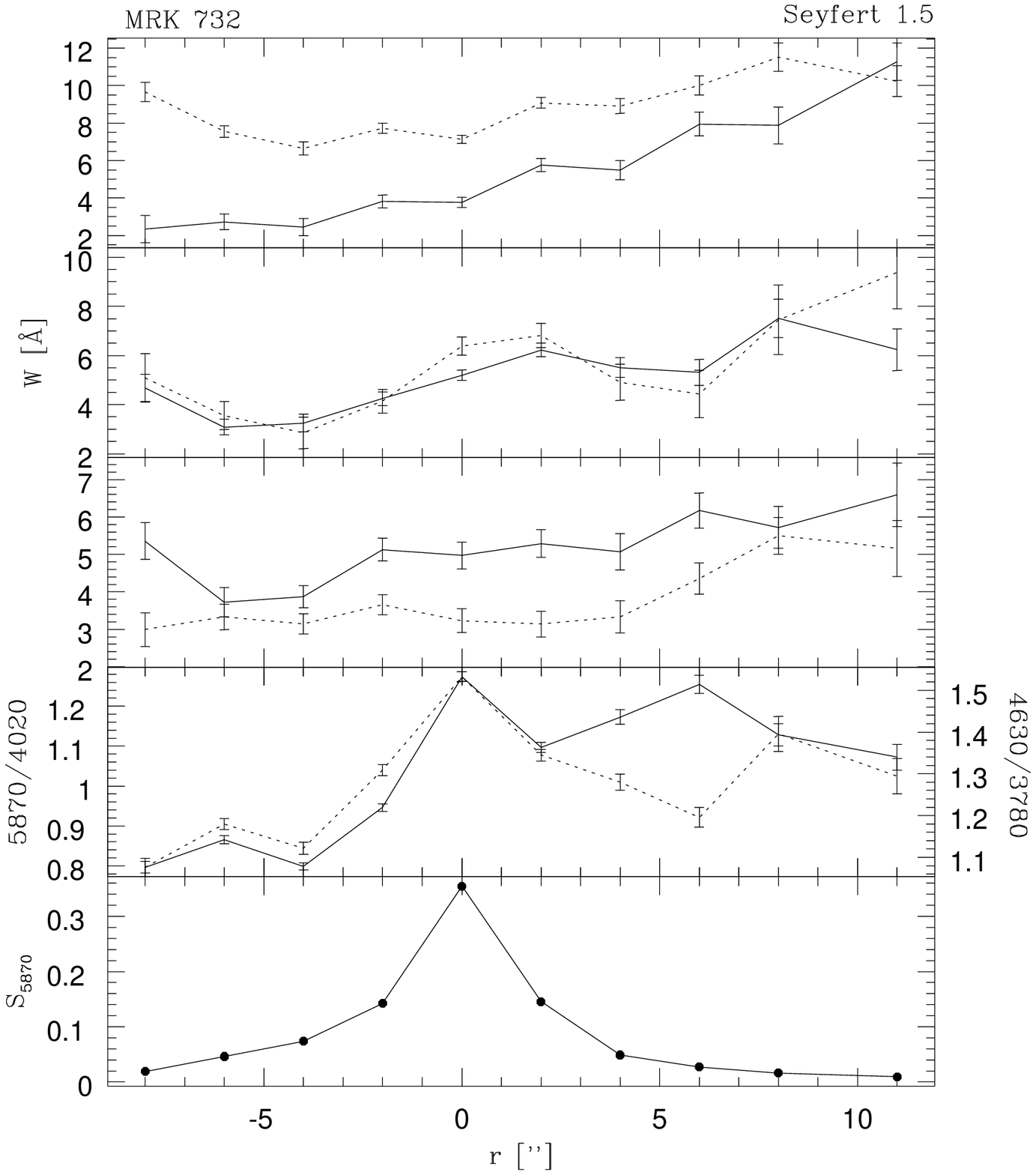}
    \caption{Same as Fig.\ 3.   \label{fig:m732} }
\end{figure}

\begin{figure}
    \cidfig{7.5cm}{30}{150}{430}{705}{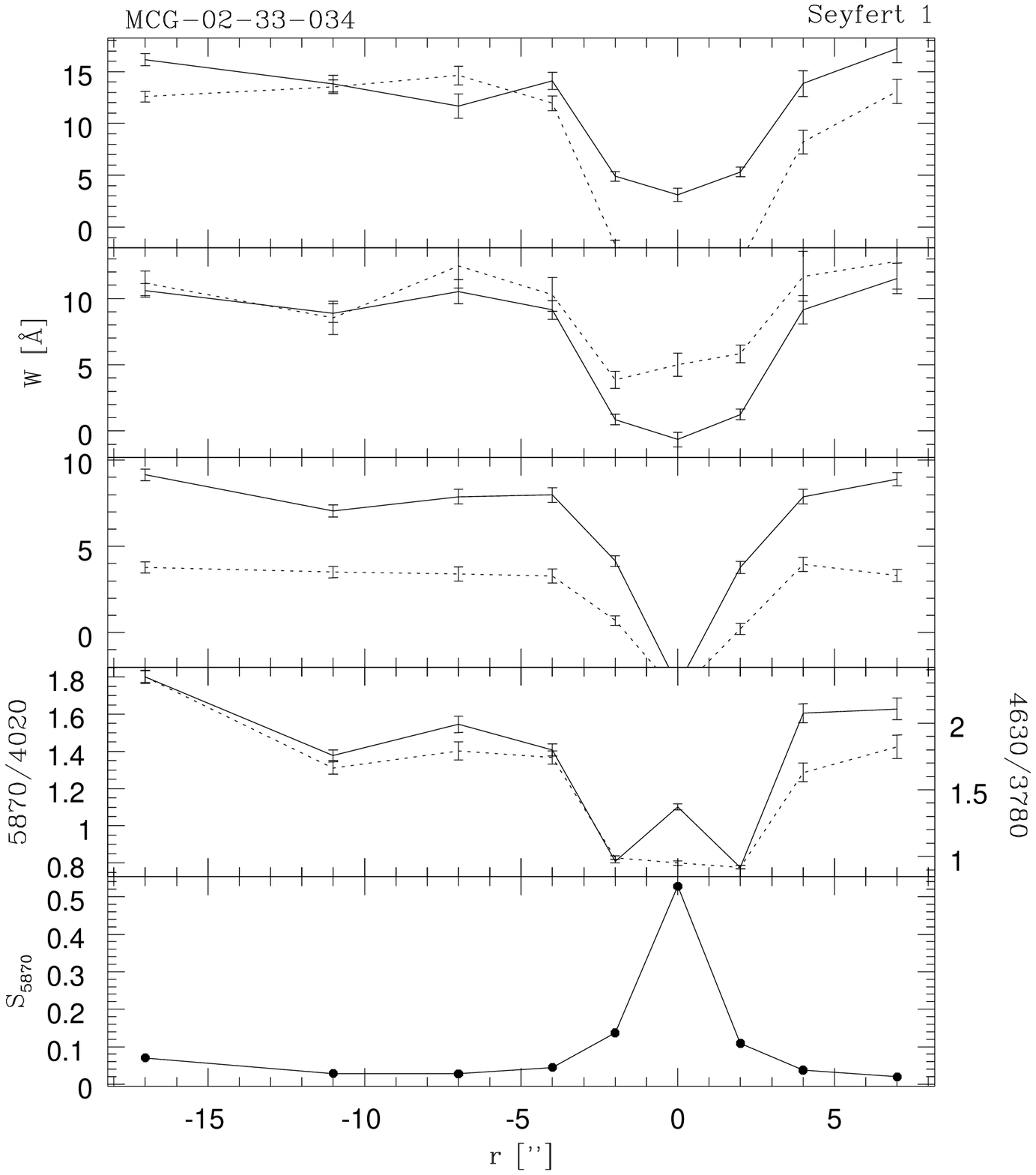}
    \caption{Same as Fig.\ 3.   \label{fig:i309} }
\end{figure}

\clearpage

\begin{figure}
    \cidfig{7.5cm}{30}{150}{430}{705}{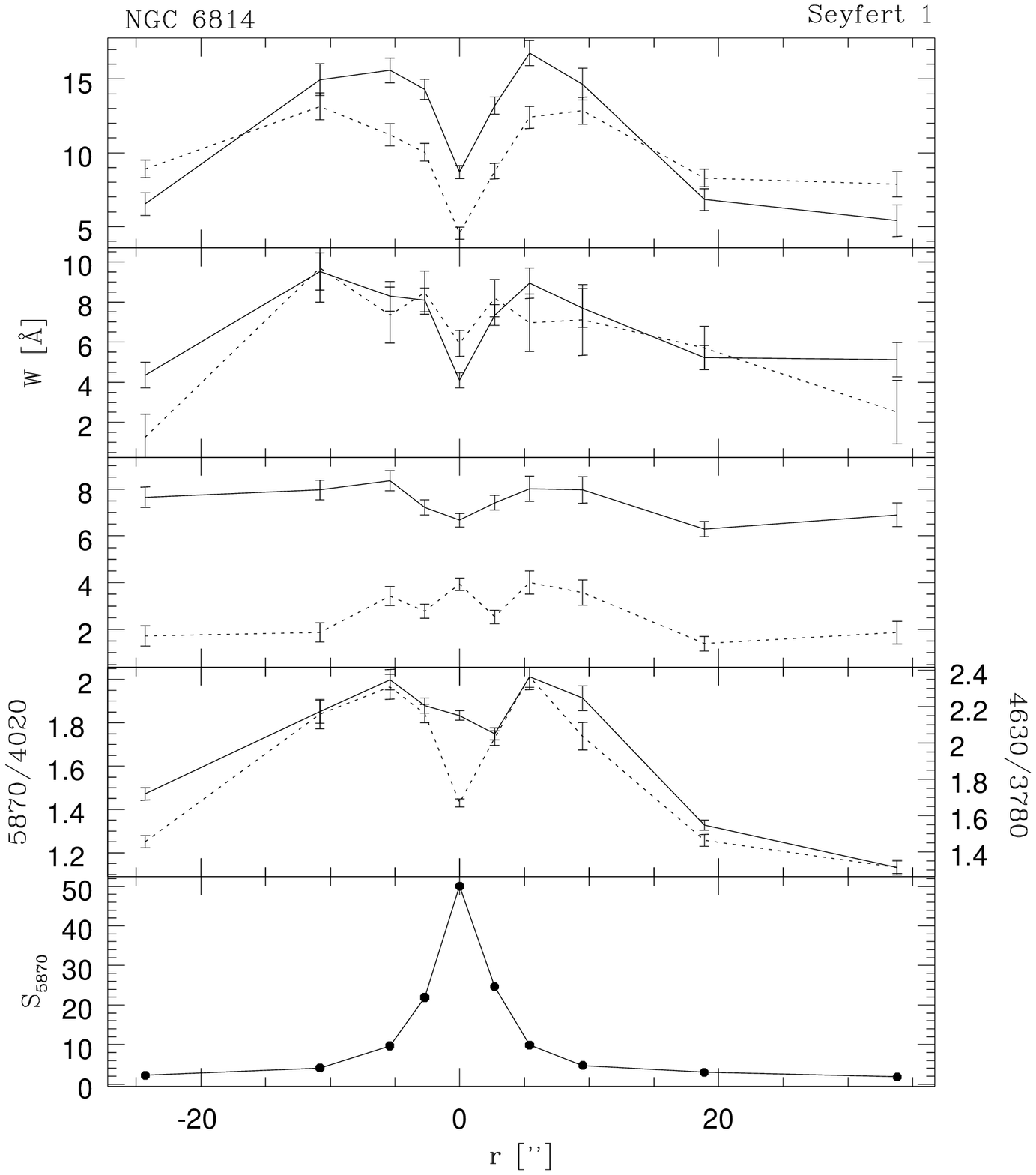}
    \caption{Same as Fig.\ 3.   \label{fig:n6814} }
\end{figure}

\begin{figure}
    \cidfig{7.5cm}{30}{150}{430}{705}{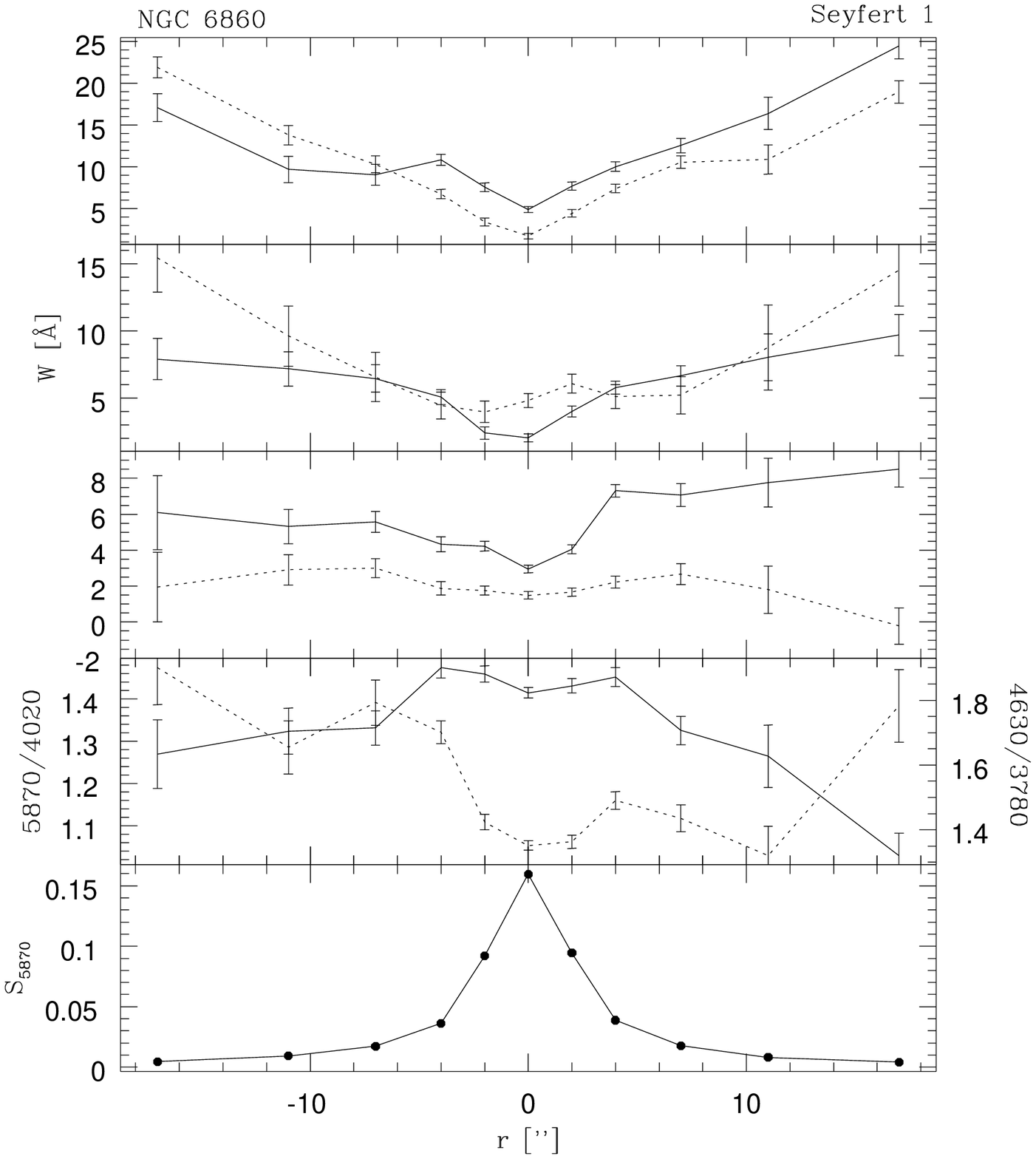}
    \caption{Same as Fig.\ 3.   \label{fig:n6860} }
\end{figure}

\begin{figure}
    \cidfig{7.5cm}{30}{150}{430}{705}{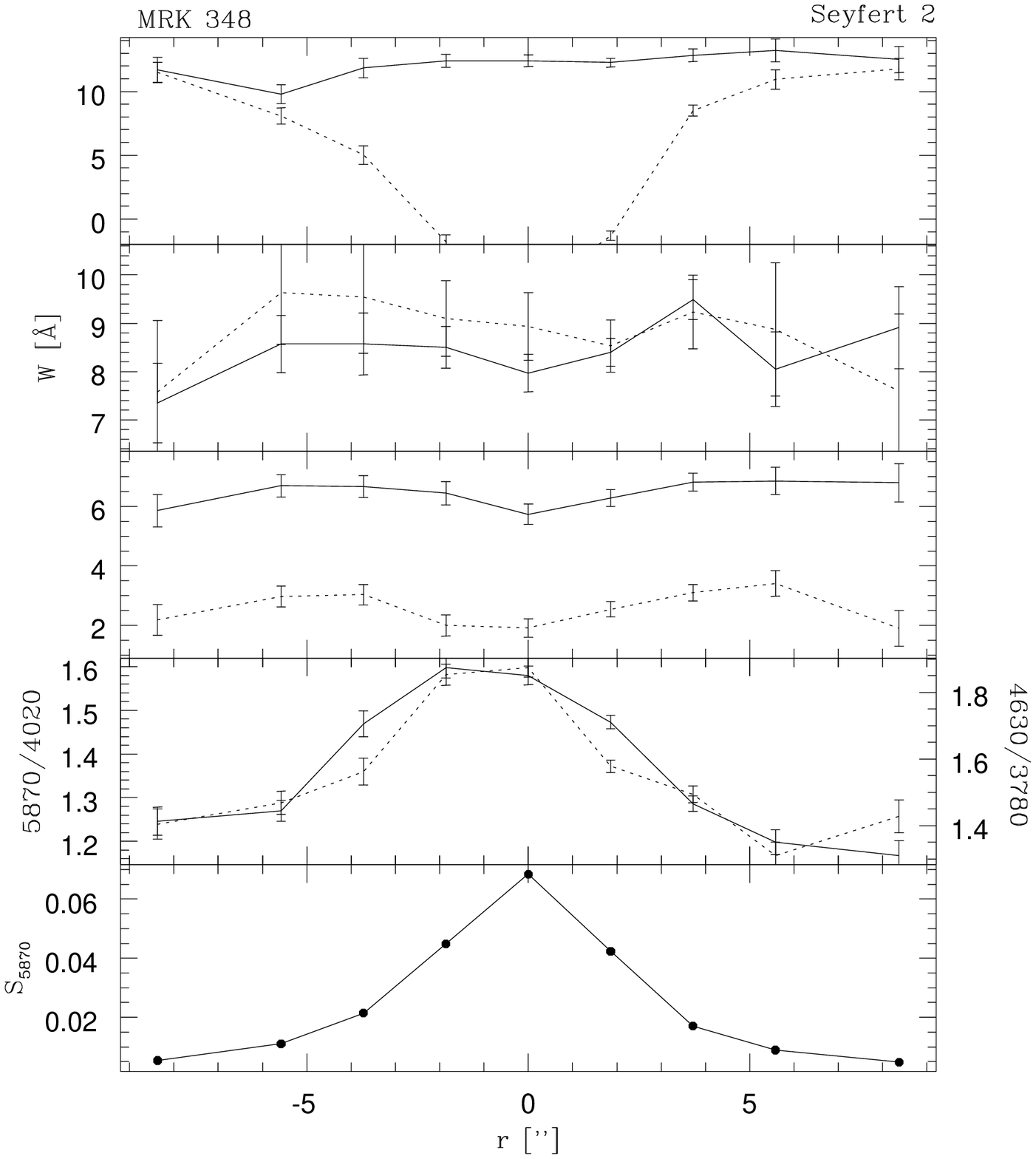}
    \caption{Same as Fig.\ 3.   \label{fig:m348} }
\end{figure}

\begin{figure}
    \cidfig{7.5cm}{30}{150}{430}{705}{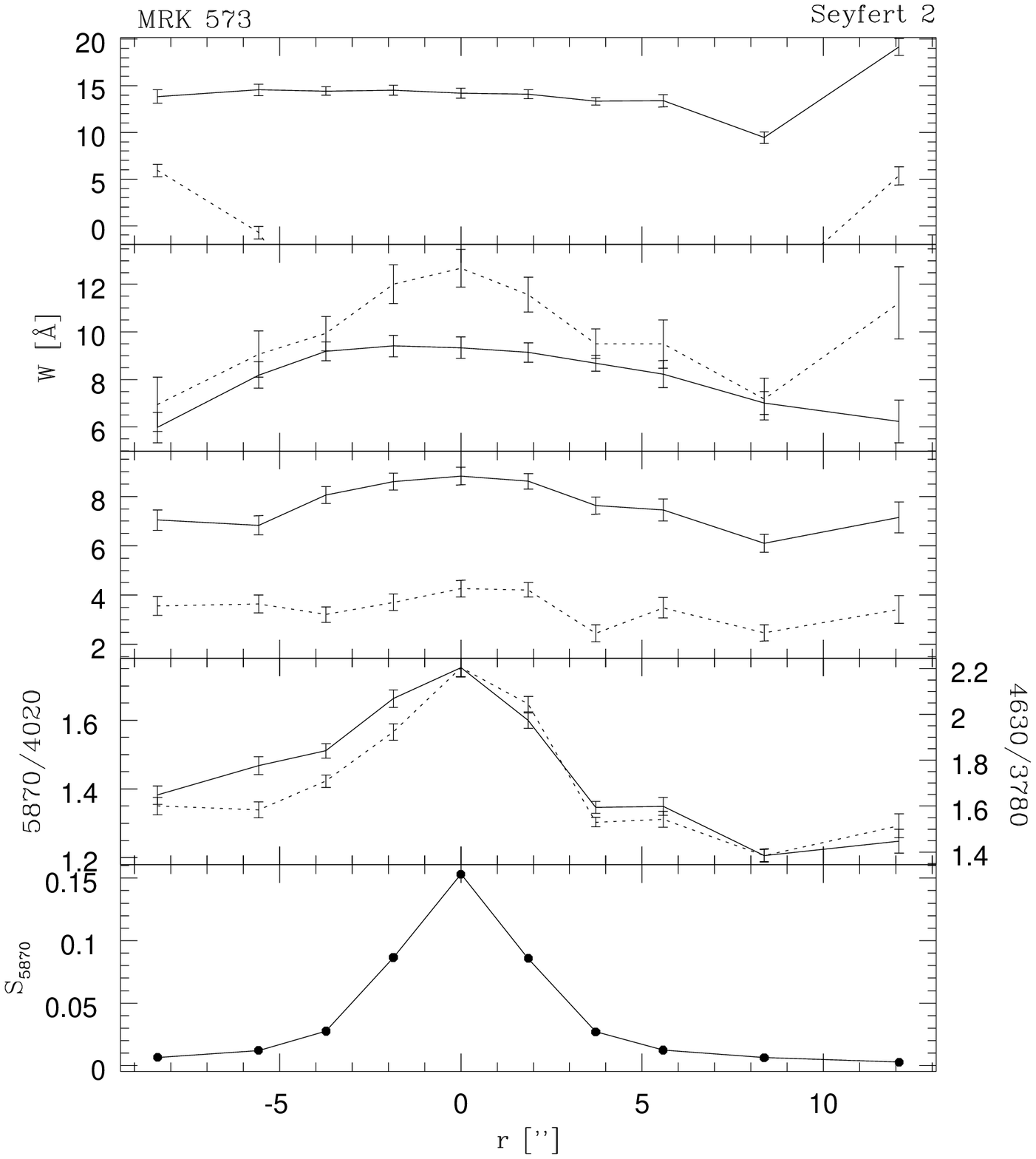}
    \caption{Same as Fig.\ 3.   \label{fig:m573} }
\end{figure}

\clearpage

\begin{figure}
    \cidfig{7.5cm}{30}{150}{430}{705}{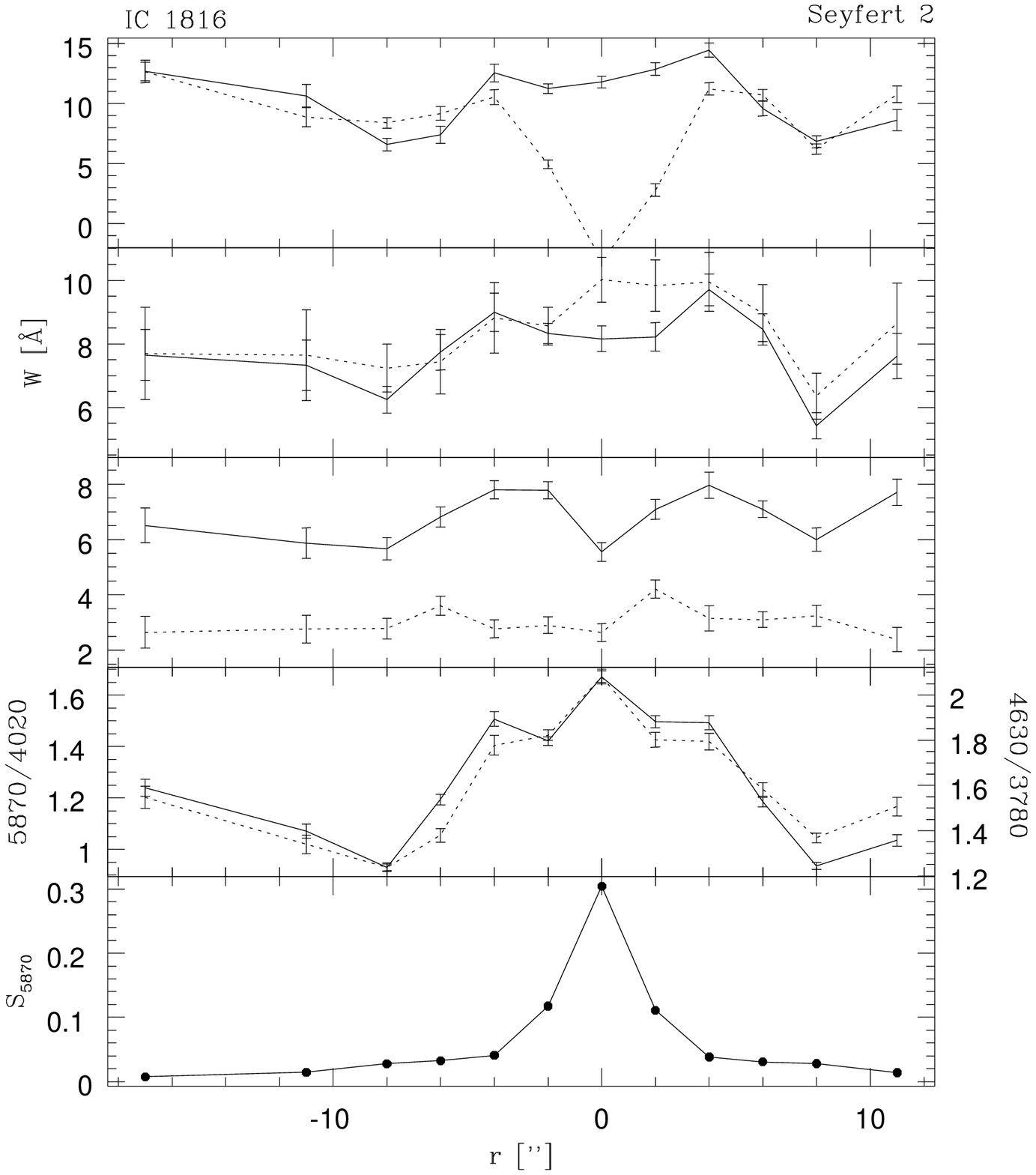}
    \caption{Same as Fig.\ 3.   \label{fig:i1816} }
\end{figure}

\begin{figure}
    \cidfig{7.5cm}{30}{150}{430}{705}{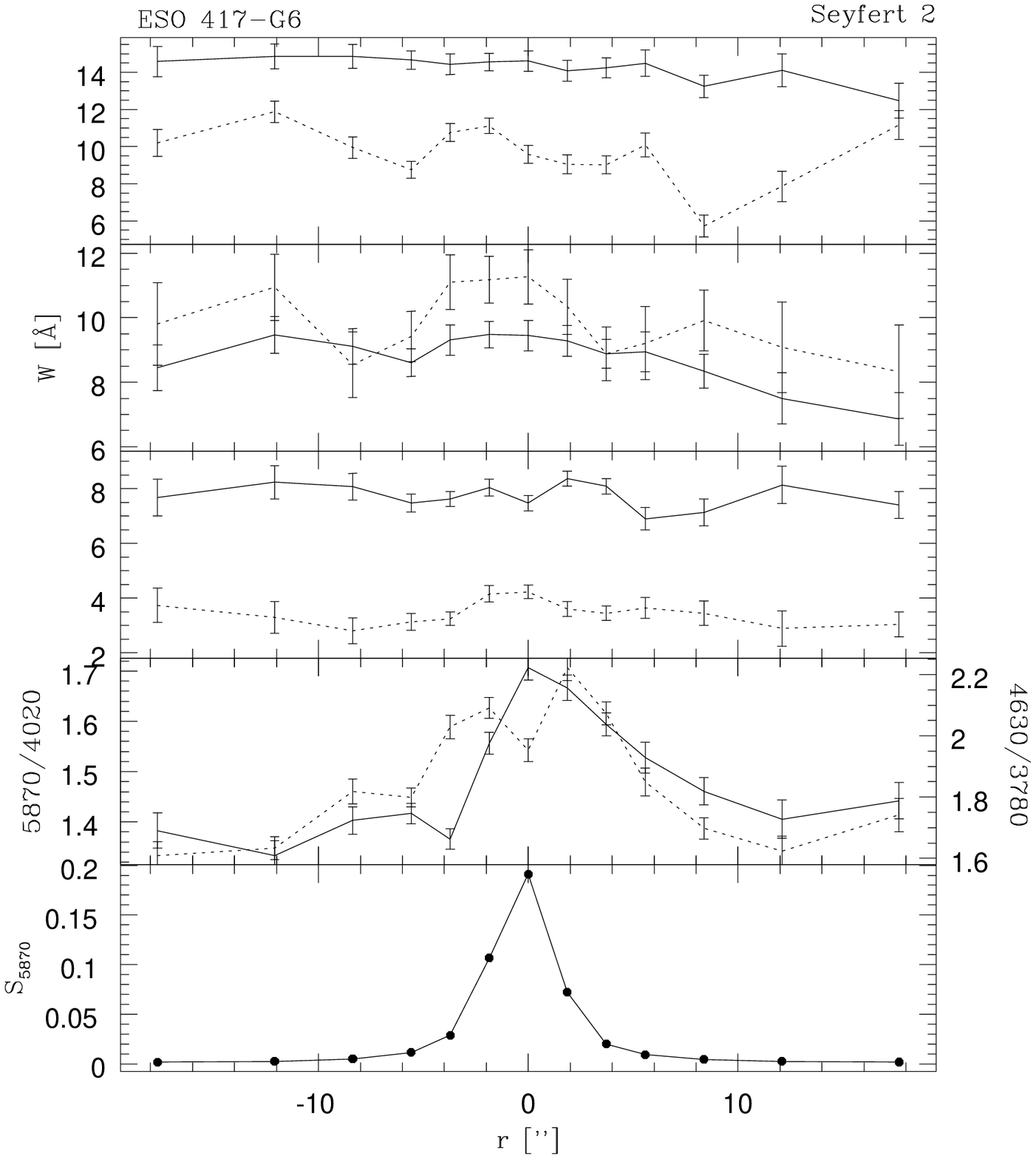}
    \caption{Same as Fig.\ 3.   \label{fig:e417} }
\end{figure}

\begin{figure}
    \cidfig{7.5cm}{30}{150}{430}{705}{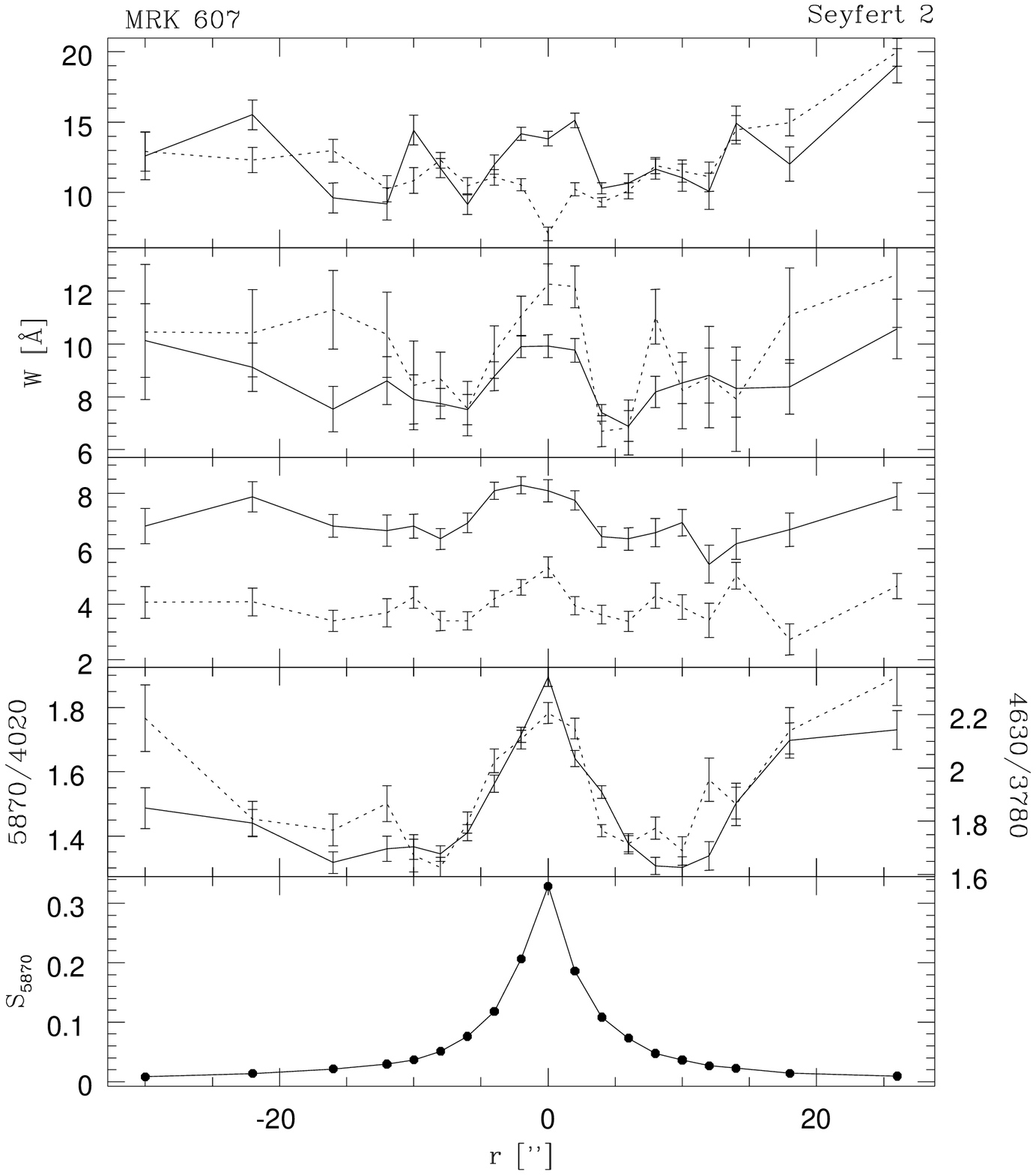}
    \caption{Same as Fig.\ 3.   \label{fig:m607} }
\end{figure}

\begin{figure}
    \cidfig{7.5cm}{30}{150}{430}{705}{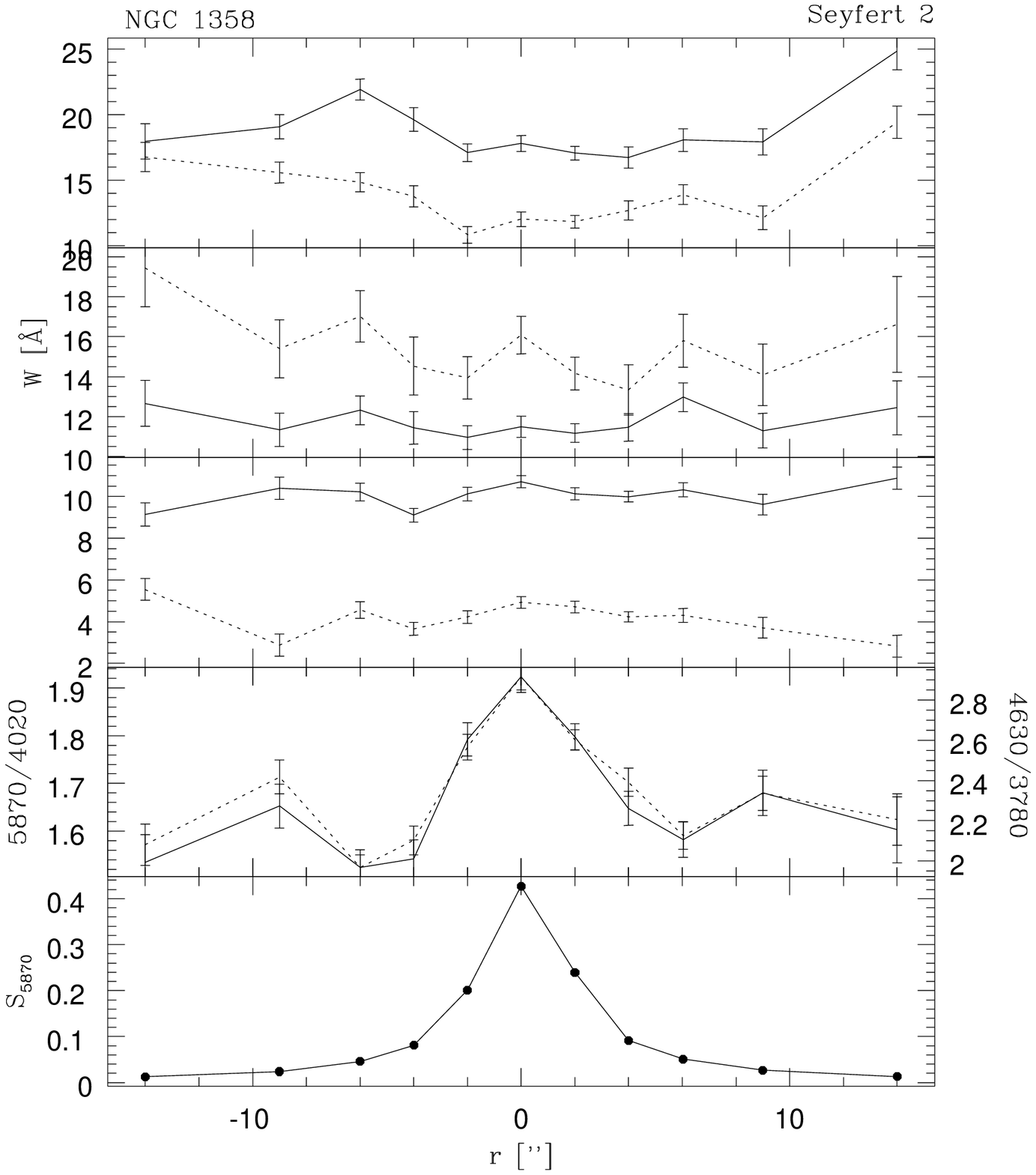}
    \caption{Same as Fig.\ 3.   \label{fig:n1358} }
\end{figure}

\clearpage

\begin{figure}
    \cidfig{7.5cm}{30}{150}{430}{705}{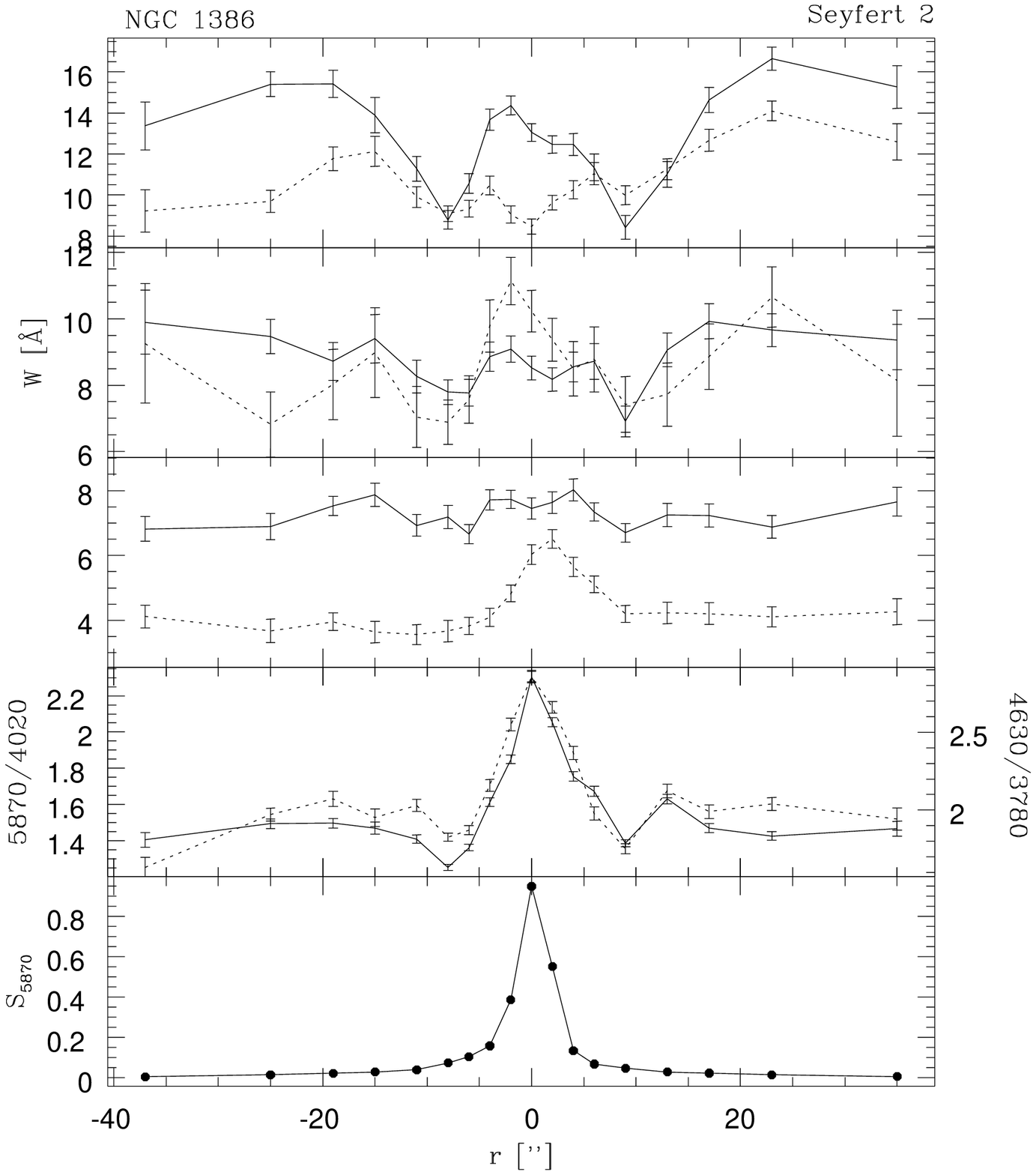}
    \caption{Same as Fig.\ 3.   \label{fig:n1386} }
\end{figure}

\begin{figure}
    \cidfig{7.5cm}{30}{150}{430}{705}{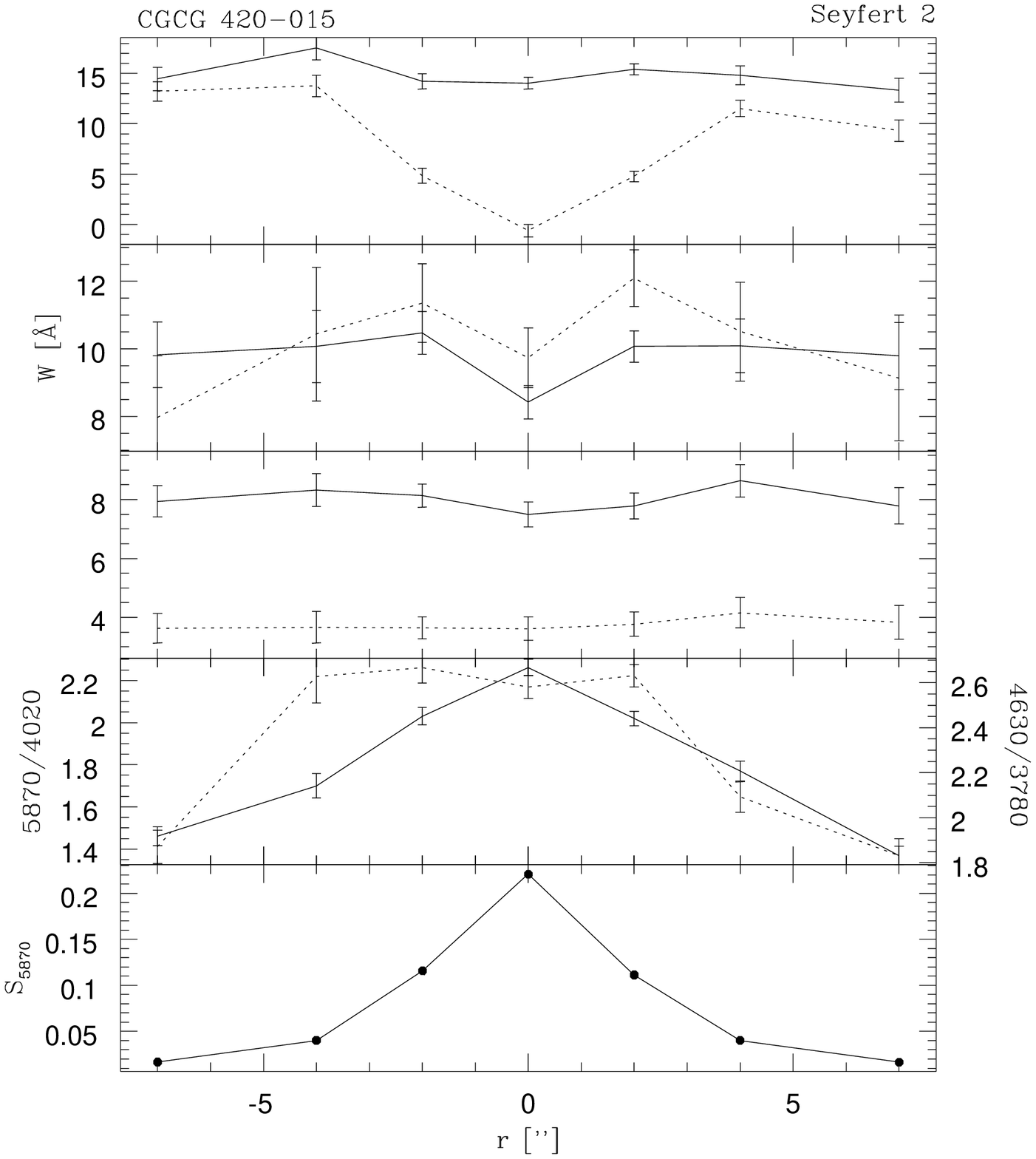}
    \caption{Same as Fig.\ 3.   \label{fig:i156} }
\end{figure}

\begin{figure}
    \cidfig{7.5cm}{30}{150}{430}{705}{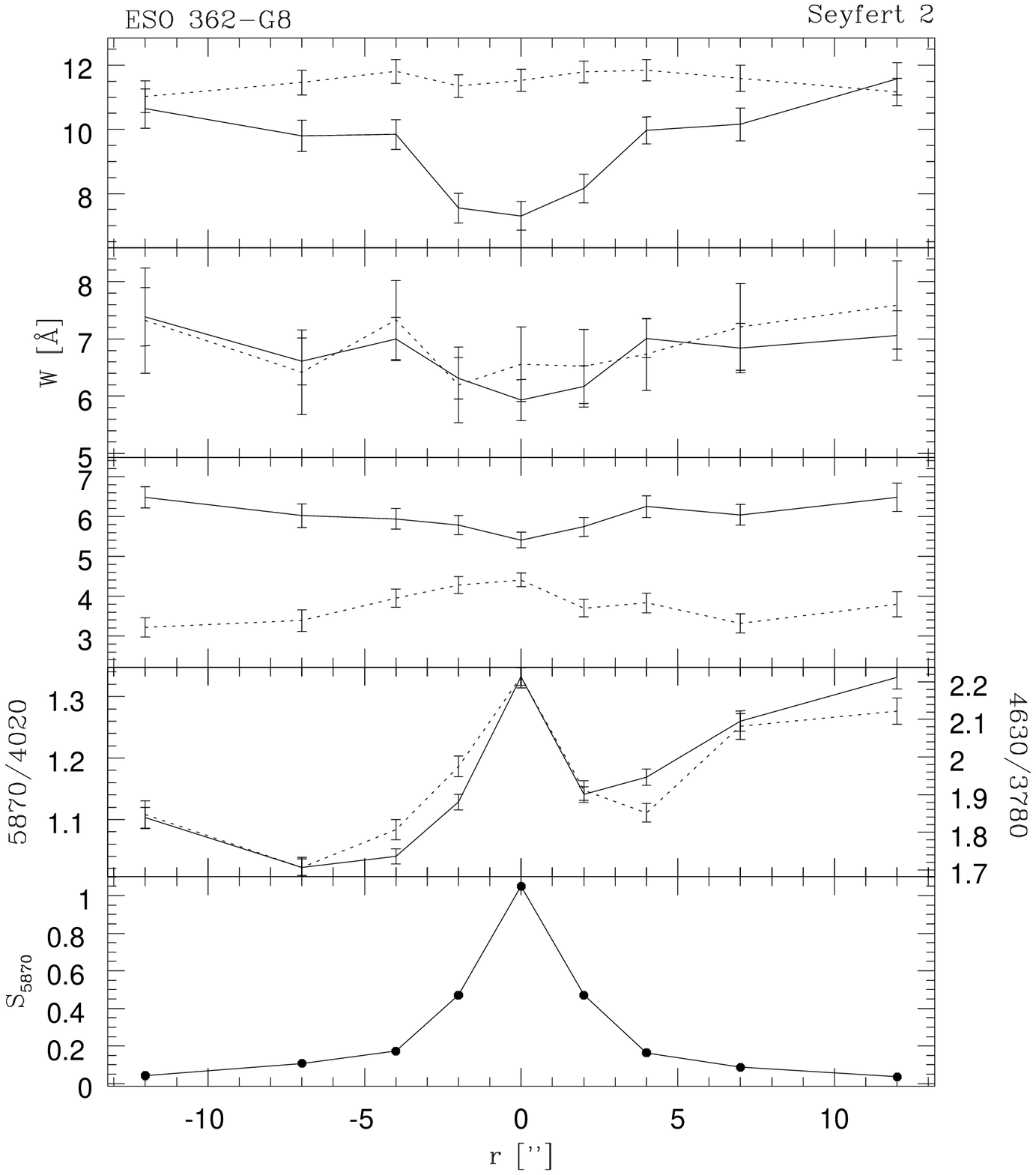}
    \caption{Same as Fig.\ 3.   \label{fig:e8} }
\end{figure}

\begin{figure}
    \cidfig{7.5cm}{30}{150}{430}{705}{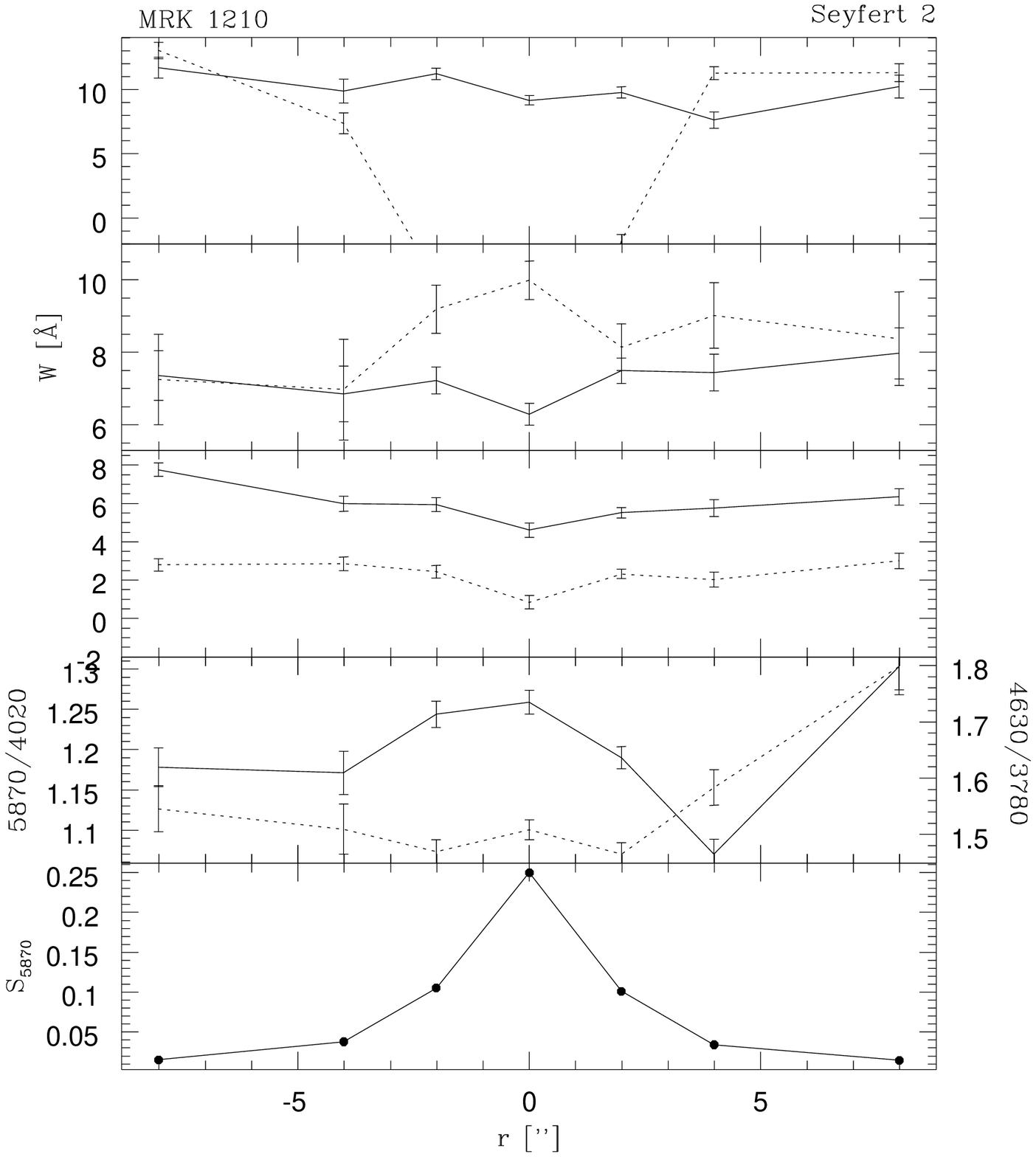}
    \caption{Same as Fig.\ 3.   \label{fig:m1210} }
\end{figure}

\clearpage

\begin{figure}
    \cidfig{7.5cm}{30}{150}{430}{705}{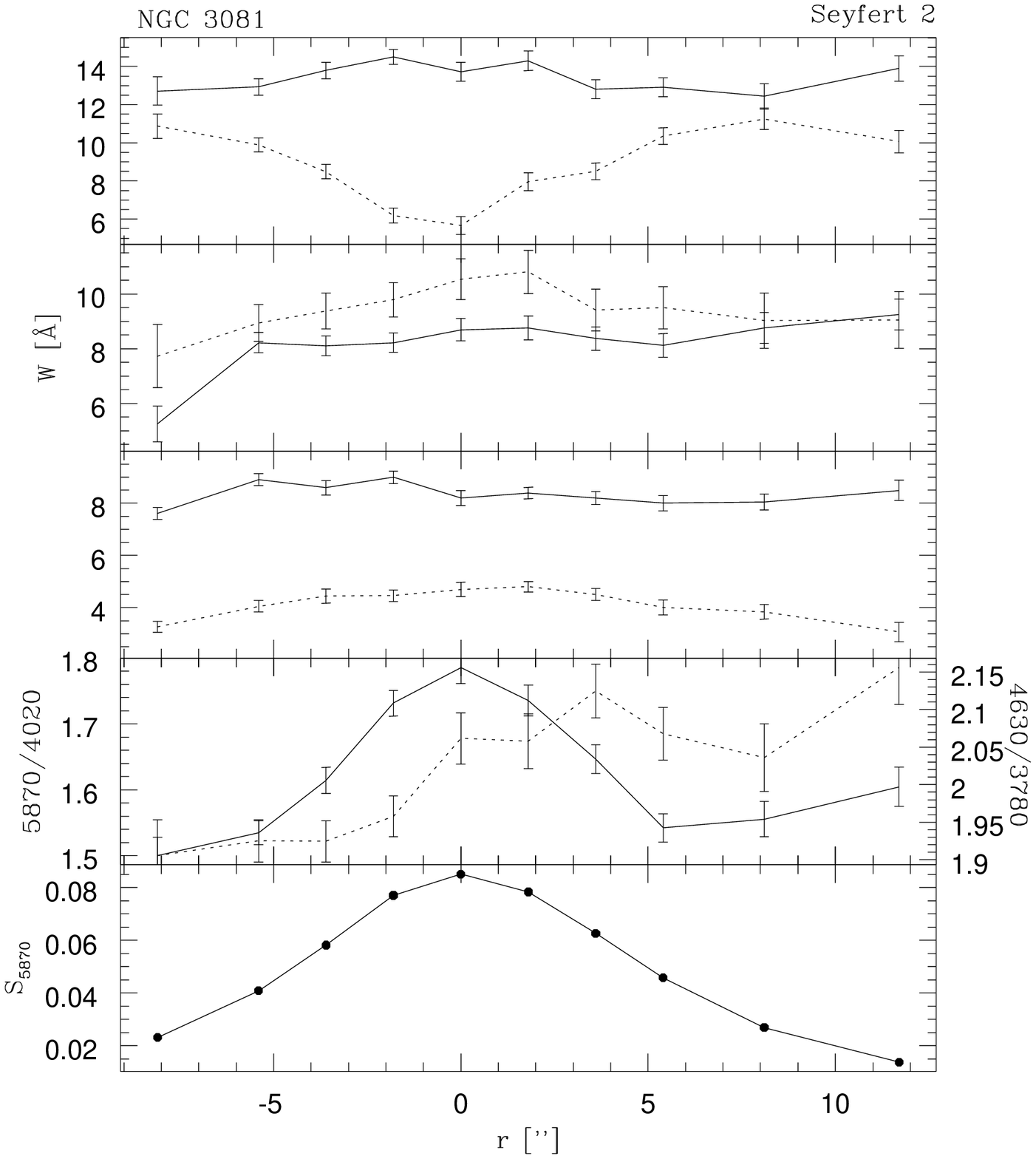}
    \caption{Same as Fig.\ 3.   \label{fig:n3081} }
\end{figure}

\begin{figure}
    \cidfig{7.5cm}{30}{150}{430}{705}{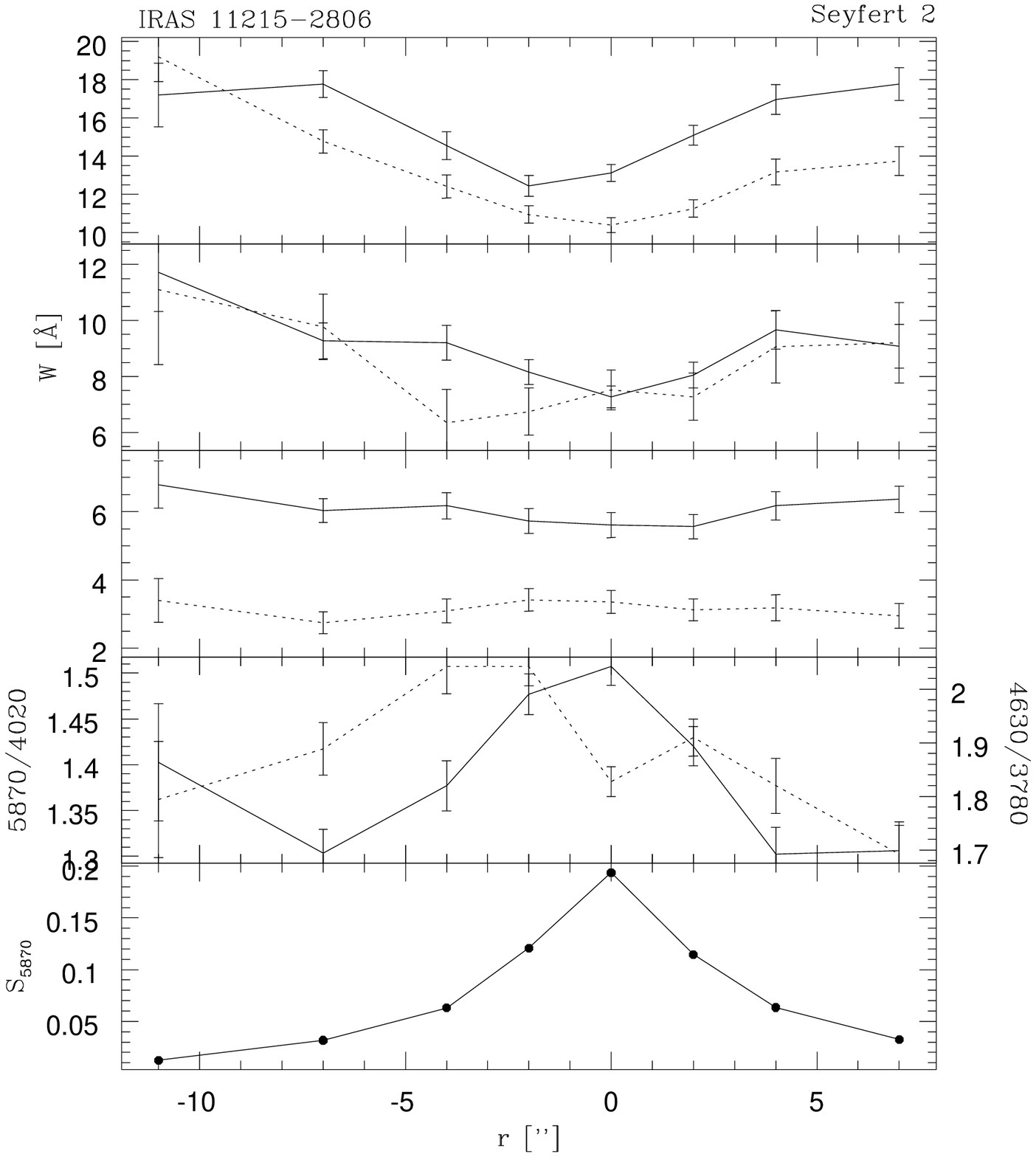}
    \caption{Same as Fig.\ 3.   \label{fig:i281} }
\end{figure}

\begin{figure}
    \cidfig{7.5cm}{30}{150}{430}{705}{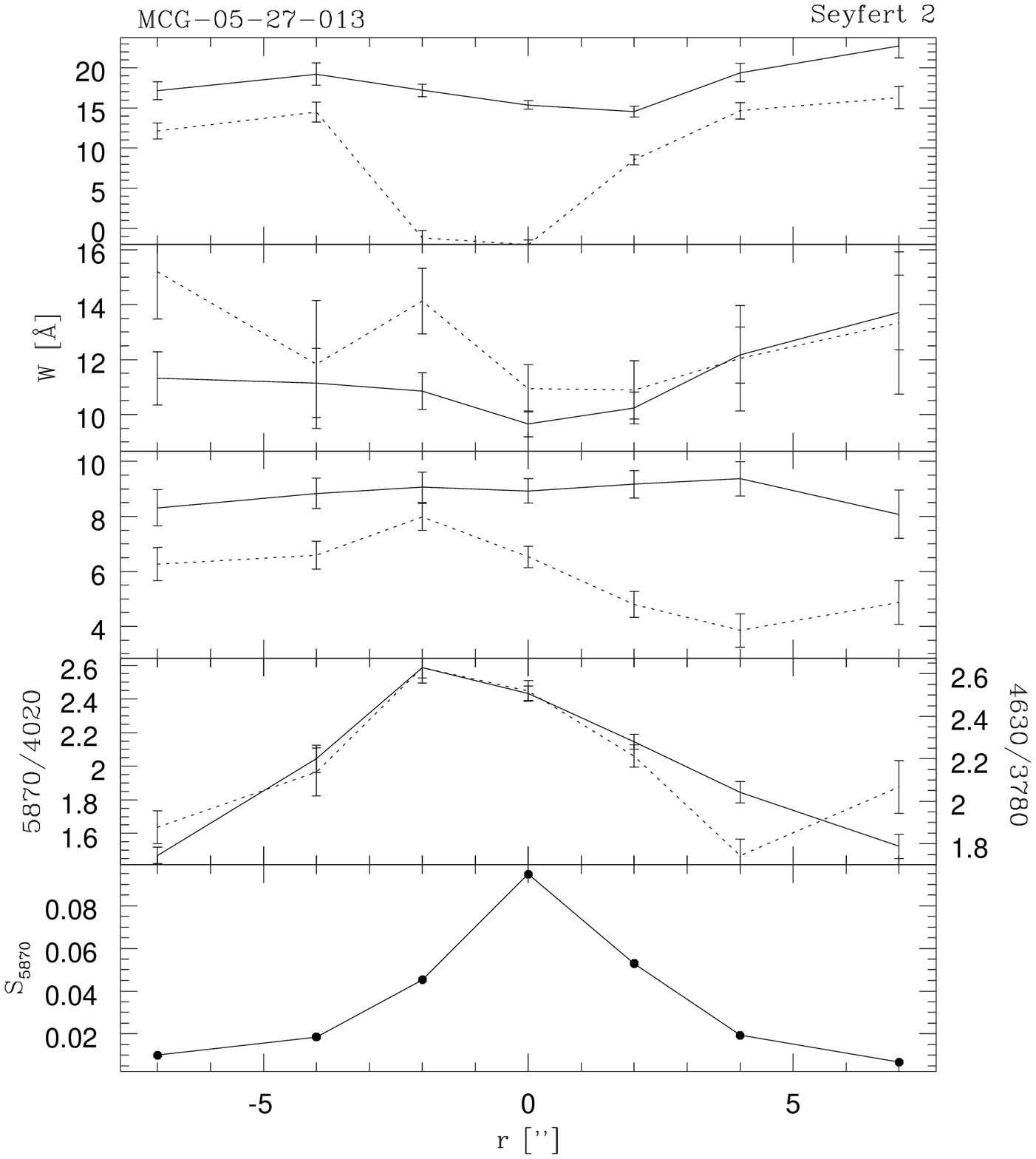}
    \caption{Same as Fig.\ 3.   \label{fig:i282} }
\end{figure}

\begin{figure}
    \cidfig{7.5cm}{30}{150}{430}{705}{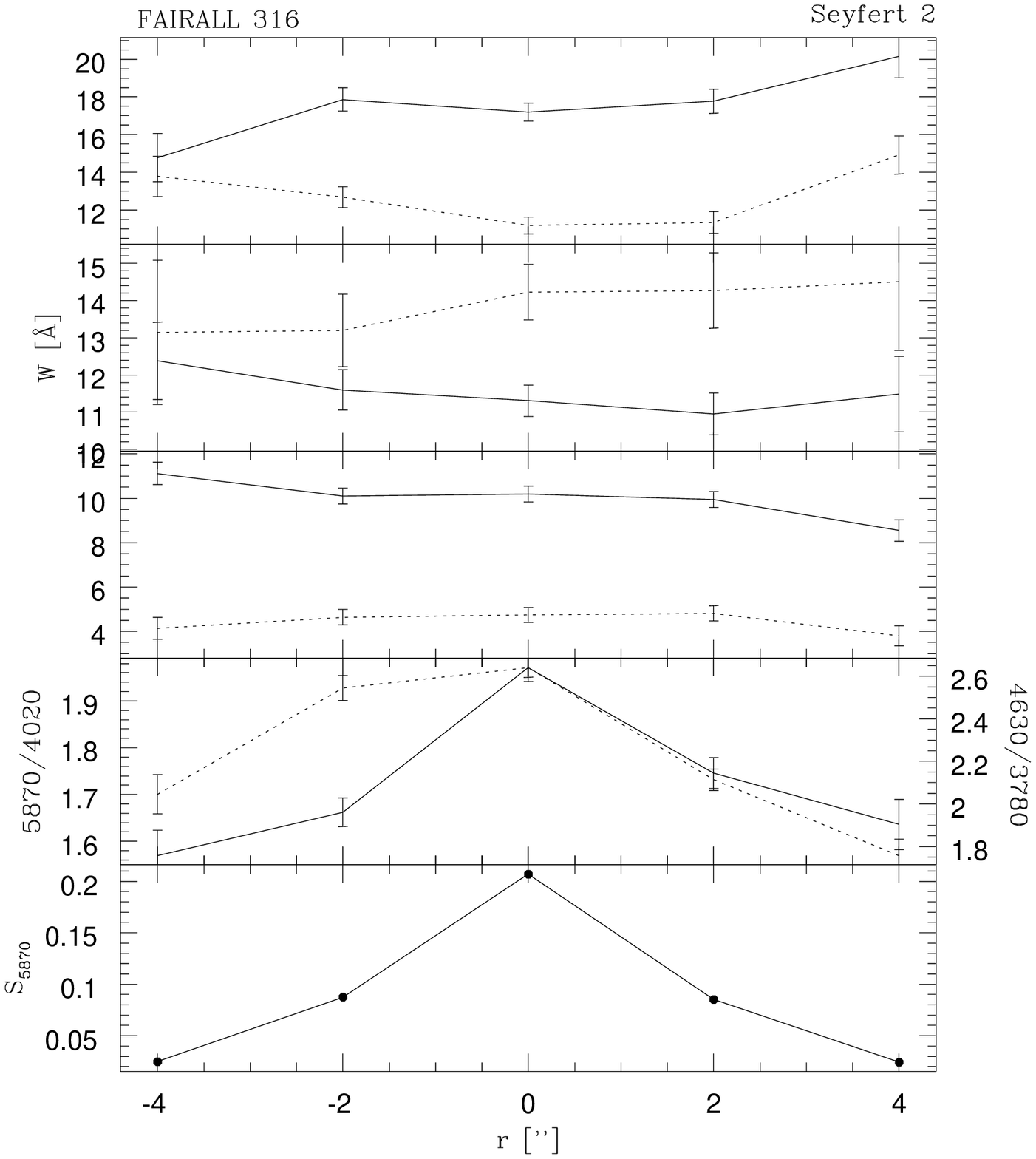}
    \caption{Same as Fig.\ 3.   \label{fig:f316} }
\end{figure}

\clearpage

\begin{figure}
    \cidfig{7.5cm}{30}{150}{430}{705}{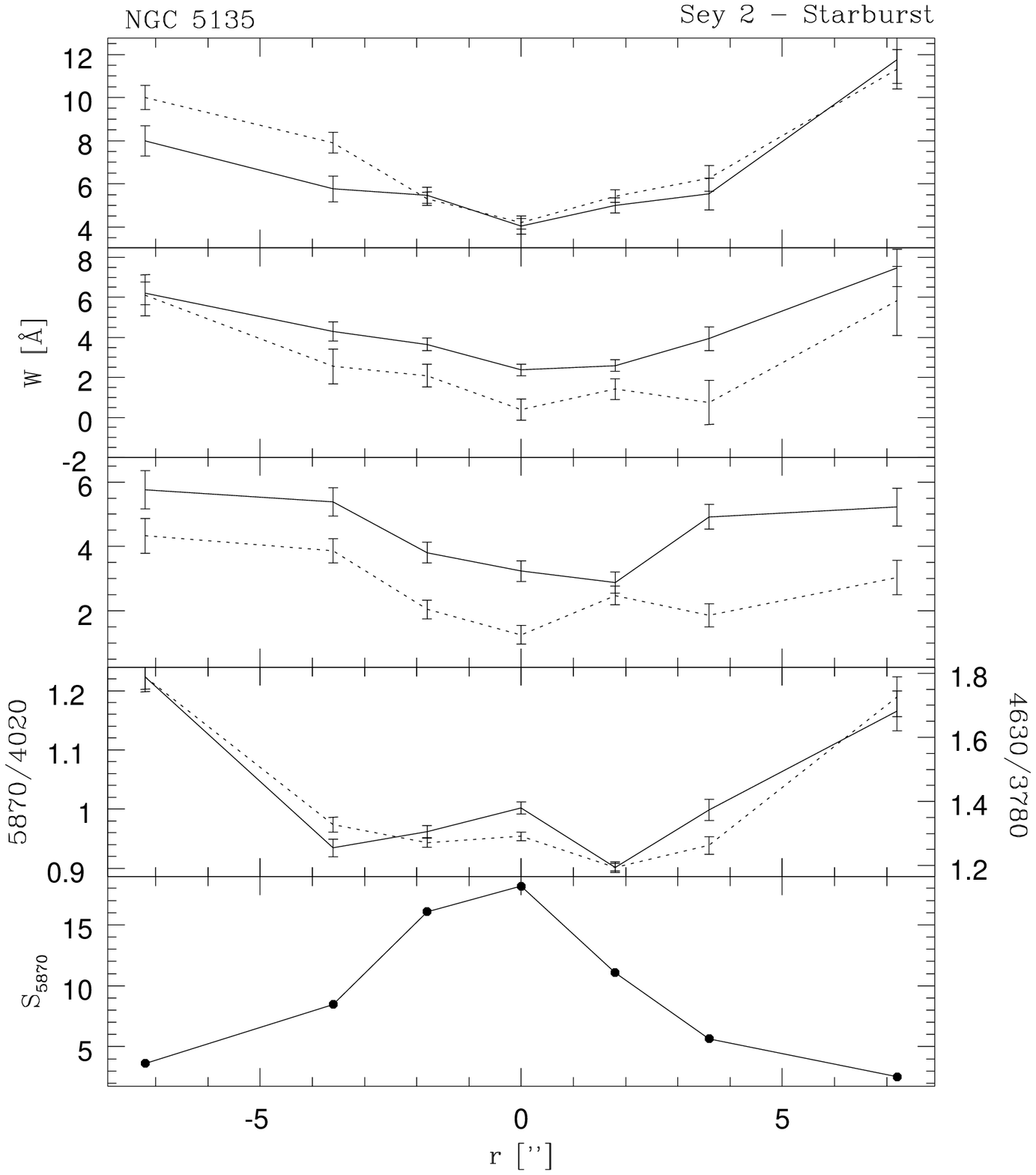}
    \caption{Same as Fig.\ 3.   \label{fig:n5135} }
\end{figure}

\begin{figure}
    \cidfig{7.5cm}{30}{150}{430}{705}{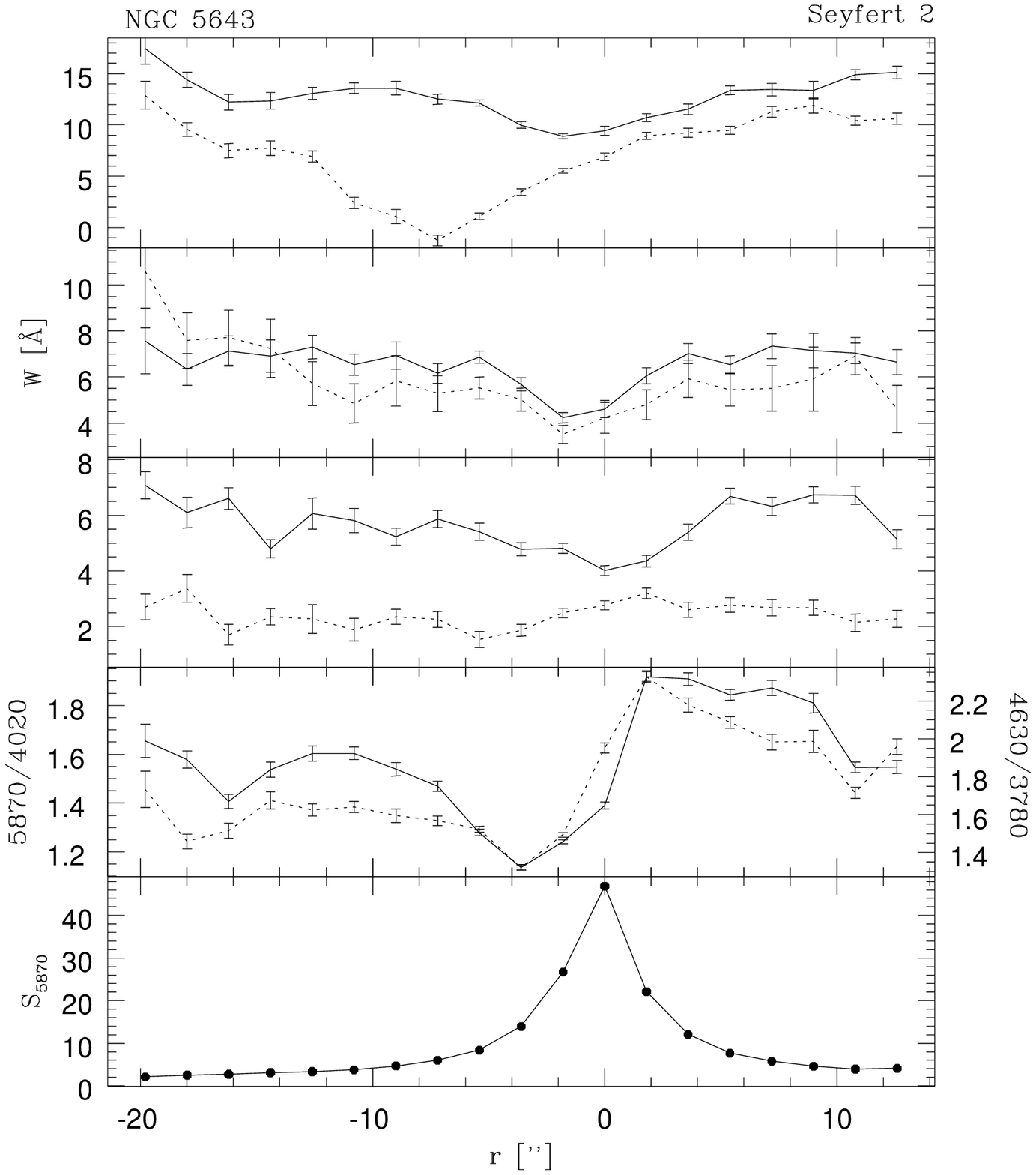}
    \caption{Same as Fig.\ 3.   \label{fig:n5643} }
\end{figure}

\begin{figure}
    \cidfig{7.5cm}{30}{150}{430}{705}{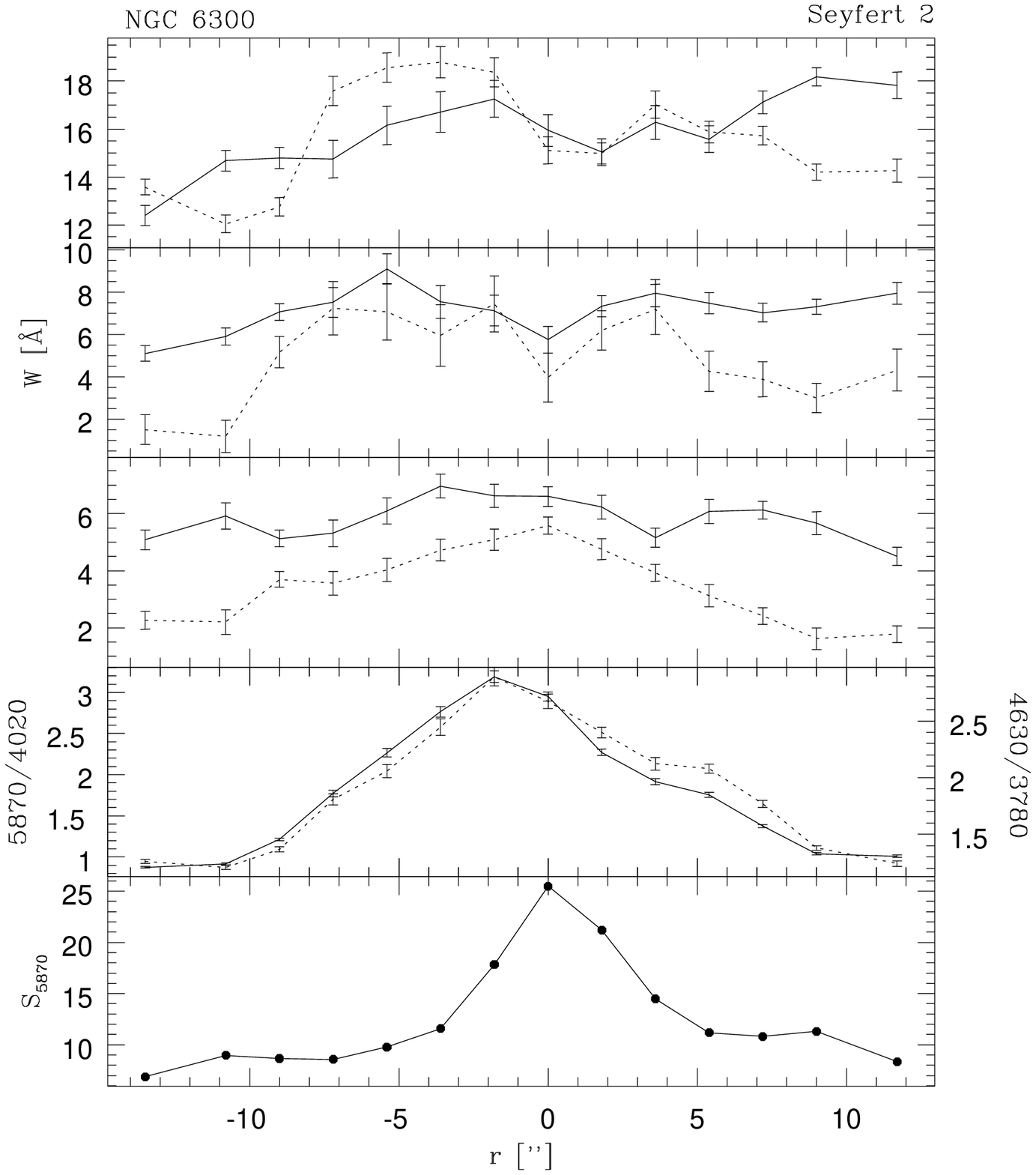}
    \caption{Same as Fig.\ 3.   \label{fig:n6300} }
\end{figure}

\begin{figure}
    \cidfig{7.5cm}{30}{150}{430}{705}{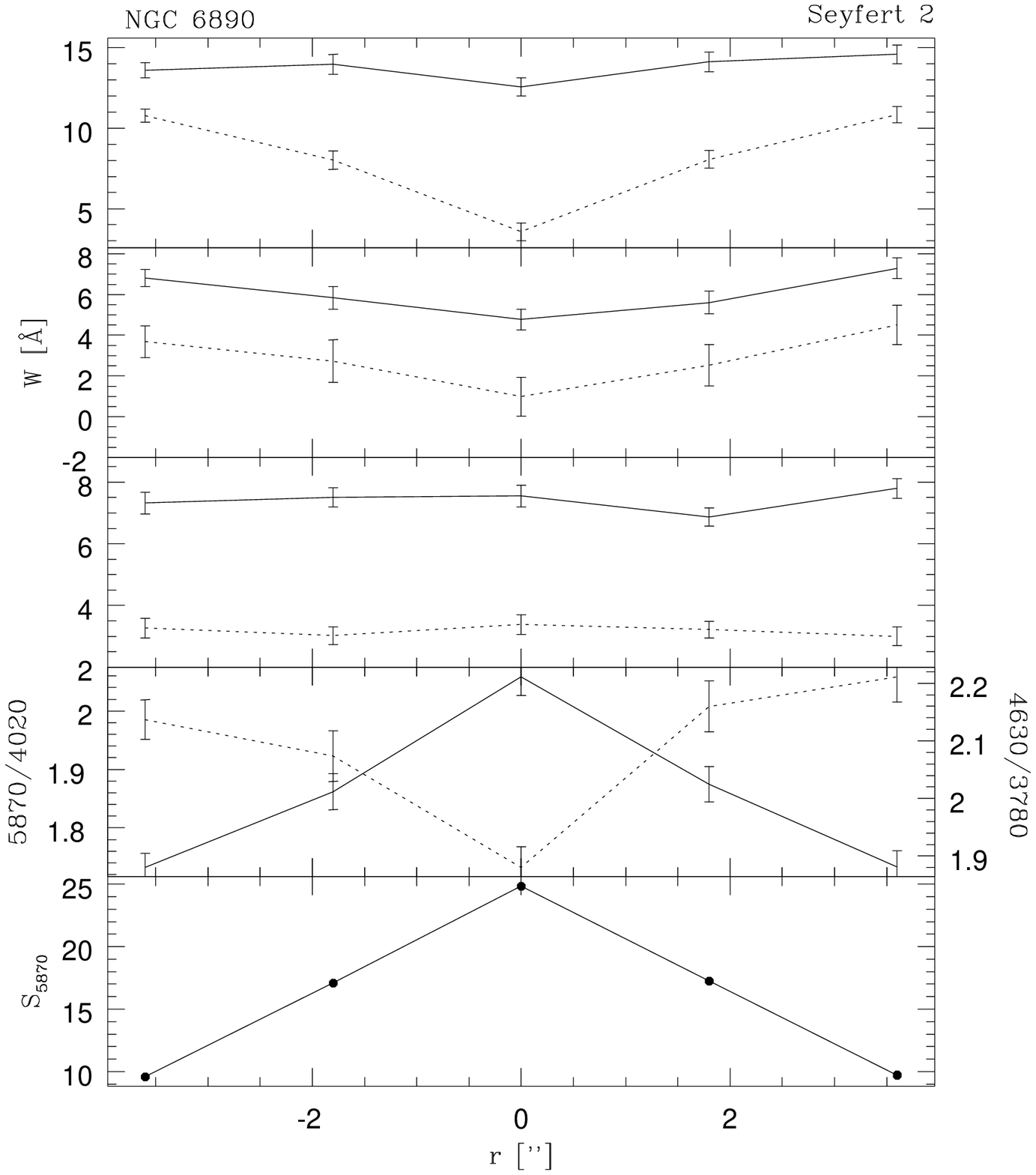}
    \caption{Same as Fig.\ 3.   \label{fig:n6890} }
\end{figure}

\clearpage

\begin{figure}
    \cidfig{7.5cm}{30}{150}{430}{705}{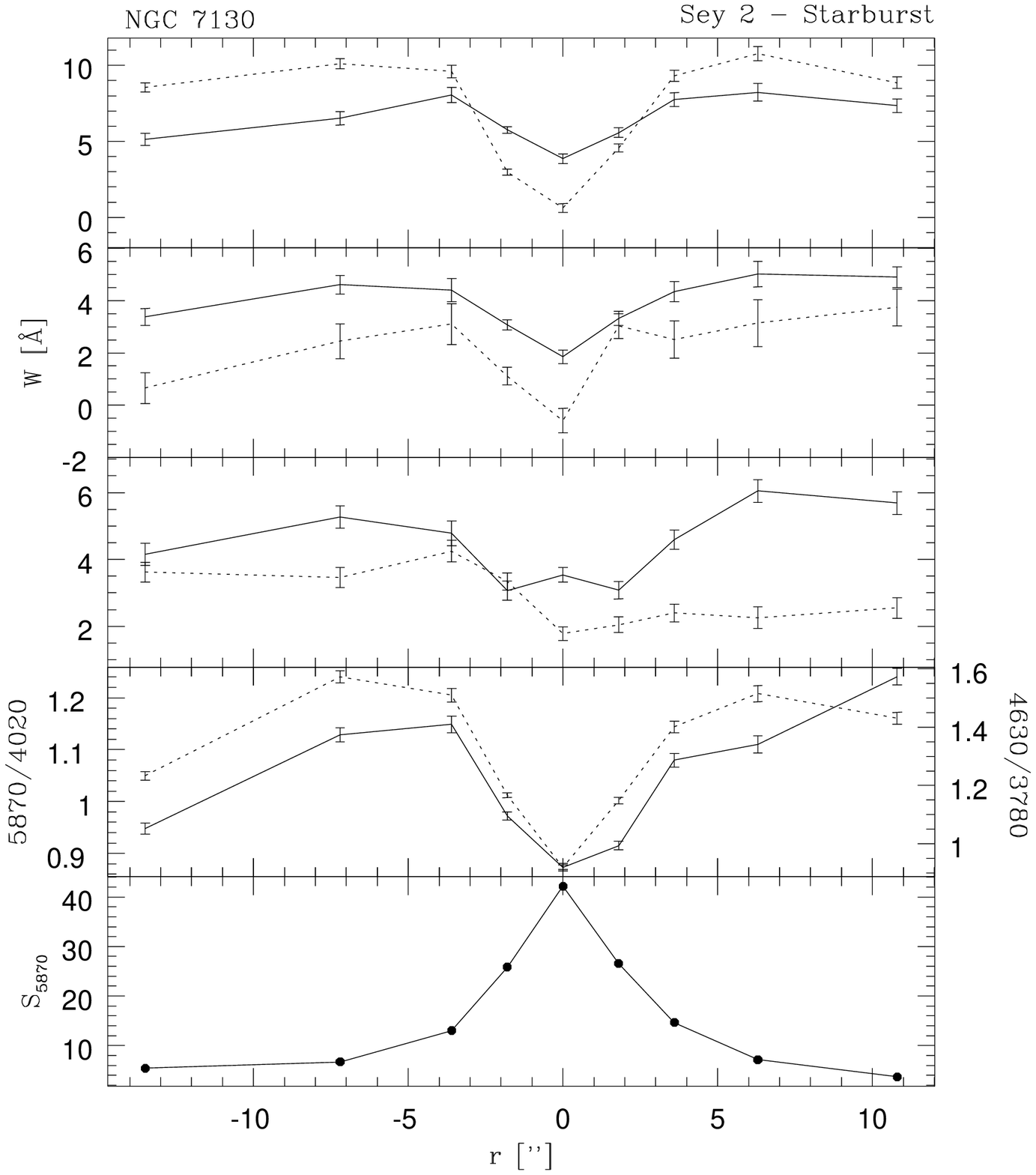}
    \caption{Same as Fig.\ 3.   \label{fig:n7130} }
\end{figure}

\begin{figure}
    \cidfig{7.5cm}{30}{150}{430}{705}{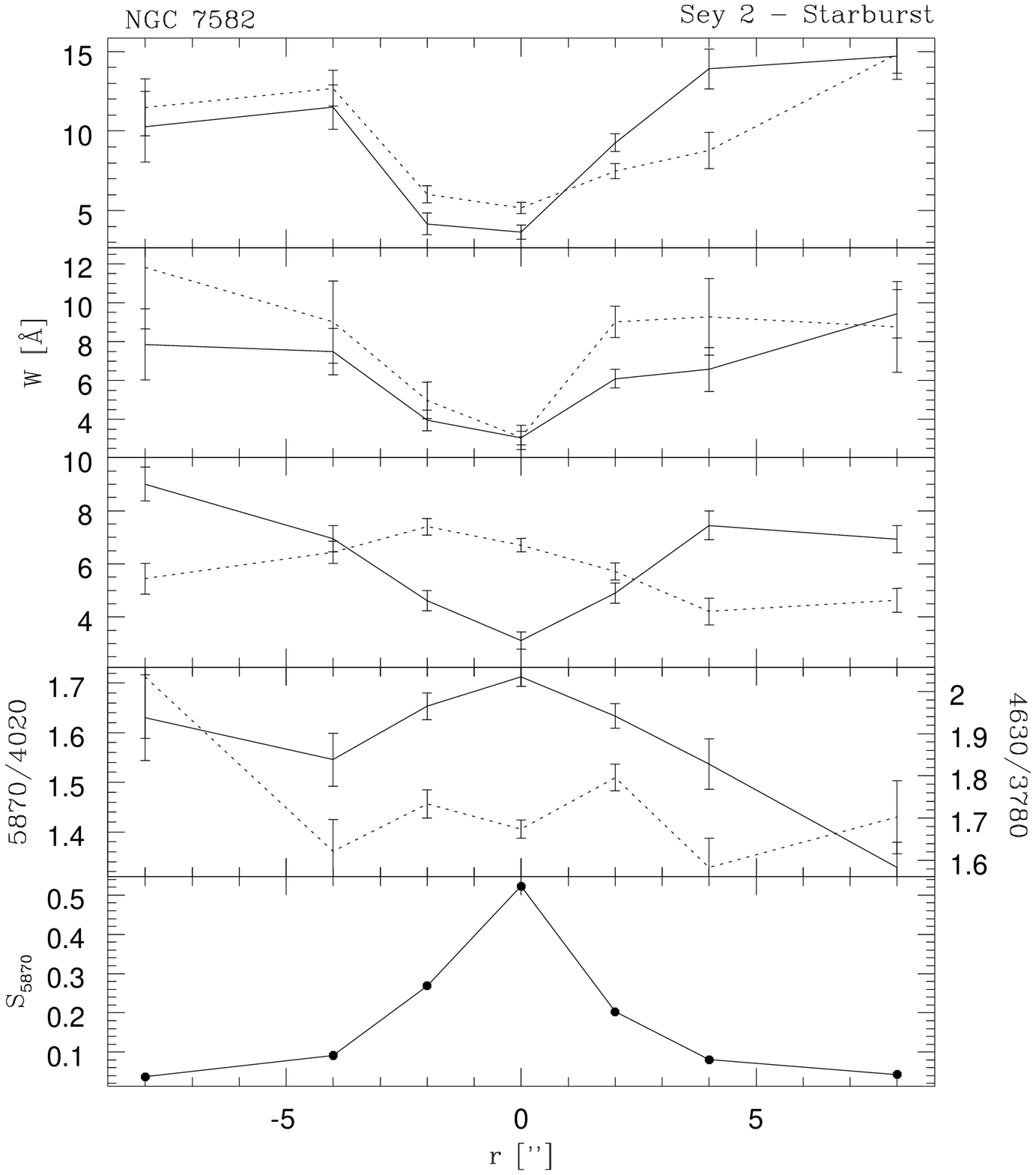}
    \caption{Same as Fig.\ 3.   \label{fig:n7582} }
\end{figure}

\begin{figure}
    \cidfig{7.5cm}{30}{150}{430}{705}{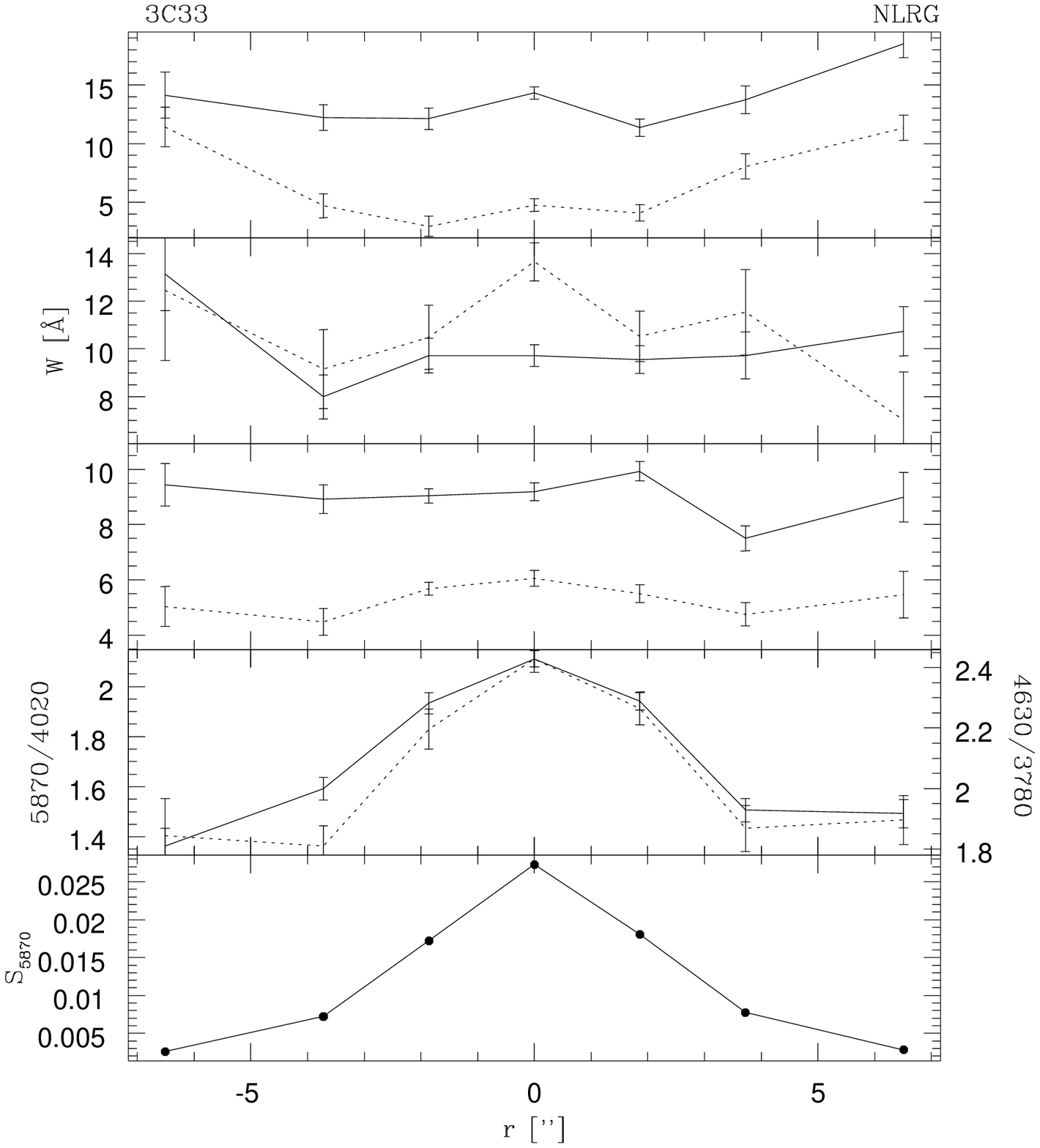}
    \caption{Same as Fig.\ 3.   \label{fig:pc33} }
\end{figure}

\begin{figure}
    \cidfig{7.5cm}{30}{150}{430}{705}{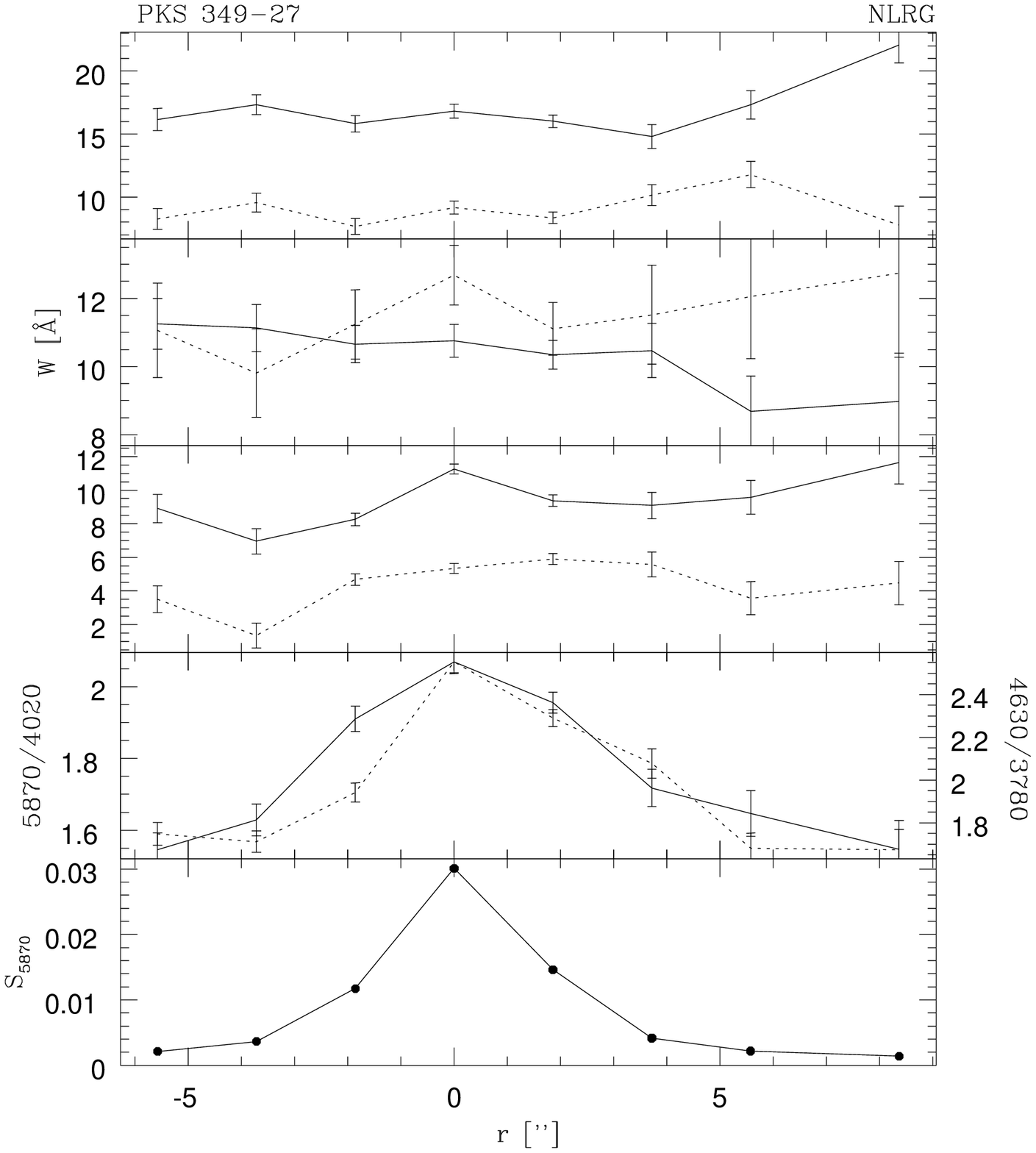}
    \caption{Same as Fig.\ 3.   \label{fig:p349} }
\end{figure}

\clearpage

\begin{figure}
    \cidfig{7.5cm}{30}{150}{430}{705}{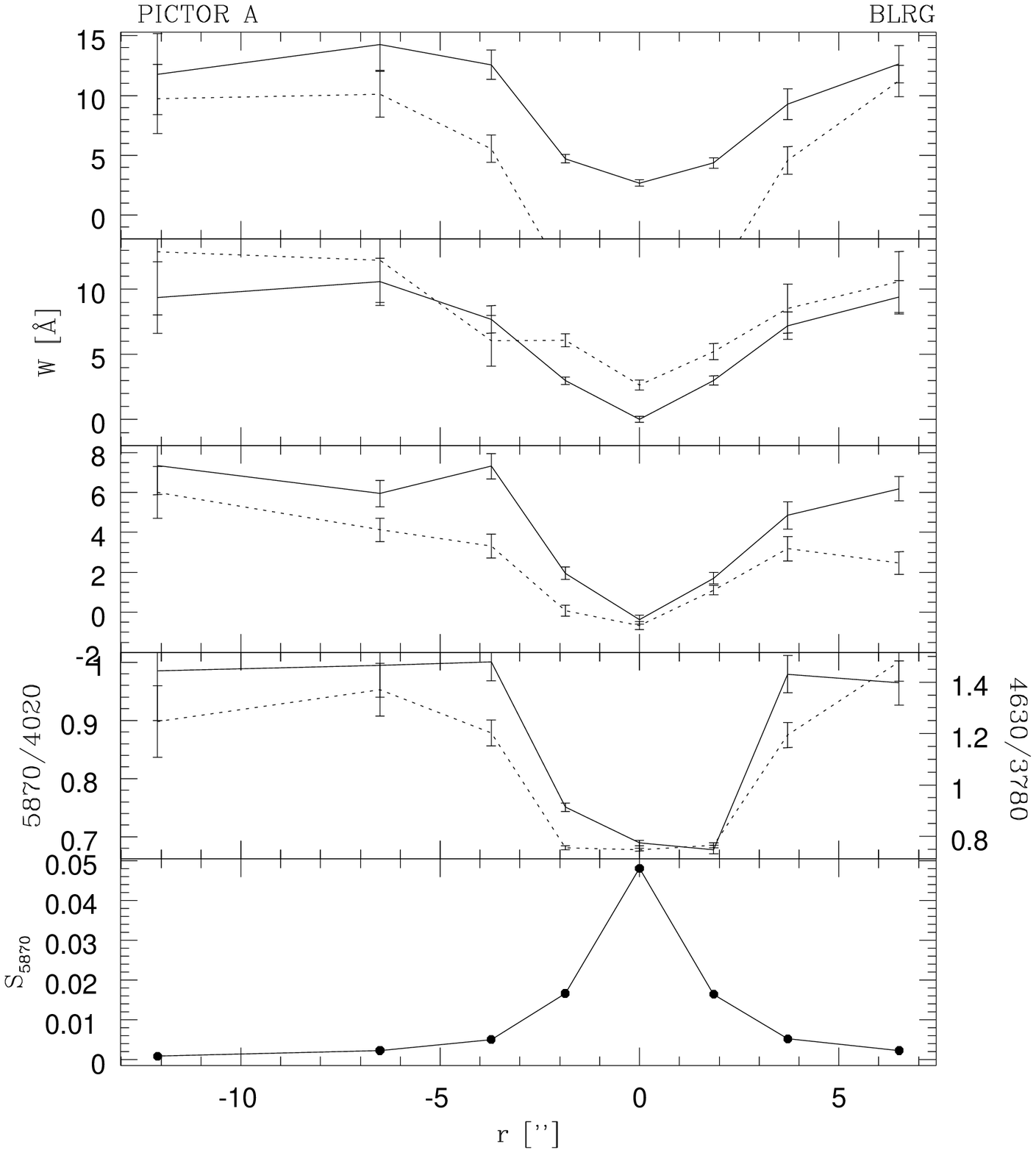}
    \caption{Same as Fig.\ 3.   \label{fig:pic} }
\end{figure}

\begin{figure}
    \cidfig{7.5cm}{30}{150}{430}{705}{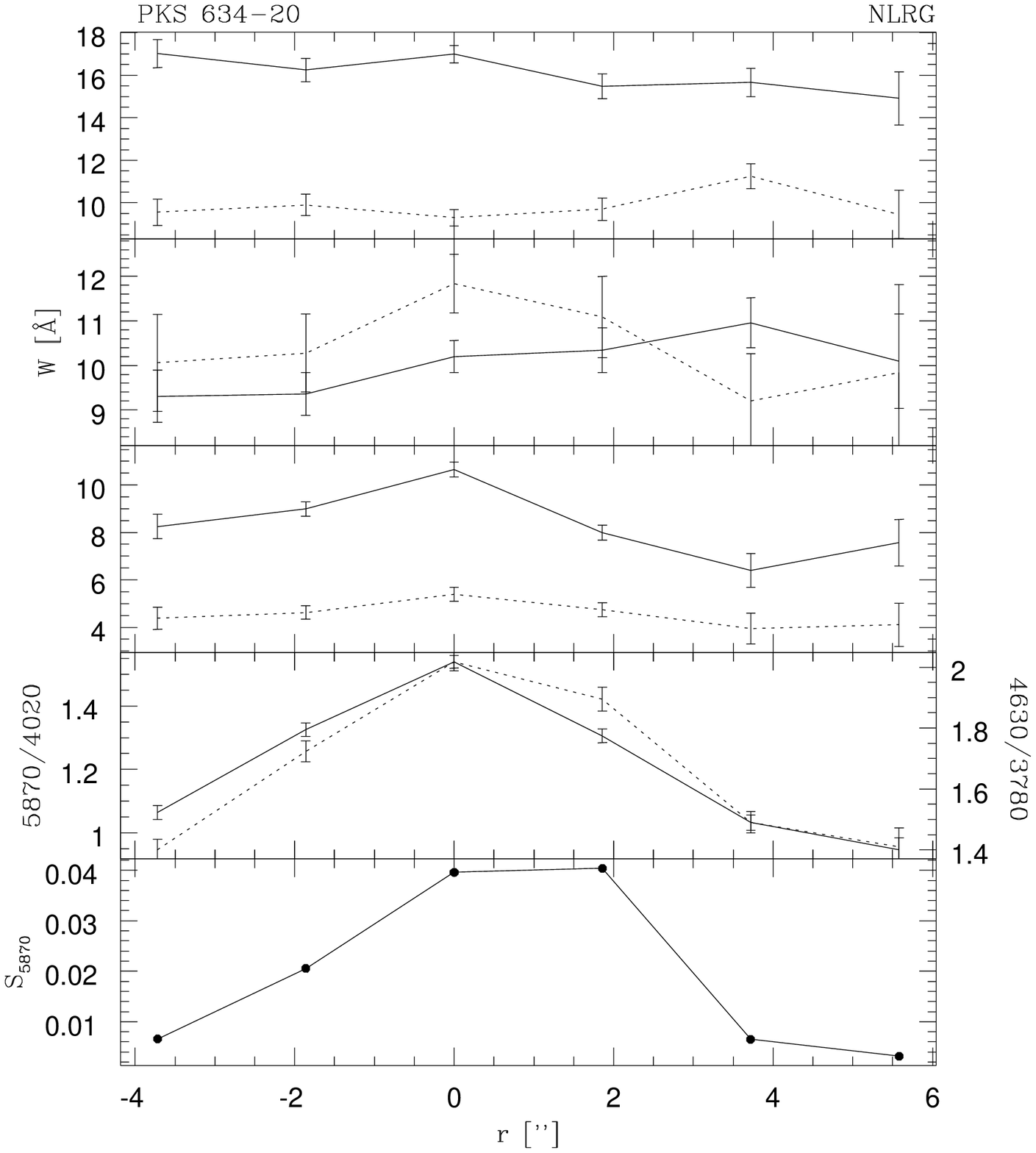}
    \caption{Same as Fig.\ 3.   \label{fig:p634} }
\end{figure}

\begin{figure}
    \cidfig{7.5cm}{30}{150}{430}{705}{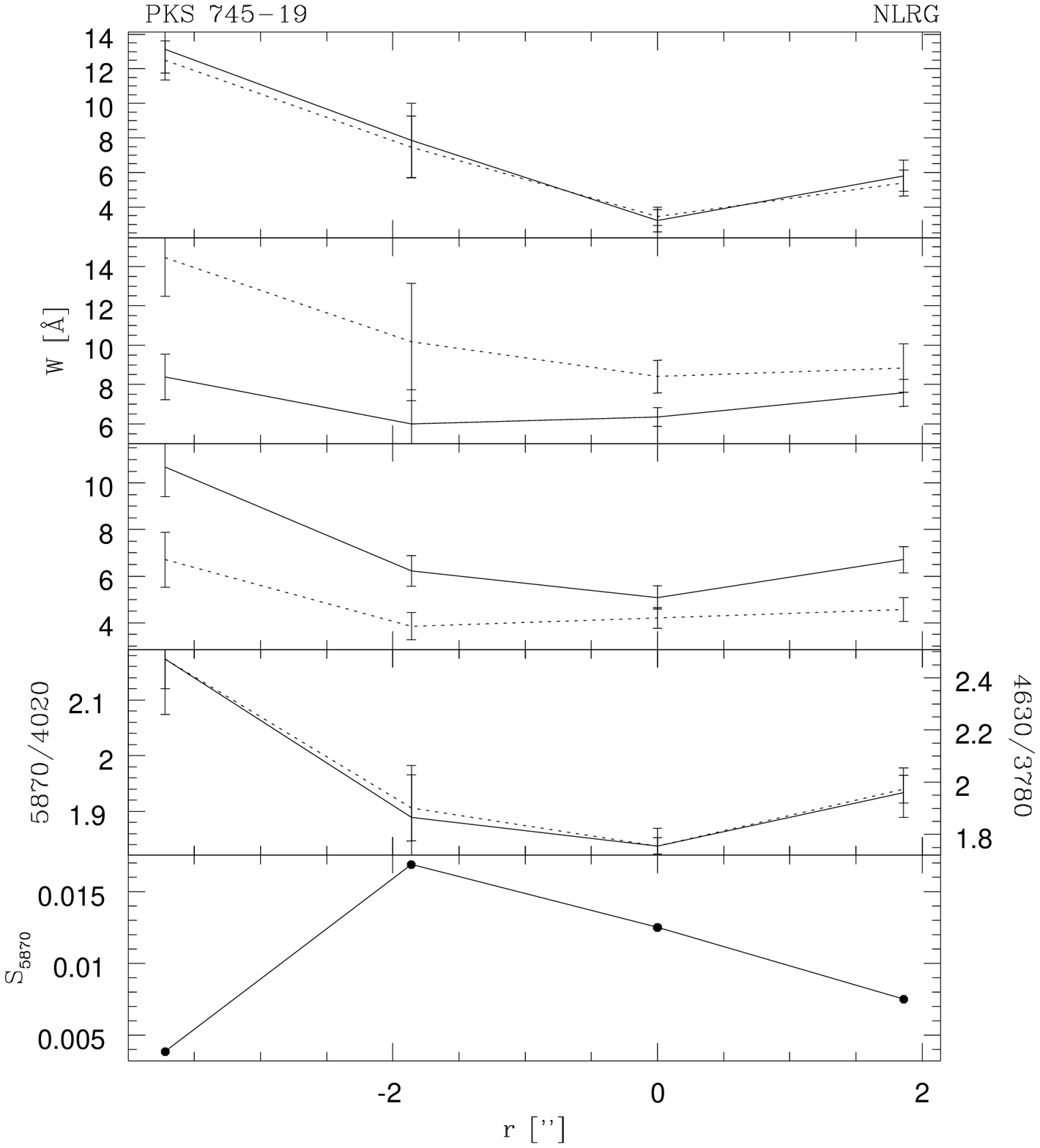}
    \caption{Same as Fig.\ 3.   \label{fig:p745} }
\end{figure}

\begin{figure}
    \cidfig{7.5cm}{30}{150}{430}{705}{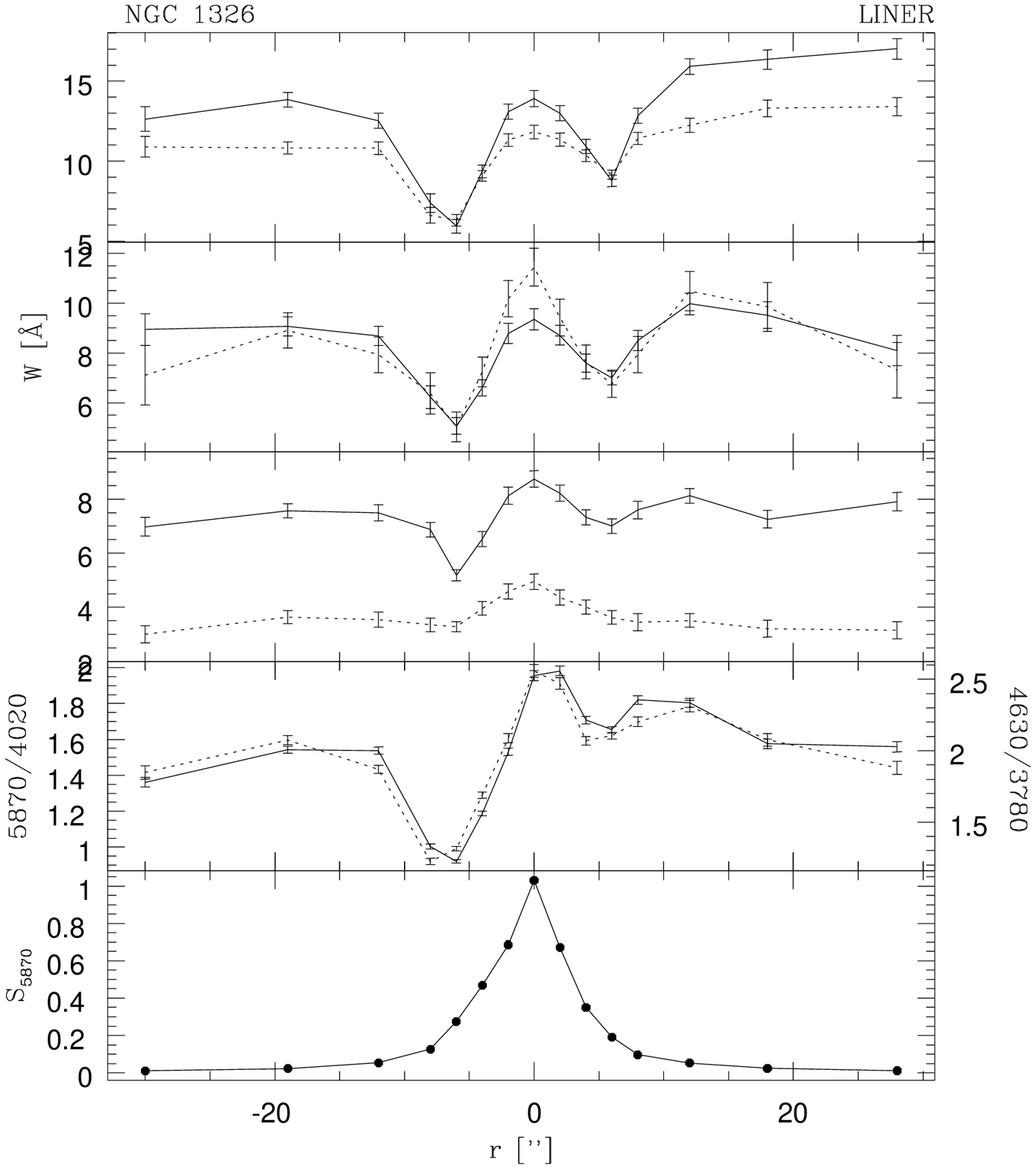}
    \caption{Same as Fig.\ 3.   \label{fig:n1326} }
\end{figure}

\clearpage

\begin{figure}
    \cidfig{7.5cm}{30}{150}{430}{705}{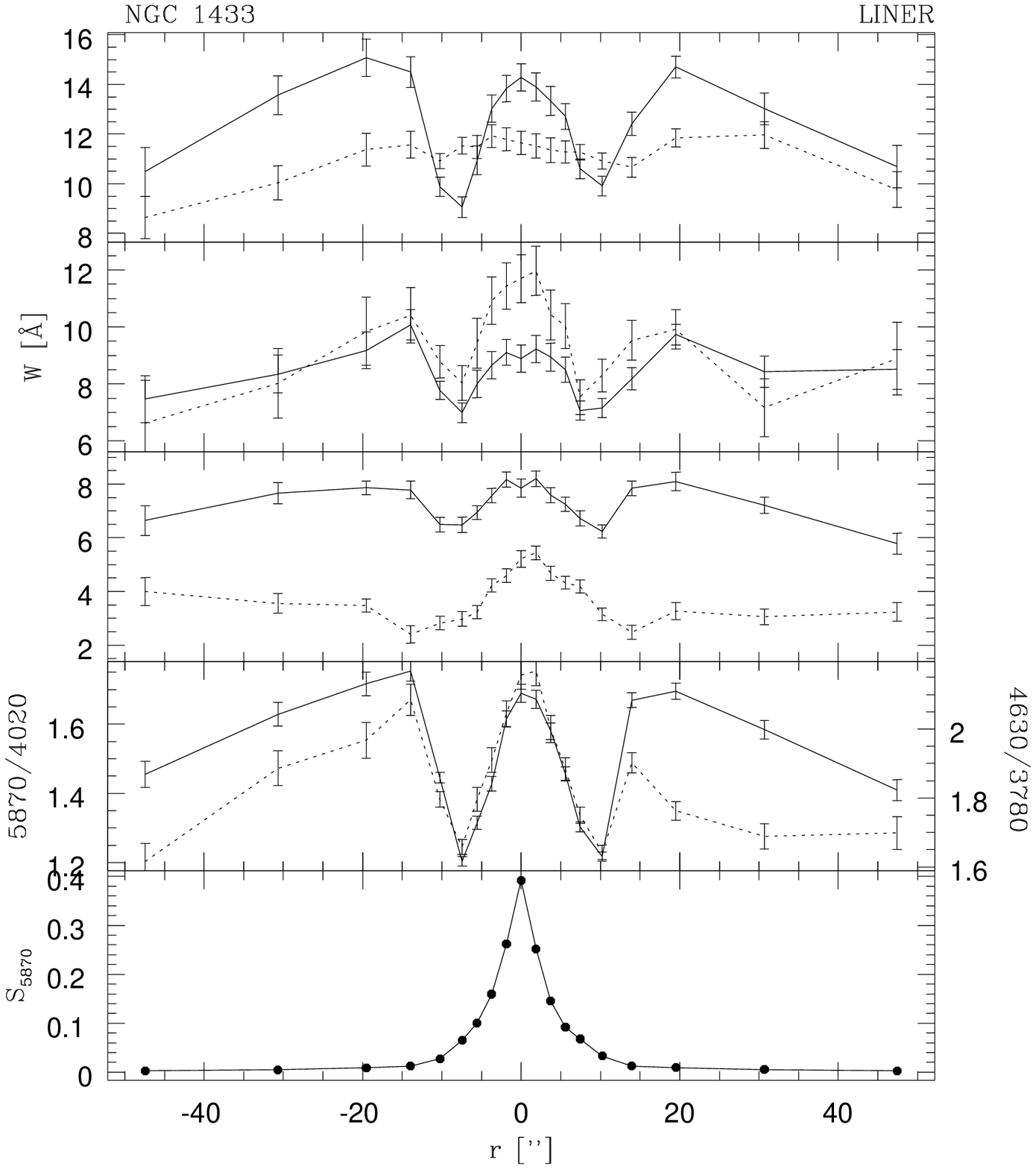}
    \caption{Same as Fig.\ 3.   \label{fig:n1433} }
\end{figure}

\begin{figure}
    \cidfig{7.5cm}{30}{150}{430}{705}{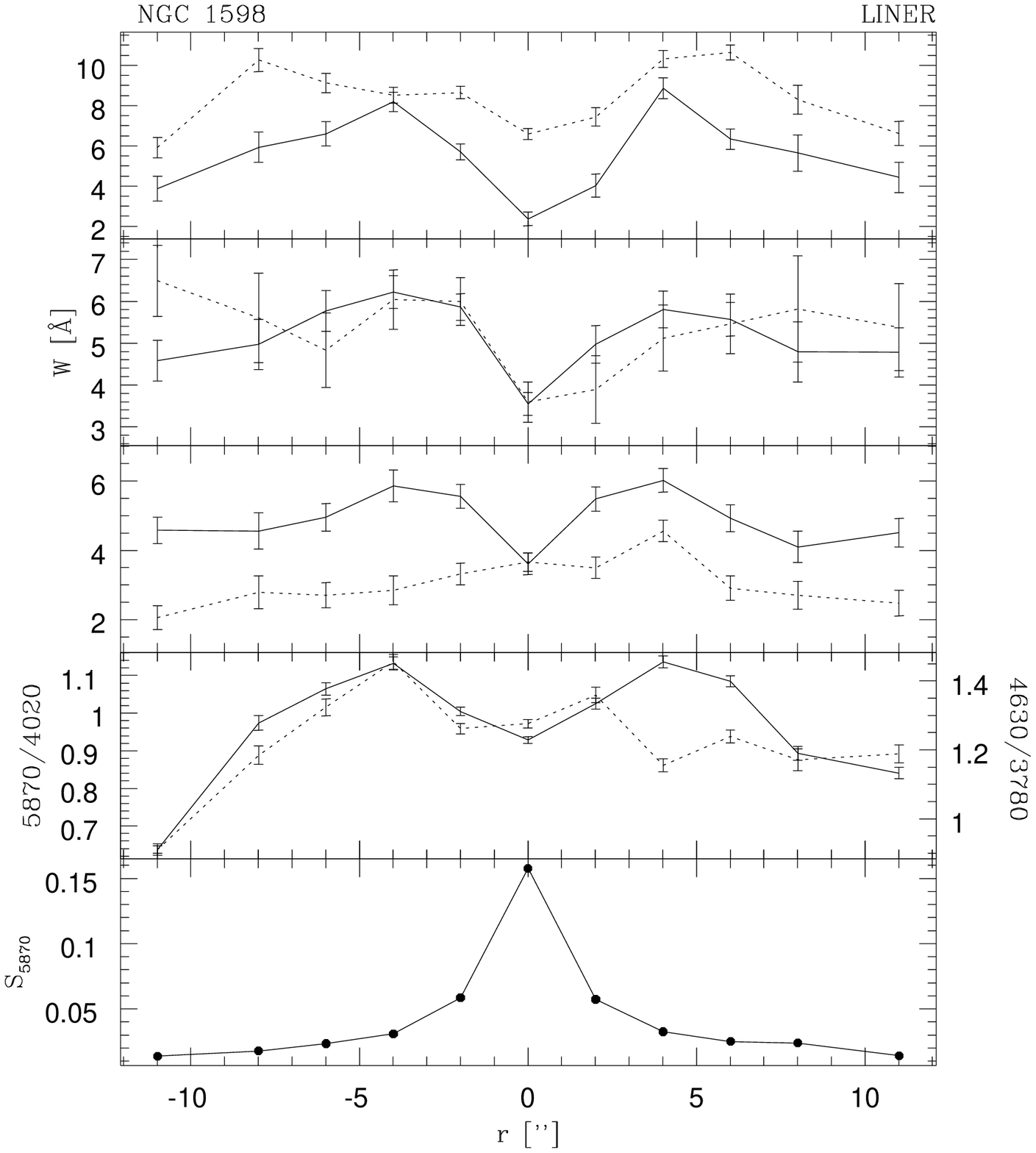}
    \caption{Same as Fig.\ 3.   \label{fig:n1598} }
\end{figure}

\begin{figure}
    \cidfig{7.5cm}{30}{150}{430}{705}{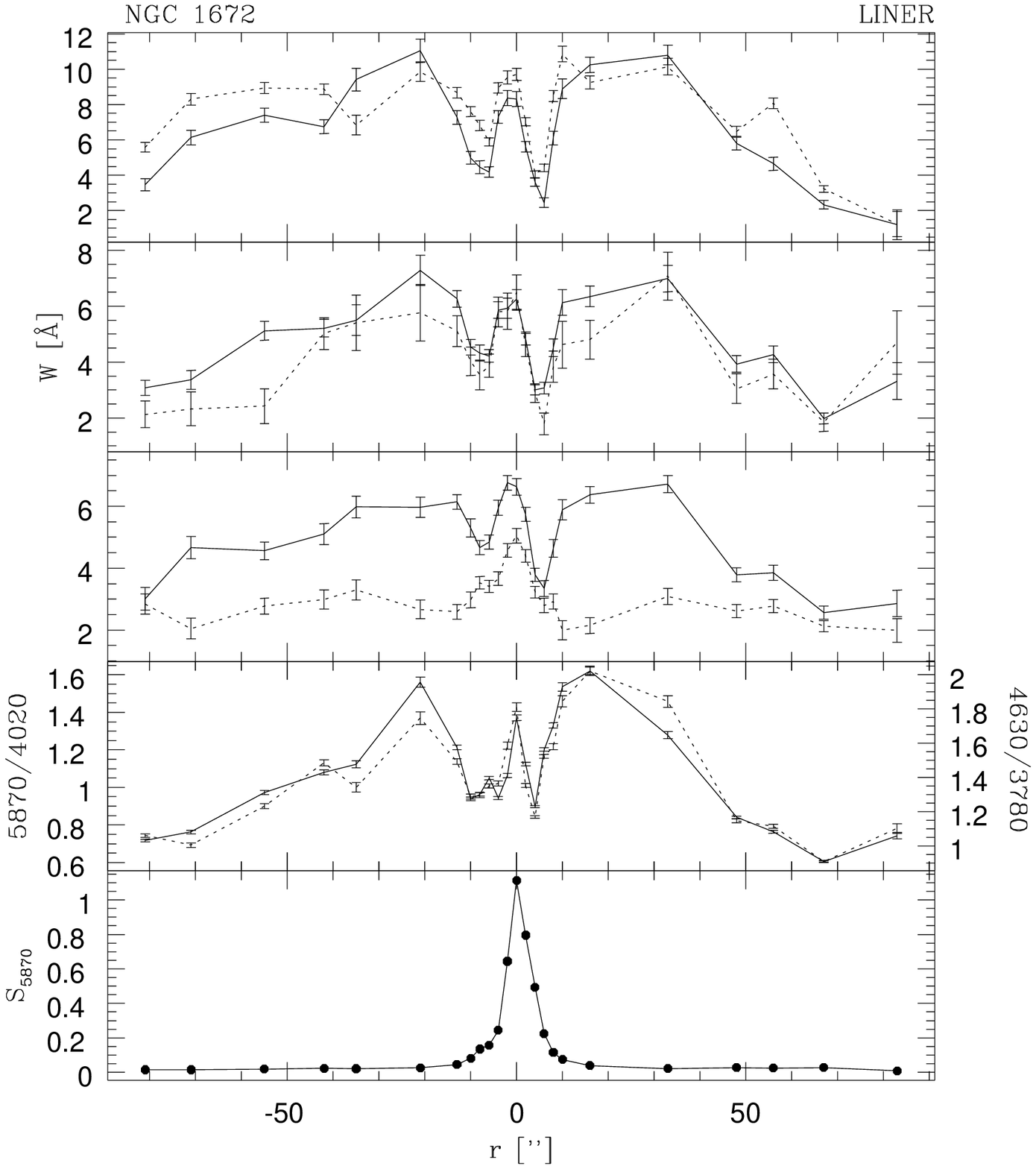}
    \caption{Same as Fig.\ 3.   \label{fig:n1672} }
\end{figure}

\begin{figure}
    \cidfig{7.5cm}{30}{150}{430}{705}{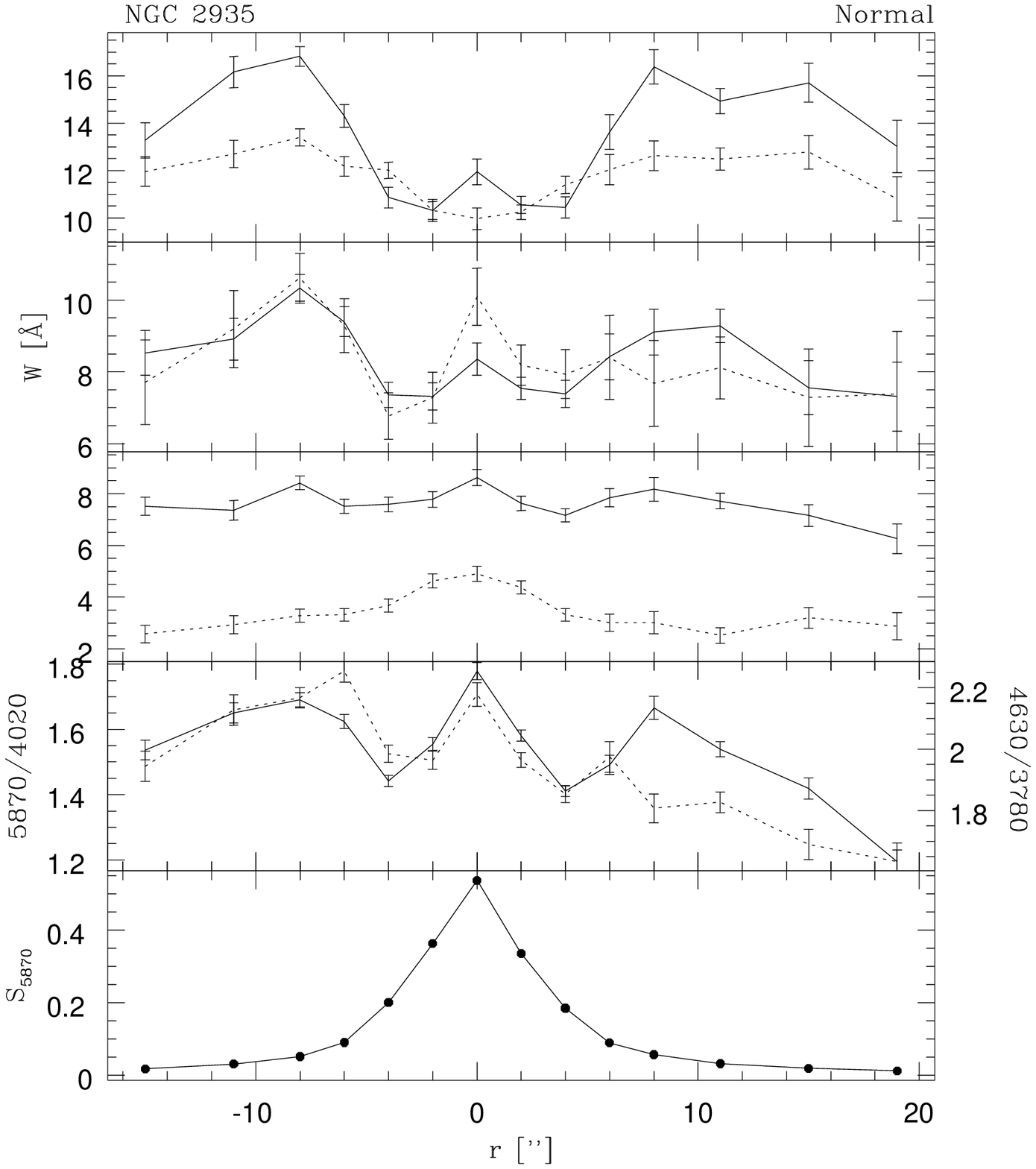}
    \caption{Same as Fig.\ 3.   \label{fig:n2935} }
\end{figure}

\clearpage

\begin{figure}
    \cidfig{7.5cm}{30}{150}{430}{705}{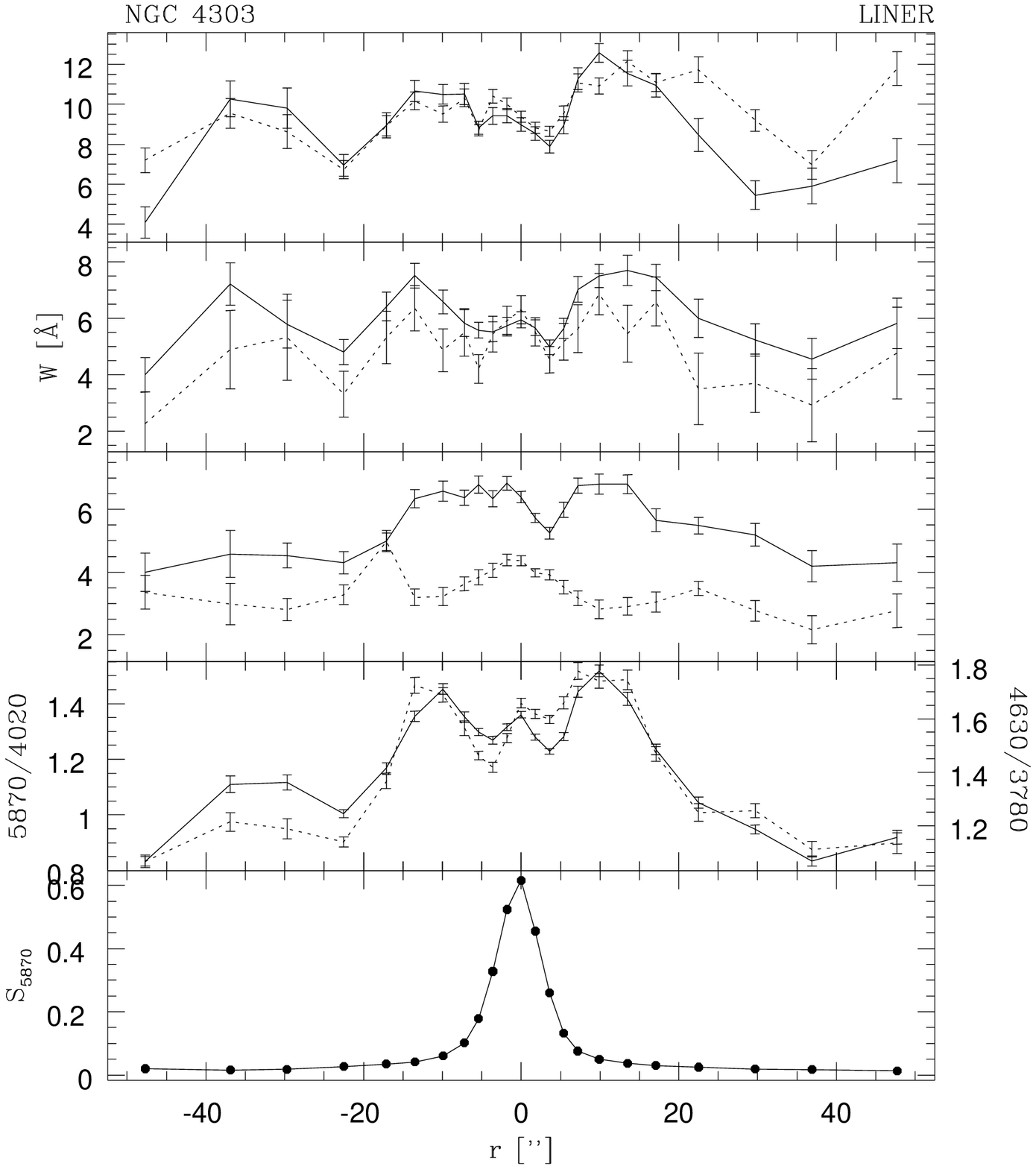}
    \caption{Same as Fig.\ 3.   \label{fig:n4303} }
\end{figure}

\begin{figure}
    \cidfig{7.5cm}{30}{150}{430}{705}{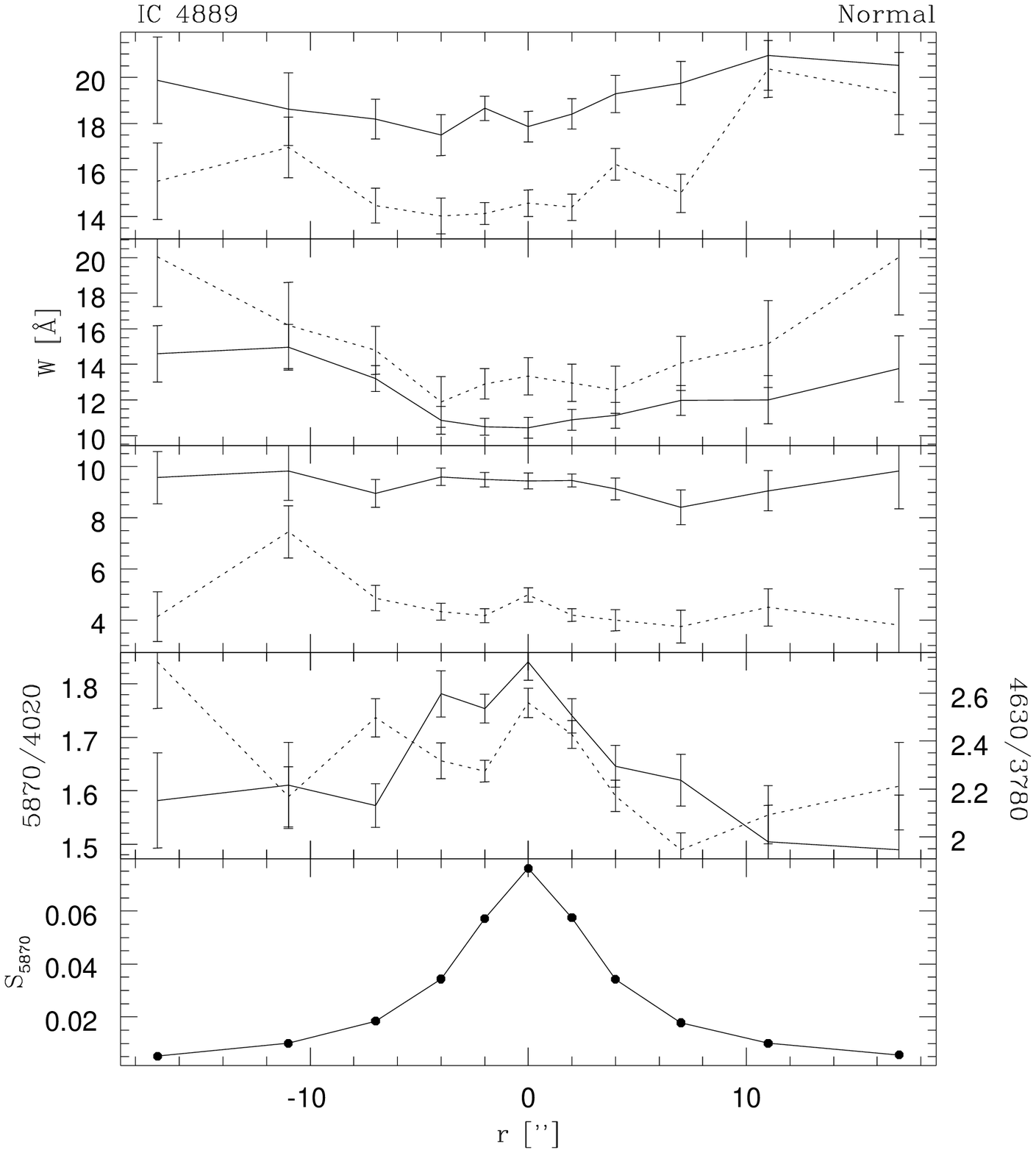}
    \caption{Same as Fig.\ 3.   \label{fig:i4889} }
\end{figure}

\begin{figure}
    \cidfig{7.5cm}{30}{150}{430}{705}{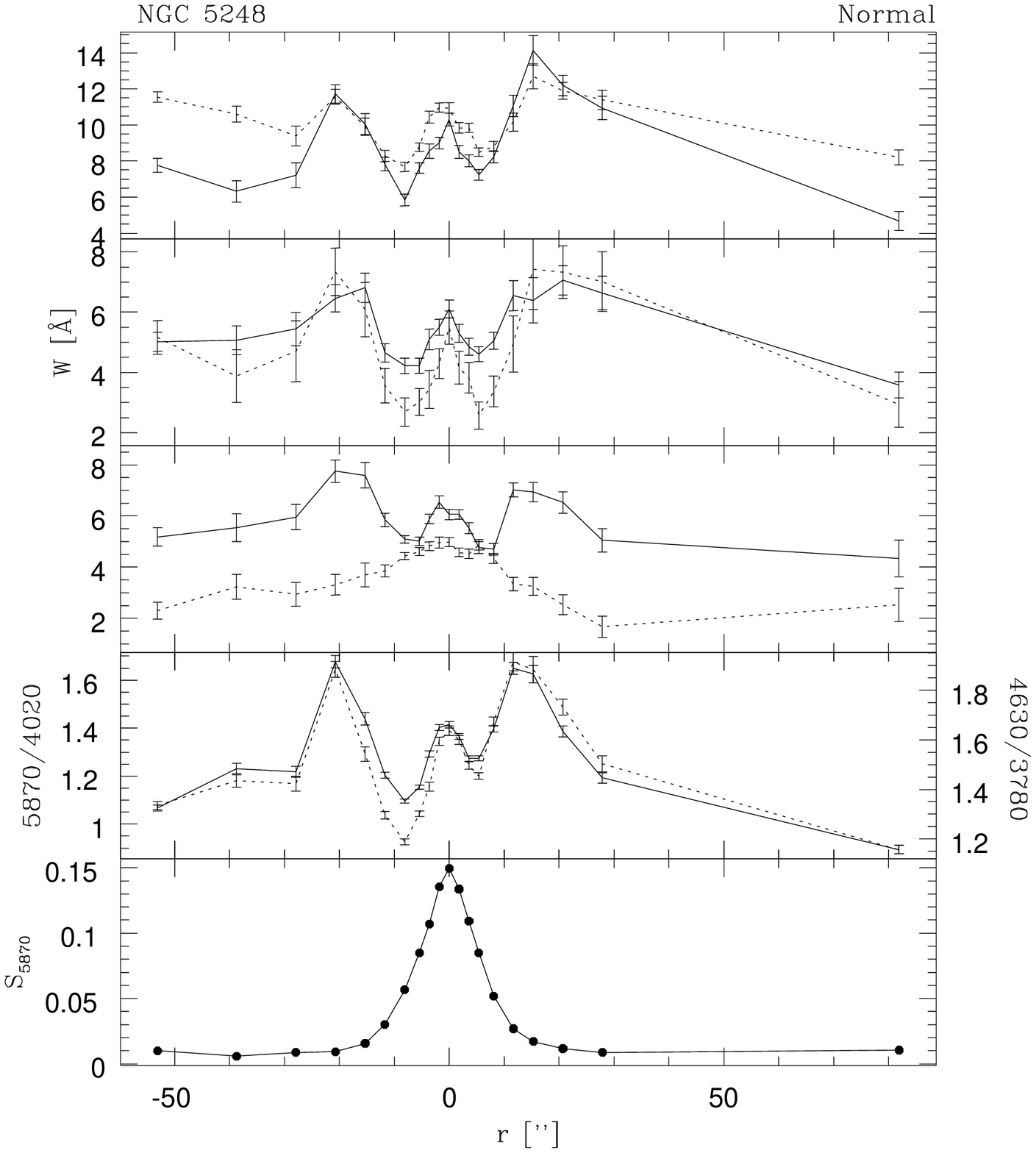}
    \caption{Same as Fig.\ 3.   \label{fig:n5248} }
\end{figure}

\begin{figure}
    \cidfig{7.5cm}{30}{150}{430}{705}{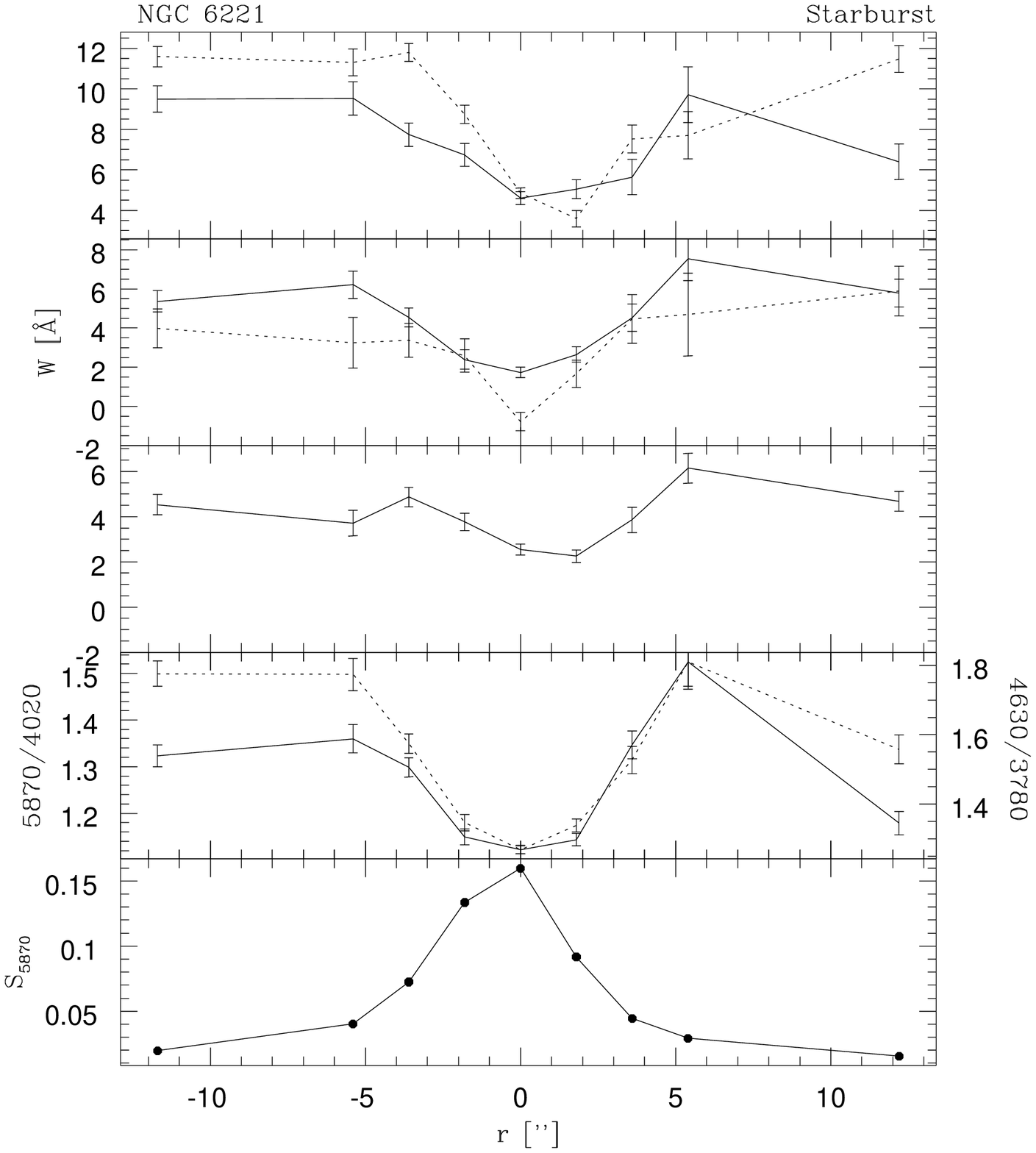}
    \caption{Same as Fig.\ 3.   \label{fig:n6221} }
\end{figure}

\clearpage

\begin{figure}
    \cidfig{7.5cm}{30}{145}{515}{700}{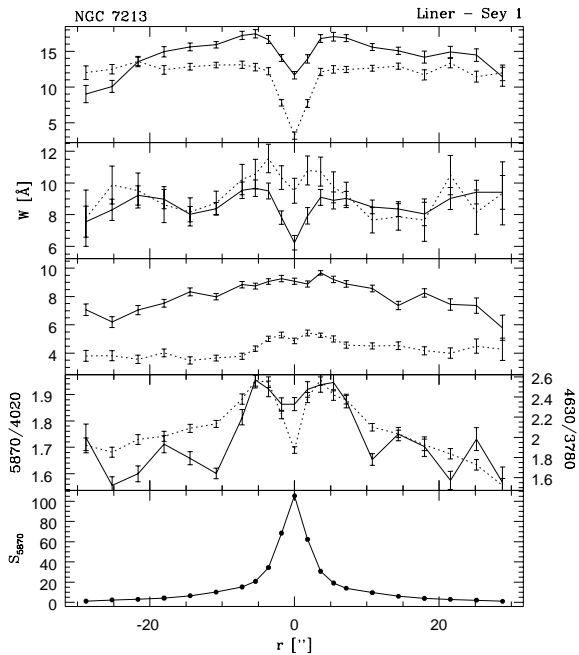}
    \caption{Same as Fig.\ 3.   \label{fig:n7213} }
\end{figure}

One of the clearest examples of dilution by an AGN continuum in our
data is NGC 6814 (Figs.~\ref{fig:n6814} and
\ref{fig:ZOOMED_spectra}c).  The high degree of variability both in
the lines and continuum of NGC 6814 (Yee 1980, Sekiguchi \& Menzies
1990) are an unequivocal signature of the presence of an AGN in this
galaxy. The presence of a FC is reflected in our W profiles, which
show substantial dilution in the nucleus.  Comparing the values of
the nuclear Ws with those between $|r|=$5 and 11\sec\ we obtain
dilution factors of 47\% for Ca~K, 28\% for
CN, 53\% for the G-band and 22\% for Mg. The larger dilution in the
G-band is due to broad H$\gamma$\ entering the line window
(Fig.~\ref{fig:ZOOMED_spectra}c). The FC also affects the nuclear
colors, which are bluer than in the neighboring regions
(Fig.~\ref{fig:n6814}).  The Ws and continuum ratios  at 5\sec\ from
the nucleus correspond to a S2--S3 template, changing to S5--S6
outwards. The attribution of a spectral template to the nucleus is
not meaningful in this case given the obvious presence of a
non-stellar continuum.

An interesting case of a galaxy showing signs of dilution is that of
the LINER/Seyfert 1 NGC~7213 (Phillips 1979, Filippenko \& Halpern
1984). Our W profiles show a clear drop in the Ws of Ca K, CN and
G-band, but not in Mg. (Ca H is also diluted, but mostly due to
H$\epsilon$.) This indicates that the FC is not strong enough in the
5100 \AA\ region to dilute the Mg line, but its contrast with the
stellar component increases towards shorter wavelengths, resulting
in a dilution of $\sim 31$\% of Ca~K. This value is very
similar to that obtained by Halpern \& Filippenko (1984).

Apart from their dilution at the nucleus of the NGC~7213, the Ws of Ca
K and H, CN and G-band indicate a S2--S3 template at 5\sec\ and
outwards. Mg has a W similar to that of a S1 template at the nucleus,
gradually changing to S4 in the outer regions. The continuum ratios are
similar to a S2 template in the inner 5\sec\ radius, decreasing to
S3 in the outwards. The stellar population template obtained by Bonatto
\et (1989) is S2, similar to the values we obtained for the regions
outside the nucleus.

\setcounter{figure}{44}
\begin{figure*}
    \cidfig{18cm}{35}{175}{585}{545}{\DIRFIGS/FIG_ZOOMab.ps}
    \caption{
    Spectra of five of the sample galaxies for several positions
along the slit, zoomed into the regions of the absorption features
discussed in this work. (a) NGC~1672, (b) NGC~7213 (c) NGC 6814,
(d) Mrk~573 and (e) 3C33. The radial positions are indicated on the 
right side of the plots.
    }
    \label{fig:ZOOMED_spectra}
\end{figure*}

\setcounter{figure}{44}
\begin{figure*}
    \cidfig{18cm}{35}{175}{585}{515}{\DIRFIGS/FIG_ZOOMcd.ps}
    \caption{(cont.)}
\end{figure*}

\setcounter{figure}{44}
\begin{figure}
    \cidfig{8cm}{35}{175}{305}{510}{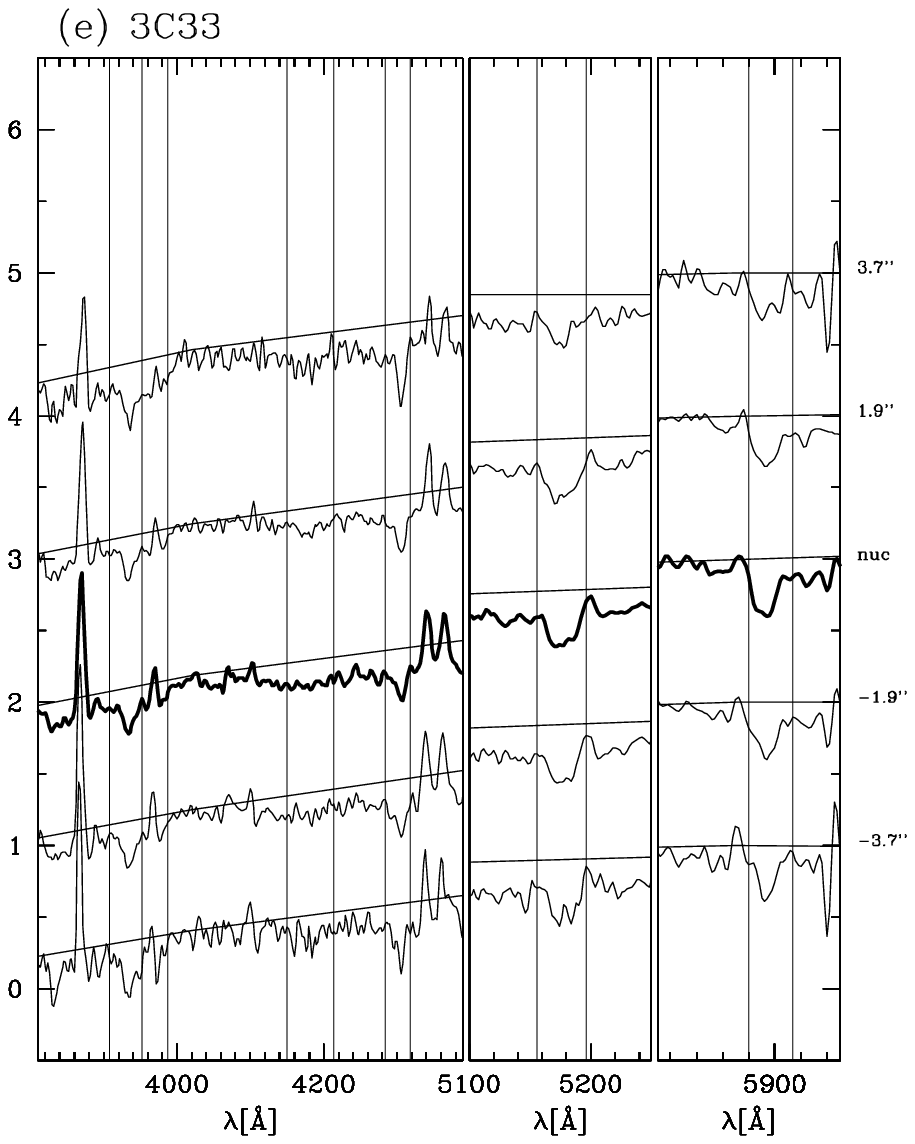}
    \caption{(cont.)}
\end{figure}

In Fig.~\ref{fig:ZOOMED_spectra}b we show the spatially resolved,
zoomed spectra of NGC 7213. The correspondence between the Ws
variations in Fig.~\ref{fig:n7213} and the features in the individual
spectra is clearly seen. The dilution of Ca K and H, CN and G band is
observed as shallower absorption lines in the nuclear spectrum, while
the lack of dilution in Mg can be noticed by the similar depths of the
line in different extractions.

\subsection{Individual Objects}

\label{sec:Indiv}

The five cases discussed above fully illustrate the potential of
long slit spectroscopy as a tool to probe stellar population
gradients and the presence of a featureless AGN-like continuum.  In
this section we provide a description of our results for the
remaining objects in our sample, separated by activity class
(Seyfert 1s, Seyfert 2s, Radio Galaxies, LINERs and Normal
Galaxies).

\subsubsection{Seyfert 1s}

{\bf NGC~526a}:  This is the brightest galaxy of a strongly
interacting pair, which presents an optical emission line region
elongated in the direction of the companion galaxy (NW-SE, Mulchaey
\et 1996). Both Ws and continuum ratios of this galaxy present a
mild gradient (Fig.~\ref{fig:n526a}), but opposite to that expected
from dilution by a blue continuum. The Ws are larger at the nucleus,
indicating S3--S4 templates, decreasing outwards, with values of
S4--S5 templates at 7\sec\ from the nucleus. The continuum ratios
indicate the same templates outside the nucleus but a redder ratio
at the nucleus, corresponding to a S1 template.

{\bf ESO~362-G18}:  According to Mulchaey \et (1996), this Seyfert 1
is a strongly perturbed galaxy, with the strongest [OIII] emission
concentrated in the nucleus and showing an extension to the SE,
suggestive of a conical morphology. The Ws and continuum ratios show
a strong dilution by a FC in the nuclear region
(Fig.~\ref{fig:e18}). At 5\sec\ and farther out, both Ws and
continuum ratios indicate a S5 template.

{\bf Mrk~732}: The individual spectra
(Fig.~\ref{fig:example_spectra}) show signatures of young stars to
the NE of the nucleus ($r < 0$), where the Ws consistently indicate
an S7 template. The nuclear Ws indicate a dilution when compared to
the SW side ($r > 0$), which shows Ws of a S5 template
(Fig.~\ref{fig:m732}). The continuum ratios indicate the same
templates.

{\bf MCG-02-33-034}:  This galaxy was classified as Seyfert 1 by
Osterbrock \& De Robertis (1985), based on the appearence of
permitted lines broader than the forbidden ones and strong FeII
emission. Mulchaey \et (1996) show that it presents evidence of two
nuclei, both emitting in [OIII] and H$\alpha$. We can see the
dilution of Ws and color changes due to a blue FC in the nuclear
region (Fig.~\ref{fig:i309}). The Ws change to values of a S3
template at 4\sec\ and outwards, while the continuum ratios have
values similar to a S4 at 4\sec\ and farther out. As noted in
\S\ref{sec:Method}, though dilution clearly occurs, the complex
nuclear spectrum (Fig.~\ref{fig:example_method}d) prevents an
accurate determination of both Ws and continuum at $r = 0$\ for this
galaxy.

{\bf NGC~6860}: According to L\'{\i}pari, Tsvetanov \& Macchetto
(1993) the H$\alpha$\ image of this galaxy shows bright emission
line regions associated with the nucleus and a circumnuclear ring of
star-formation. The emission line spectrum is typical of a Seyfert
1.5 and variable. The Ws (Fig.~\ref{fig:n6860}) are diluted by a FC
at the nucleus. Their values change to those of a S2 template at
4\sec\ and outwards. The continuum ratios show a gradient in the
opposite direction, decreasing from S4 in the inner 5\sec\ radius to
S5 outwards, bluer than the values indicated by the Ws in the outer
regions.

\subsubsection{Seyfert 2s}

{\bf Mrk~348}: This is a Seyfert 2 galaxy with evidence of broad
H$\alpha$\ (De Robertis \& Osterbrock 1986b), HeI 10830 \AA\ and
Pa$\beta$\ lines (Ruiz, Rieke \& Schmidt 1994). Koski (1978)
observed this galaxy through a 2.7\sec$\times$4\sec\ aperture,
obtaining a 14\% contribution from a FC at 5000 \AA. Kay (1994)
estimated a FC contribution of 35\% to the flux at 4400 \AA, while
Tran (1995a) finds 27\% at 5500 \AA. Mrk 348 is one of the Seyfert 2
galaxies which had a hidden Seyfert 1 nucleus revealed by
spectropolarimetry (Miller \& Goodrich 1990).

The Ws and continuum ratios of this galaxy are shown in
Fig.~\ref{fig:m348}. The Ws have values similar to those of a S4
template, without any noticeable gradient in the inner 8\sec.  From
Koski (1978), Kay (1994) and Tran (1995a) results, we would expect
the Ws to be substantially diluted in the nuclear region, but this
is not detected. One possibility would be that the FC is extended,
diluting {\it all} the Ws. However, the NLR of this galaxy is
extended by less than 2\sec\ (Schmitt \& Kinney 1996), and the FC
would have to be extended by $\approx 8$\sec.

The continuum ratios show a gradient, having values similar to those
of a S4 template at the nucleus and becoming bluer outwards,
reaching values similar to those of S5--S6 templates. The continuum
behavior of Mrk~348 is intriguing, due to the fact that outside the
nucleus it indicates bluer templates than the ones expect from the
analisys of the Ws. This behavior is also seen in many other objects
in the present sample, particularly Seyfert 2s.

{\bf Mrk~573}: Pogge \& De Robertis (1993) found excess near-UV
emission, spatially extended along the [OIII] emission, interpreted
as scattered nuclear continuum.  Koski (1978) estimated that the FC
contributes 12\% of the light at 5000 \AA, observed through a
2.7\sec$\times 4.0$\sec\ aperture, while Kay (1994) finds 20\% FC at
4400 \AA\ within an aperture of $\approx 2 \times 6$\sec. She also
estimated, based on spectropolarimetry, that the FC is polarized by
5.6\% in the 3200--6300 \AA\ range, in agreement with the values
found by Martin \et (1983).

The Ws (Fig.~\ref{fig:m573}) show a mild gradient, from a S3
template at the nucleus, to a bluer one, S4 at 5\sec.  The continuum
ratios show a gradient from a S3--S4 template at the nucleus to a
bluer one, S5--S6 at 5\sec.  Koski (1978) and Kay (1994) estimations
of FC continuum contribution to the nuclear spectrum are not
confirmed by our data, which do not show any detectable dilution in
the Ws.  This galaxy is another case in which the continuum ratios
outside the nucleus have values bluer than we would expect from the
Ws analysis.

{\bf IC~1816}: This galaxy was identified as a Seyfert 1 by Fairall
(1988), but the lack of variability lead Winkler (1992) to question
this classification. Our spectrum does not show broad emission
lines, which lead us to classify it as a Seyfert 2. The Ws of this
galaxy (Fig.~\ref{fig:i1816}) have values typical of S3 templates in
the inner 4\sec, with an {\it apparent} dilution in Mg caused by
contamination by [NI] emission.  The Ws decrease to values of a S5
template at 8\sec\ from the nucleus.  The continuum ratios have
values corresponding to templates bluer than the Ws, S5 at the
nucleus, changing to S6 at 8\sec.

{\bf ESO~417-G6}: Our spectrum was obtained with the slit oriented
along the extended emission (NW-SE direction, according to Mulchaey
\et 1996). The Ws show no gradient (Fig.~\ref{fig:e417}), having
values typical of a S3 template. The {\it apparent} dilution in the
Mg band is due to contamination by [NI] emission. The continuum
ratios show a gradient, changing from S3 at the nucleus, to a bluer
S5 template at $\approx 5$\sec\ from the nucleus.

{\bf Mrk~607}:  According to De Robertis \& Osterbrock (1986a) the
nuclear spectrum of this galaxy shows high-excitation lines, like
[FeVII] and [FeX], but only narrow permitted lines. Kay (1994)
obtained a 10\% FC contribution to the nuclear flux, and  estimates
a 4.3\% continuum polarization in the 3200--6300 \AA\ range.

The Ws and continuum ratios (Fig.~\ref{fig:m607}) indicate similar
templates, S2--S3 at the nucleus, S5 at 6\sec, suggesting a weak
starforming ring, and S3 outwards. Our spectrum of this edge-on
galaxy was obtained with the slit positioned along the major axis,
which may explain the large variations in the Ws in such small
scales. Considering these variations, we cannot evaluate whether the
nuclear Ws are diluted or not.

{\bf NGC~1358}: The nuclear FC contributes with 17\% of the flux and
is polarized by 1.7\% in the 3200--6300 \AA\ range, according to Kay
(1994). The Ws of NGC~1358 (Fig.~\ref{fig:n1358}) show no systematic
gradient and have values typical of a S1 template, the same result
obtained by Storchi-Bergmann \& Pastoriza (1989). The only exception
is Ca K, which reaches values larger than S1 at $r = -6$\sec. The
continuum ratios show a gradient from values typical of a S1
template at the nucleus to bluer values typical of a S5 template
outwards. This galaxy is another case in which the continuum ratios
outside the nucleus have values bluer than the ones expected from
the analysis of the Ws.

{\bf NGC~1386}: The Ws are typical of a S3 template at the nucleus
(Fig.~\ref{fig:n1386}), S4--S5 at 10\sec\ from the nucleus, due to
the presence of spiral arms (Storchi-Bergmann \et 1996a), and S3
outwards. The continuum ratios indicate a S1 template at the
nucleus, with a gradient to values typical of S5--S6 at 10\sec\ and
outwards.

{\bf CGCG~420-015}:  This galaxy was classified as Seyfert 2 by de
Grijp \et (1992). The W values are indicative of a S3 template at
the nucleus, with G-band and Mg showing a small gradient to S2
outwards, consistent with a small dilution in the nucleus
(Fig.~\ref{fig:i156}). The continuum ratios indicate a S1 template
at the nucleus, redder than the value predicted from the Ws, with a
gradient to S3--S4 at 4\sec\ and farther out.  Care must be taken
when analyzing the Ws and continuum ratios of this galaxy, because
it was observed with the slit positioned almost perpendicular to the
parallatic angle, introducing large differential refraction effects.

{\bf ESO~362-G8}: The Ws are similar to a S5--S6 template at the
nucleus, with a gradient to S4--S5 at 4\sec\ and farther out
(Fig.~\ref{fig:e8}). The continuum ratios are similar to a S5
template at the nucleus, changing to bluer values, typical of S6--S7
templates outwards. The Ws of this galaxy indicate bluer templates
in the nuclear region, which is due to the presence of young stars,
revealed by strong HI absorption features. This galaxy also has
continuum ratios outside the nucleus bluer than the values predicted
from the Ws.

{\bf Mrk~1210}: Tran, Miller \& Kay (1992) found polarized broad
H$\alpha$\ and H$\beta$\ components in the spectrum of this galaxy.
Later Tran (1995a,b,c) confirmed this result and determined that the
FC contributes to 25\% of the nuclear flux at 5500 \AA, while Kay
(1994) estimates a 64\% FC contribution at 4400 \AA. Nevertheless,
the Ws show almost no gradient indicative of such a dilution
(Fig.~\ref{fig:m1210}). The Ws correspond to S4 and S5 templates.
The continuum ratios are similar to S5--S6 at the nucleus, with a
gradient to S6 outwards, which is bluer than the template estimated
from the Ws. There is an {\it apparent} dilution of the Mg and Na
features in the nuclear spectrum due to emission line contamination
by [NI] and HeI respectively.

{\bf NGC~3081}:  This galaxy was classified as a Seyfert 2 by
Phillips, Charles \& Baldwin (1983). According to Pogge (1989a), the
[OIII] image is symmetrical but more extended than the stars
profile. The W profiles (Fig.~\ref{fig:n3081}) do not show any
evidence of a stellar population gradient, with values similar to
S2--S3 templates. The continuum ratios show a gradient from a S2--S3
template at the nucleus, to S4--S5 at 5\sec\ and outwards, bluer
than the values indicated by the Ws.

{\bf IRAS~11215-2806}:  This galaxy was classified by de Grijp et
al. (1992) as a Seyfert 2. The Ca~K and G-band Ws
(Fig.~\ref{fig:i281}) present a gradient from a S4 template at the
nucleus to S2 in the outer regions, indicating dilution by a blue
continuum. Mg presents almost no gradient, with values typical of
S3--S4 templates. The continuum ratios present a gradient in the
opposite direction, from S4 at the nucleus, to S5 outwards.

{\bf MCG-05-27-013}:  This galaxy was classified as Seyfert 2 by
Terlevich \et (1991). The Ws of Ca~K are typical of a S3 template at
the nucleus, changing to S1 outwards, which suggests dilution by a
FC (Fig.~\ref{fig:i282}). The G-band has a smaller gradient, with
values similar to a S2 template at the nucleus and S1 outwards,
while for Mg the values are similar to a S2 template, with no
gradient. The continuum ratios have values larger than S1 at the
nucleus, due to strong reddening, changing to values similar to S4
in the outer regions, bluer than the values predicted by the
analysis of the Ws.

{\bf Fairall 316}:  This is a Seyfert 2 galaxy discovered by Fairall
(1981). The Ws show no gradient (Fig.~\ref{fig:f316}), with Ca~K
having values typical of S2, while G-band and Mg have values typical
of S1. The continuum ratios have values of S1 template at the
nucleus, decreasing outwards to values bluer than those predicted by
the Ws, similar to a S3 template.

{\bf NGC~5135}: NGC5135 was described by Phillips, Charles \&
Baldwin (1983) as having a {\it composite} nucleus with
characteristics of both Seyfert 2 and Starburst.  Thuan (1984)
obtained IUE spectra of this galaxy, confirming the dual nature of
the nucleus, which presents both emission lines typical of Seyfert
2s and absorption lines typical of Starbursts, the latter component
contributing with 25\% of the total UV emission.  

The Ws (Fig.~\ref{fig:n5135}) show a gradient from values typical of
a S7 template at the nucleus to S6--S5 at 7\sec. The continuum
ratios also show a gradient from S6 at the nucleus to S5 at 7\sec.
This gradient is due to the presence of young stars in the nuclear
region, as revealed by strong HI absorption features.

{\bf NGC~5643}:  This galaxy was classified by Phillips \et (1983)
as a low luminosity Seyfert 2. Schmitt, Storchi-Bergmann \& Baldwin
(1994) presented [OIII] and H$\alpha$\ images, as well as optical
spectra of NGC5643. The images show the high-excitation gas to be
extended along the E-W direction by $\approx 20$\sec. The stellar
population is moderately old in the central region, showing evidence
of absorption lines diluted by a blue continuum, supposed to be
scattered nuclear light.

The Ws of this galaxy present a mild gradient to the E of the
nucleus (i.e., towards negative $r$\ in Fig.~\ref{fig:n5643}),
suggesting the presence of scattered light up to at least 4\sec\
from the nucleus. In the outer regions, the Ws correspond to
templates S3--S4, while the continuum ratios correspond to S3 to the
E of the nucleus and S1 to the W. Bonatto \et (1989) obtained that
the nuclear stellar population of this galaxy can be represented by
a S3 template, similar our result.

{\bf NGC~6300}:  Storchi-Bergmann \& Pastoriza (1989) observed the
nuclear stellar population of this galaxy to be old, with a small
contribution from intermediate age stars. The Ws do not show a clear
gradient (Fig.~\ref{fig:n6300}), with Ca~K varying between S2 and S3
templates, while G-band and Mg vary between S4 and S5. The continuum
ratios decrease from values much larger than those of a S1 template
at the nucleus to values typical of a S6 template at 10\sec\ and
farther out, bluer than the values predicted from the Ws in this
region.

{\bf NGC~6890}: Storchi-Bergmann, Bica \& Pastoriza (1990) studied
the nuclear optical spectrum of this galaxy, finding that the
stellar population is old and that H$\alpha$\ may have a broad
component. The Ws show a gradient in Ca K and G band, which change
from S4 at the nucleus to S3 in the outer regions, but not in Mg,
which has values typical of a S3 template, with no gradient
(Fig.~\ref{fig:n6890}). The continuum ratios have values  larger
than S1 at the nucleus, decreasing to S2 in the outer regions. These
results are similar to the ones obtained by Storchi-Bergmann, Bica
\& Pastoriza (1990).

{\bf NGC~7130}: Like NGC 5135, this galaxy has a Seyfert 2 nucleus
surrounded by a Starburst (Phillips \et 1983). Thuan (1984) presents
IUE spectra which show both high-excitation emission lines typical
of Seyfert 2s and absorption lines typical of Starburst. The
starburst component dominates the UV emission, with $\approx 75$\%
of the flux. Shields \& Filippenko (1990) showed that the narrow
line region have two kinematical components and emission line ratios
which vary from those of AGN in the nucleus to those typical of HII
regions outwards.

Both Ws and continuum ratios show a marked gradient
(Fig.~\ref{fig:n7130}).  The Ws change from values similar to S7 at
the nucleus to S6 outwards, while the continuum ratios change from
S6--S7 at the nucleus to S5--S6 outwards. This gradient is due to
the presence of a circumnuclear starforming region.

{\bf NGC~7582}: Morris \et (1985) found circumnuclear H$\alpha$
emission in this Seyfert 2 galaxy, suggesting a ring of rotating HII
regions in the galaxy plane. The Ws of this galaxy show a dilution,
changing from S7 at the nucleus to S3--S4 in the outer regions
(Fig.~\ref{fig:n7582}). The continuum ratios are similar to S3 at
the nucleus and NE ($r < 0$), but decrease to values similar to
S5--S6 towards $r > 0$. The continuum ratios to the SW are bluer
than what we would expect from the Ws results. The dilution measured
in this object is probably due to its inner star-forming ring.

\subsubsection{Radio Galaxies}

{\bf 3C~33}:  This is a FR~II radio source. Koski (1978) estimated
that the FC contributes to 19\% of the flux at 5000 \AA, observed
through a 2.7\sec$\times$4\sec\ aperture. Antonucci (1984) obtained
spectropolarimetric data, which shows wavelength independent
polarization, but no broad polarized emission lines.

The Ws of this radio-galaxy (Fig.~\ref{fig:pc33}) have values similar
to an E3 or E6 template, though W(MgI) is more similar to an E1
template. The continuum is redder at the nucleus, with values of an E2
template, decreasing to values typical of an E3 or E6 template
outwards. The FC contribution estimated by Koski (1978) would produce
a dilution of the Ws in the nuclear region. Our data do not support
this result, as the W-profiles are essentially flat.

{\bf Pks~0349-27}: The Ws (Fig.~\ref{fig:p349}) do not show a
gradient and have values typical of an E1--E2 template, whereas the
continuum ratios show a gradient from an E2 template at the nucleus
to a bluer, E3--E4 template outwards.

{\bf Pictor A}:  This is a FR~II Broad Line Radio Galaxy with a
strong double lobed radio source oriented along the E-W direction
(Christiansen \et 1977). Halpern \& Eracleous (1994) detected
double-peaked Balmer lines, not present in previous spectra. The Ws
and continuum ratios (Fig.~\ref{fig:pic}) show a strong dilution by
the FC at the nucleus. Their values correspond to a S4 template
outwards, except for the G-band, where the values are similar to
those of a S1 template in the outer regions. The remarks about the
uncertain continuum and W measurements in the nucleus of
MCG-02-33-034 also apply to Pictor A.

{\bf Pks~0634-20}:  This is a FR~II radio source. Simpson, Ward \&
Wilson (1995) did not detect a broad Pa$\alpha$\ line and show that
the K-L color, [OIII] and soft X-ray fluxes are consistent with the
spectral energy distribution of a quasar absorbed by A$_V \approx
30$\ mag. The Ws (Fig.~\ref{fig:p634}) are consistent with an E2
stellar population, without a gradient. The continuum ratios
indicate bluer templates, E3 at the nucleus, decreasing to values
typical of E6 outwards.

{\bf Pks~0745-19}:  According to Baum \& Heckman (1989), this narrow
line radio galaxy does not have a well defined FR class. The Ws
indicate an E8 template at the nucleus, with a gradient to values
typical of E4 outwards (Fig.~\ref{fig:p745}), suggesting dilution by
a FC. The continuum ratios indicate a redder, E2 stellar population.
However, it should be noticed that these continuum ratios were not
corrected by foreground reddening, because this galaxy lies too
close to the Galactic plane (${\rm b} < 4^{\circ}$).

\subsubsection{LINERs and Normal Galaxies}

{\bf NGC~1326}: Storchi-Bergmann \et (1996a) obtained narrow band
[OIII] and H$\alpha$\ images, which reveal a circum-nuclear ring at
$|r| \approx 6$\sec. As for NGC 1097 and other galaxies containing
star-forming rings, the Ws and color profiles of this galaxy
(Fig.~\ref{fig:n1326}) confirm the presence of a ring. The radial
behavior of the Ws and continuum ratios do not evidence the presence
of a FC in the nucleus. The Ws have values typical of a S3 template
at the nucleus, S6 at the ring and S2--S3 outwards. The continuum
ratios indicate a S1 template at the nucleus (redder than the one
indicated by the Ws), S6 at the ring, and S3--S4 outwards.

{\bf NGC~1598}:  According to Phillips \et (1984), the nuclear
spectrum of this galaxy is similar to that of LINERs, with evidence
of young stars at the nucleus, which is also surrounded by a ring of
HII regions. The Ws and continuum ratios (Fig.~\ref{fig:n1598})
indicate similar templates: S7 at the nucleus, in agreement with the
presence of young stars, and S5 from 4 to 6\sec\ reaching values
typical of a S6 template outwards.

{\bf NGC~2935}:  This is a normal galaxy with a nuclear ring of star
formation (Buta \& Crocker 1992). The Ws of Ca~K and G-band are
similar to S4 at the nucleus, S5 at the ring (5\sec) and S2
outwards, while Mg is similar to S2 at the nucleus, S3 at the ring
and S2 outwards (Fig.~\ref{fig:n2935}). The continuum ratios are
similar to S3 at the nucleus, S5 at the ring and S3 outwards.

{\bf NGC~4303}:  This is a Sersic-Pastoriza galaxy, classified as a
LINER by Huchra, Wyatt \& Davis (1982). Filippenko \& Sargent (1985)
observations show that HII regions are prominent in the nucleus and
circumnuclear region. They also found a substantially broader
component in each line and propose we are seeing a faint Seyfert 2
or LINER nucleus hidden by HII regions. 

The Ws are similar to a S5 template at the nucleus, decreasing to
values similar to a S6 template at the ring (5\sec;
Fig.~\ref{fig:n4303}). From 7\sec\ to 15\sec\ from the nucleus the
Ws are similar to a S4 template, decreasing to S6--S7 in the outer
regions.  The continuum ratios behave like the Ws, with the
exception of the nucleus and ring, where they indicate redder
templates, S4 and S5 respectively. In the other regions Ws and
continuum ratios indicate similar templates. Our results are in good
agreement with those of Bonatto \et (1989).

{\bf IC~4889}:  Phillips \et (1986) found H$\alpha$, [NII] and [OII]
emission in this normal elliptical galaxy.  The Ws show no
significant gradient, with values typical of an E1 template, while
the continuum ratios are E1 at the nucleus and decrease to values
similar to those of an E3 template in the outer regions
(Fig.~\ref{fig:i4889}).

{\bf NGC~5248}: Both Ws and continuum ratios show the presence of a
ring at $|r| = 10$\sec\ from the nucleus (Fig.~\ref{fig:n5248}). The
values of Ca K and G-band are typical of a S5 template at the
nucleus, S6 at the ring, S4 at 15\sec\ from the nucleus, decreasing
to values typical of S6--S7 templates outwards. Mg and continuum
ratio values indicate redder templates, S4 at the nucleus,  S5 at
the ring, S2--S3 at 15\sec, decreasing to S5--S6 outwards. Our
result for the nuclear stellar population is similar to the one
obtained by Bonatto \et (1989).

{\bf NGC~6221}:  This southern spiral galaxy was classified as a
Seyfert 2 by V\'eron, V\'eron \& Zuiderwijk (1981) and Pence \&
Blackman (1984), based on the detection of a faint broad H$\beta$
component, an [OIII] line broader than H$\beta$\ and to the
proximity to a hard X-ray source (Marshall \et 1979, Wood \et 1984).
 However, Schmitt \& Storchi-Bergmann (1995) recently proposed that
the identification of NGC6621 with the X-ray source is probably
wrong and that it would rather be related to the high-excitation
Seyfert 2 galaxy ESO138-G01. Pence \& Blackman (1984) data also show
that the nucleus is surrounded by a Starburst.

Our W profiles (Fig.~\ref{fig:n6221}) show a gradient from a S7
template at the nucleus to S5 at 5\sec\ and farther out, in agreement
with the presence of a nuclear starburst, confirmed by the gradient
also observed in the continuum ratios, which correspond to S6 at the
nucleus and to S5 at 5\sec\ and farther out.

\section{Discussion}

\label{sec:Discussion}

The present set of high quality long-slit spectra contains a wealth
of information on the properties of AGN and of the stellar
population of their host galaxies. In this section we discuss some
of the global results revealed by our analysis and their possible
implications.

\subsection{Stellar Populations}

\label{sec:StellarPop}

The values of the continuum colors and Ws of the stellar lines found
in our analysis cover a wide range, as can be seen in Figs.\ 3--44.
Even in inner regions of the bulge not affected by circumnuclear
star-formation or by a non-stellar FC, the galaxies exhibit
substantial differences in the Ws and colors, reflecting a variety
of stellar population properties. This was confirmed by our
characterization of the spectra in terms of Bica's (1988) templates,
which showed that, despite a predominance of S2 and S3
templates, every spectral type from S1 to S7 appears in the
representation of the spectra in the inner regions of our sample
galaxies.

This observed {\it variety} of stellar population properties
illustrates the inadequacy of using a single starlight template to
evaluate and remove the stellar component from AGN spectra.  The
importance of an accurate evaluation of the starlight component
cannot be underestimated. A `template mismatch' affects not only the
derived emission line fluxes, but also, and more strongly, the
determination of the strength and shape of a residual FC. We have
seen, for instance, that many of the nuclei in the present sample
show no evidence for dilution of the stellar lines in comparison
with off-nuclear spectra of the same galaxy. However, some dilution
would most probably be measured if a different galaxy was used as a
starlight template. The cases of Mrk 348, 573, 1210 and 3C33 are
illustrative in this respect. Whereas Koski (1978), Kay (1994) and
Tran (1995a) obtained large contributions of a FC comparing their
spectra with a normal galaxy template (\S\ref{sec:Indiv}), we find
little evidence for dilution of the absorption lines comparing the
nuclear Ws with those measured a few arcseconds outside the nucleus.

This brings up the question of which is the best way to estimate the
stellar component in an AGN spectrum. By far the most commonly
adopted procedure is to use the spectrum of a normal galaxy as a
starlight template. Given the variety of stellar populations found
in AGN, if a template galaxy is to be used at all then it has to be
chosen among a library contemplating a large variety of stellar
population characteristics. An alternative technique, implicitly
used in this paper, is to use an off-nucleus spectrum of the object
as the starlight template. This is  arguably a more robust method,
as the best representation of the stellar population in a given
galaxy is the galaxy itself! The extrapolation to zero radius,
however, involves the assumption that the stellar population in the
bulge of the host galaxy does not change dramatically in the inner
few arcseconds. A comparison between these two methods of starlight
evaluation will be presented in a future communication. 

\subsection{Dilution factors}

\label{sec:Dilution}

Figs.~3--44 demonstrate that several galaxies show signs of dilution
of the stellar lines in the nucleus. In this section we compare the
nuclear Ws to those a few arcseconds outside the nucleus in order to
estimate the fraction of the continuum associated with a featureless
spectral component. 

In order to obtain Ws representative of the neighbourhood of the
nucleus we have co-added typically four extractions, two in each
side of the nucleus, located at distances ranging from 2 to 12\sec\
(see Table~\ref{tab:dilution}). Extractions adjacent to the nucleus
were avoided whenever possible, since, due to seeing, they are often
contaminated by the nuclear spectrum. The typical distance of the
off-nuclear extractions was 4\sec, corresponding to physical radii
between $\sim 0.5$\ and 2 kpc for the distances of the galaxies in
the sample.

After constructing these off-nuclear templates and measuring their
Ws, the FC fraction of the total continuum spectrum at the
wavelength of an absorption line was estimated by

\begin{equation}
\label{eq:f_FC}
f_{{\rm FC}}(\lambda) = 
    \frac{ {\rm W}_{{\rm off-nuc}} - {\rm W}_{{\rm nuc}}}{
               {\rm W}_{{\rm off-nuc}} }
\end{equation}

\ni  where W$_{{\rm nuc}}$\ and W$_{{\rm off-nuc}}$\ are the Ws in
the nucleus and in the off-nuclear template respectively.

\begin{table*}
\begin{centering}
\begin{tabular}{lrrrrr}
\multicolumn{6}{c}{ DILUTION FACTORS} \\ \hline 
Object & Type  & $r$\ [\sec] & 
    \multicolumn{3}{c}{ $f_{{\rm FC}}(\lambda)$\ [\%] } \\ 
    &  & & 3930 \AA & 4301 \AA & 5176 \AA \\ \hline
NGC 1097     & Lin-1 & 4.0       &  6$\pm$5 & -            & -            \\
NGC 1598$^c$ &   Lin & 4.0       & 73$\pm$4 & 39$\pm$6     & 39$\pm$6     \\
NGC 7213     & Lin-1 & 3.6--7.2  & 31$\pm$4 & 33$\pm$6$^b$ & -            \\ 
ESO 362-G18  &     1 & 7.0--12.0 & 46$\pm$6 & 59$\pm$5$^b$ & 22$\pm$6$^a$ \\ 
NGC 6814     &     1 & 5.0--10.8 & 47$\pm$4 & 53$\pm$6$^b$ & 22$\pm$5$^a$ \\ 
MCG-02-33-034&     1 & 4.0--7.0  & 76$\pm$5 & 100$^b$      & 100$^a$      \\ 
NGC 6860     &     1 & 4.0--7.0  & 51$\pm$4 & 64$\pm$6$^b$ & 50$\pm$5     \\ 
Pictor A     &  BLRG & 3.7--6.5  & 77$\pm$3 & 100$^b$      & 100$^a$      \\ 
Mrk 348      &     2 & 1.9--3.7  &  1$\pm$5 & 9$\pm$6      &  9$\pm$7$^a$ \\
Mrk 1210     &     2 & 4.0--8.0  & 11$\pm$6 & 12$\pm$7     & 26$\pm$7$^a$ \\ 
IC 1816      &     2 & 4.0       & 13$\pm$5 & 15$\pm$6     & 30$\pm$5$^a$ \\ 
ESO 362-G8$^c$&    2 & 4.0--7.0  & 25$\pm$6 & 15$\pm$7     & 11$\pm$5     \\ 
NGC 1358     &     2 & 4.0--6.0  & 0$\pm$5  & 1$\pm$7      & -            \\ 
NGC 5643     &     2 & 3.6--5.4  & 19$\pm$4 & 33$\pm$6     & 33$\pm$4$^a$ \\ 
NGC 6890     &     2 & 1.8--3.6  & 17$\pm$5 & 39$\pm$7     & -            \\
CGCG~420-015 &     2 & 2.0--4.0  &  4$\pm$5 & 15$\pm$6     &  5$\pm$7$^a$ \\ 
IRAS~11215-2806&   2 & 4.0--7.0  & 24$\pm$4 & 34$\pm$5     & 30$\pm$5$^a$ \\ 
MCG-05-27-013&     2 & 4.0--7.0  & 10$\pm$5 & 12$\pm$7     & -            \\ 
NGC 5135$^c$ &  2-SB & 3.6--7.2  & 37$\pm$8 & 51$\pm$7     & 36$\pm$8$^a$ \\ 
NGC 7130$^c$ &  2-SB & 3.6--7.2  & 47$\pm$5 & 62$\pm$6     & 35$\pm$5$^a$ \\ 
NGC 7582$^c$ &  2-SB & 4.0--8.0  & 70$\pm$4 & 60$\pm$6     & 57$\pm$5$^a$ \\ 
NGC 6221$^c$ &    SB & 3.6--5.4  & 52$\pm$4 & 73$\pm$4     & 49$\pm$6$^a$ \\ 
PKS 0745-19  &  NLRG & 1.9       & 55$\pm$13& 3$\pm$20     & 15$\pm$11$^a$\\ \hline
\end{tabular}
    \caption{
    Column 3 lists the range in distances from the nucleus used to
compute W$_{{\rm off-nuc}}$. Columns 4--6 list the FC fractions
corresponding to the Ca~K, G-band and Mg lines respectively. 
    \newline $^a$ contamination by [FeVII] and/or [NI].
    \newline $^b$ contamination by broad H$\gamma$.
    \newline $^c$ dilution due to young stars in the nucleus.
    \label{tab:dilution}
    }
\end{centering}
\end{table*}

Table~\ref{tab:dilution} presents the results of this analysis.
Galaxies not listed in Table~\ref{tab:dilution} show no signs of
dilution, and even for some objects in the table (e.g., NGC 1097,
Mrk 348) the evidence for dilution is only marginal---by `marginal'
we mean within $2 \sigma$\ of a spurious dilution caused by the
errors in the W measurements. We recall that dilution of the Ws by
less than 10\% cannot be reliably detected with this method due to
the errors in the Ws. We concentrated this analysis in the Ca~K,
G-band and Mg lines. However, as indicated in the Table, the Mg line
is very often contaminated by [Fe~VII] and/or [NI] in the nuclear
spectrum, producing an {\it apparent} dilution. In such cases the
$f_{{\rm FC}}(5176\AA)$\ fractions have to be taken as upper limits
of the actual dilution. The line which is less affected by such
spurious effects is Ca~K. Empty slots in Table~\ref{tab:dilution}
occur whenever an increase in the W, instead of dilution, was
observed.

Of the broad-lined objects in our sample, only NGC 526a does not
exhibit a systematic dilution of the Ws in the nucleus. The other
possible exception is NGC 1097, which does show some dilution in the
Ca~K line, but at a marginal level. The highest values of  $f_{{\rm
FC}}$\ occur for MCG-02-33-034 and Pictor A, where it reaches nearly
80\% at the wavelength of Ca~K. However, as discussed in \S3, these
values are not as reliable as for other Seyfert 1s due to the
uncertain positioning of the continuum for these two objects.
Indeed, their nuclear spectra are so complex that the G-band and Mg
lines are completely filled by emission, preventing a determination
of $f_{{\rm FC}}$\ at the corresponding wavelengths. 

Perhaps the most intriguing aspect of Table~\ref{tab:dilution} is
the low degree of dilution detected in Seyfert 2s. Of the 20 Seyfert
2s in our sample, 13 are listed in Table~\ref{tab:dilution}, and
only 9 of these (IC 1816, ESO 362-G8, NGC 5135, 5643, 6890, 7130,
7582, MCG -05-27-013 and IRAS 11215-2806) have dilution convincingly
detected. As reviewed in \S\ref{sec:Indiv}, NGC 5135, 7130, 7582 and
ESO 362-G8 are composite nuclei, showing features of both Seyfert 2s
and Starbursts. In these cases, as for the LINER NGC 1598
(Fig.~\ref{fig:n1598}), the dilution is almost certainly dominated
by the young stellar component, not a genuine AGN continuum.  This
lowers the count to 5 objects with detected dilution out of 16
`pure' Seyfert 2s. Marginal indications of dilution were found in
Mrk 348, Mrk 1210, NGC 1358 and CGCG 420-015. In the other objects,
the FC, if present at all, contributes 10\% or less in the optical.
This result is further discussed below.

\subsubsection{Implications for FC2}

\label{sec:FC2}

The low levels or lack of dilution in Seyfert 2s found in the
present analysis is an interesting result in light of the current
debate over the existence and origin of a FC in Seyfert 2s and its
implications for unified models. It has been shown that the FC in
Seyfert 2s cannot be simply the continuum of a hidden AGN scattered
into our line of sight, otherwise broad emission lines should also
be seen, in contradiction with the very definition of type 2
Seyferts (Cid Fernandes \& Terlevich 1992, 1995, Heckman \et 1995).
Furthermore, the low levels of polarization (Miller \& Goodrich
1990), and the observation of reflected broad lines substantially
more polarized than the FC (Tran 1995c) are also in conflict with
the simple obscuration/reflection picture.

These facts lead to the idea that a {\it second} FC component (dubbed
FC2 by Miller 1994) must be present in Seyfert 2s. The nature of this
component remains unknown, though both a circumnuclear starburst (Cid
Fernandes \& Terlevich 1995, Heckman \et 1995) and free-free emission
from the scattering region (Tran 1995c) have been suggested.

Our observations seem to indicate that a FC (and consequently FC2)
is {\it not} an universal phenomenon in Seyfert 2s, and that
previous determinations of the FC strength have been overestimated.
A smaller FC contribution would alleviate, if not completely solve,
all problems listed above.

A good way to further investigate this issue would be to obtain
spectropolarimetry for the Seyfert 2s in the present sample. We
expect galaxies for which no dilution of the Ws was detected to show
little or no polarization. IC 1816, NGC 5643, NGC 6890, MCG
-05-27-013 and IRAS 11215-2806, on the other hand, should exhibit
large polarizations and Seyfert 1 features in the polarized spectra.
In the cases of composite nuclei, spectropolarimetry could help
disentangling the starburst and Seyfert 2 components and possibly
reveal a hidden AGN.

As a final word of caution, we note that while throughout this paper
we have attributed the dilution of absorption lines in the nucleus
of active galaxies to a FC, dilution can also be caused by young
stars, as shown to occur in circumnuclear star forming rings (e.g.,
Figs.~\ref{fig:n1097}, \ref{fig:n1433} and \ref{fig:n1672}) and in
the nuclei of the Starburst galaxies NGC 6221 and 7130
(Figs.~\ref{fig:n6221} and \ref{fig:n7130}). It is a known problem
in optical spectroscopy of AGN that it is often difficult to
distinguish between a young stellar population and a FC (e.g.,
Miller \& Goodrich 1990, Cid Fernandes \& Terlevich 1992, Kay 1994).
UV spectroscopy offers the most direct way of removing this
ambiguity  (e.g., Leitherer, Robert \& Heckman 1995, Heckman \et
1995). 

\subsubsection{Metallicity gradients}

\label{sec:Metallicity} 

In measuring the dilution of absorption lines with respect to their
strengths $\sim 4$\sec\ (or $\sim 1$~kpc) outside the nucleus we are
making the assumption that the stellar population in the bulge of
the host galaxy does not change appreciably from $r \sim 4$\sec\ to
$r = 0$. This approximation might not be valid if the galaxy
presents a strong metallicity ($Z$) gradient, leading to an increase
of the Ws towards the nucleus (e.g., D\'{\i}az 1992). If such a
gradient is present, then a flat W profile as the one found in many
of our galaxies could reflect the compensating effects of a FC
diluting the enhanced metal lines in the nucleus\footnote{We thank
Dr.\ R. Terlevich for pointing this out to us.}. In this section we
investigate this possibility by means of an approximate treatment of
the potential effects of a metallicity gradient.

To estimate the $Z$-gradient we have used $\Delta \log Z / \Delta \log
r = 0.2$, a typical value obtained for $r$ $\gapprox$ 3\sec\ in
elliptical galaxies (Davies, Sadler \& Peletier 1993 and references
therein). Assuming that the nuclear stellar population sampled in our
2\sec$\,\times\,$2\sec\ aperture is characterized by a mean radius of
0.5\sec, and considering a typical off-nuclear extraction distance of
4\sec, this corresponds to an increase of $\sim 0.2$\ dex in $Z$\ from
$r = 4$\sec\ to the nucleus, corresponding to the inner $\sim 1$\ kpc
of the galaxy. This value is also typical of $Z$-gradients mapped from
the radial variations of the O abundances in HII regions in galactic
disks (e.g., Vila-Costas \& Edmunds 1992 and references therein). To
convert this gradient to a W-gradient we used the results of Jablonka,
Alloin \& Bica (1992). We adopted their calibration for an old,
globular cluster population, corresponding to the steepest W-$Z$\
relation, thus resulting in the largest possible W-variations for a
fixed $Z$-gradient.

This exercise resulted in the following variations of the Ws due to
a $Z$-gradient: $\Delta {\rm W(Ca K)} = 1.2$\ \AA,
$\Delta$W(G-band)$~= 0.8$\ \AA\ and $\Delta {\rm W(Mg)} = 0.5$\ \AA.
Given that the assumptions above were all made in the sense of
maximizing the effects of a $Z$-gradient, these variations should be
regarded as upper limits. These are small variations, being of the
same order of the errors in W (\S\ref{sec:Error}). The net
variations with respect to the typical values of W found in our
off-nuclear extractions ($\sim 13$ \AA\ for Ca K and $\sim 8$ \AA\
for both G-band and Mg) would correspond to changes of, at most,
10\% in W(Ca K), 8\% in W(G-band) and 6\% in W(Mg).

These results, while certainly not closing the question, seem to
indicate that $Z$-gradients do not constitute a serious caveat to
the interpretation of W-profiles in AGN. Only galaxies with a FC
weaker than 6--10\% could have a dilution of the metal lines masked
by the $Z$-gradient. Whilst we do not think that a compensation of
the effects of a FC and a $Z$-gradient provides a viable
interpretation for the ubiquitous absence of dilution in our sample
galaxies, particularly in Seyferts 2, further investigations on the
strength and detectability of $Z$-gradients in active galaxies would
clearly be wellcome, as this issue also bears on the question of
whether an off-nuclear extraction provides a better starlight
template for the nuclear stellar population than a normal galaxy
(\S\ref{sec:StellarPop}).

\subsection{Color gradients}

\label{sec:Colors}

The continuum ratios in the sample galaxies exhibit an interesting
radial behavior. As seen in Figs.~3--44, the $F_{5870}/F_{4020}$\
and $F_{4630}/F_{3780}$\ ratios very often peak in the nucleus,
indicating a {\it redder} continuum. This seems to be an extremely
ubiquitous behavior. The two kinds of objects which do not follow
this rule are those containing nuclear starbursts, namely, NGC 1598,
6221, 5135, 7130, and broad-lined objects, the most extreme examples
being MCG -02-33-034 and Pictor A. In particular, {\it all} Seyfert
2s, with the exception of NGC 5643 which shows a complex color
profile, have nuclei redder than the extranuclear regions. This
effect is not an artifact of our pseudo continuum determination, as
we verified that the color-profiles are nearly identical if instead
of the pseudo-continuum one defines line-free bands close to the
pivot points to compute mean fluxes and flux ratios.

This is a somewhat surprising result, given that the general prejudice
is that AGN are bluer than normal galaxies, and as such they should
also be bluer than their own bulges. On the other hand, if taken as
evidence of reddening, the color profiles would fit in to the
currently favored view that the nuclear regions of Seyfert 2s are rich
in dust. Some support to this interpretation comes from the behavior
of the Na line (partially produced in interstellar medium), whose W
profiles in some galaxies, notably ESO 417-G6, NGC 1386, ESO 362-G8,
NGC 6300 and NGC 7582, peaks in the nucleus, mimicking the color
profiles. Examination of the radial behavior of the
H$\alpha$/H$\beta$\ emission line ratio would provide a test of this
interpretation.

Another intriguing result was found from the comparison of the
stellar population templates inferred from the Ws with those
inferred from the continuum ratios for the Seyfert 2s. With few
exceptions, the off-nuclear colors indicate a spectral template {\it
bluer} than that indicated by the absorption lines. At present we
cannot give a clear explanation for this apparent contradiction, but
we speculate on two possible interpretations. (1) The effect is due
to a strong contribution from metal rich (thus strong lined)
intermediate age (thus blue) stars, a population not well
represented by any of templates S1--S7. This would be a genuine
stellar population effect. (2) The blue continuum is due to a
scattered AGN component while the absorption lines originate in a
metal rich population. This would result in an {\it extended
dilution} of the stellar features, whose Ws would therefore be
underestimated, leading to the incorrect attribution of a spectral
template less metallic than the actual population. Detailed modeling
will be required to test these possibilities.

\section{Summary}

\label{sec:Conclusions}

We have reported the results of a high signal-to-noise ratio long
slit spectroscopy study of 39 active galaxies and 3 normal galaxies.
The run of stellar absorption lines and continuum colors with
distance from the nucleus was investigated using a consistent
methodology, aiming a global characterization of the stellar
populations and the detection of a nuclear FC through the dilution
of conspicuous absorption lines with respect to off-nuclear
extractions. Our results can be summarized as follows.

(1) Radial variations of the stellar populations were detected in
most cases. Star-forming rings, in particular, were found to leave
clear imprints in the equivalent widths and color profiles.

(2) The stellar populations in the inner arcseconds of active
galaxies are varied, as inferred from the wide spread of equivalent
widths and colors measured in this work. This fact alone raises
serious doubts as to whether it is appropriate to use a single
starlight template to evaluate and remove the stellar component from
AGN spectra.

(3) The equivalent width profiles were used to measure the dilution
of the absorption lines in the nucleus with respect to off-nuclear
spectra. Dilution by a nuclear FC was detected in most broad lined
objects in the sample.

(4) Galaxies undergoing star-formation in their nuclei, including
composite starburst $+$\ Seyfert 2 objects, also show dilution of
the nuclear absorption lines. The dilution in these cases is due to
young stars, not to a FC. These galaxies also show a clear color
gradient, their nuclei being bluer than off-nuclear regions.

(5) About 50\% of the Seyfert 2s in the sample show no signs of
dilution of their absorption lines in the nucleus, indicating that
a FC, if present, contributes less than 10\% of the total flux in
the optical range. Possible consequences to the (controversial)
nature of the FC in Seyfert 2s were discussed.

(6) All but one of the Seyfert 2s present a color gradient in the
sense that the spectrum gets {\it redder} as we approach the
nucleus, indicative of the presence of dust.

(7) There is an apparent discrepancy between the spectral templates
assigned from the equivalent widths of the absorption lines and from
the continuum colors, an effect observed in the off-nuclear spectra
of most Seyfert 2s in the present sample. Tentative interpretations
involving either a stellar population effect or an extended blue
continuum were briefly outlined.


\vskip 3mm

{\bf Acknowledgments:} We thank Dr. E. Bica for useful discussions
and remarks on an early version of this manuscript. HRS acknowledges
the hospitality of the Department of Physics at Florian\'opolis.
Support from FAPEU-UFSC is also duly acknowledged. RCF work was
partially supported by CNPq under grant 300867/95-6. This research
has made use of the NASA/IPAC Extragalactic Database (NED) which is
operated by the Jet Propulsion Laboratory, CALTECH, under contract
with the National Aeronautics and Space Administration.



\end{document}